\documentclass[12pt,letterpaper]{article}
\usepackage{graphicx}
\usepackage{amsmath}
\usepackage{amssymb}
\let\mathbbm\relax
\usepackage{hyperref}

\usepackage[all]{xy}
\CompileMatrices
\def\sp#1{\ar@{-}[#1]}
\def\ns#1{\ar@{.}[#1]}
\def\rigid#1{\fbox{$#1$}}

\usepackage[table]{xcolor}

\usepackage{tikz}

\numberwithin{equation}{section}

\def\bC{\mathbb{C}}
\def\bH{\mathbb{H}}
\def\bR{\mathbb{R}}
\def\bZ{\mathbb{Z}}

\def\cC{\mathcal{C}}
\def\cN{\mathcal{N}}
\def\fg{{\mathfrak{g}}}
\def\fh{{\mathfrak{h}}}
\def\fl{{\mathfrak{l}}}
\def\fz{{\mathfrak{z}}}

\def\SU{\mathrm{SU}}
\def\su{\mathfrak{su}}
\def\SO{\mathrm{SO}}
\def\OO{\mathrm{O}}
\def\so{\mathfrak{so}}
\def\Sp{\mathrm{Sp}}
\def\Spin{\mathrm{Spin}}
\def\U{\mathrm{U}}
\def\u{\mathfrak{u}}
\def\coxeter{h^\vee}
\def\Higgs{\text{\textsc{Higgs}}}
\def\Coulomb{\text{\textsc{{Coulomb}}}}
\def\Pfaff{\mathop{\mathrm{Pfaff}}}
\def\Ind{\mathop{\mathrm{Ind}}\nolimits}
\def\diag{\mathop{\mathrm{diag}}}
\def\rank{\mathop{\mathrm{rank}}}
\def\grade{\mathop{\mathrm{grade}}}
\def\tr{\mathrm{tr}}
\def\sl{\mathfrak{sl}}
\def\op#1{\displaystyle{\mathop{\mbox{\large $\oplus$}}_{#1}}}

\def\Type{J}
\def\type{\mathfrak{j}}
\def\emptyset{\varnothing}
\def\dual#1{{#1^\vee}}
\def\Sommers{\cC}

\def\Flavour{{F}}
\def\flavour{\mathfrak{f}}
\def\dimH{\scalebox{.8}{$\dim\Higgs$}}
\def\dimC{\scalebox{.8}{$\dim\Coulomb$}}
\def\pz{\phantom{0}}
\let\plusminus\pm
\def\pm{\phantom{-}}% this was a bad choice.

\begin{document}

\begin{titlepage}

\begin{flushright}
ICTP-SAIFR/2012-002\\
IPMU-12-0026\\
UTTG-04-12\\
TCC-004-12
\end{flushright}
\vskip 1cm

\begin{center}
{\Large \bfseries
Nilpotent orbits and  codimension-two \\[3mm]
defects of  6d $\cN{=}(2,0)$ theories
}

\vskip 1.2cm

Oscar Chacaltana$^\natural$,
 Jacques Distler$^\sharp$, 
and Yuji Tachikawa$^\flat$

\bigskip
\bigskip

\begin{tabular}{ll}
$^\natural$ & ICTP South American Institute for Fundamental Research,\\
& Instituto de F\'isica Te\'orica, Universidade Estadual Paulista,\\
& 01140-070 S\~{a}o Paulo, SP, Brazil
\\[.5em]
$^\sharp$ & Theory Group and Texas Cosmology Center,\\
&
Department of Physics, University of Texas at Austin, Austin, TX 78712, USA \\[.5em]
$^\flat$  & Institute for the Physics and Mathematics of the Universe, \\
& University of Tokyo,  Kashiwa, Chiba 277-8583, Japan
\end{tabular}

\vskip 1.5cm

\textbf{abstract}
\end{center}

\medskip
\noindent
We study the local properties of a class of codimension-2 defects of the 6d $\cN{=}(2,0)$ theories of type $\Type=A,D,E$ 
labeled by nilpotent orbits of a Lie algebra $\fg$, where $\fg$ is determined by $\Type$ and the outer-automorphism twist around the defect.
This class is a natural generalisation of the defects of the 6d theory of type $\SU(N)$ labeled by a Young diagram with $N$ boxes.
For any of these defects, we 
determine its contribution to the dimension of the Higgs branch,
to the Coulomb branch operators and their scaling dimensions,
to the 4d central charges $a$ and $c$, and to the flavour central charge $k$.
\bigskip
\vfill
\end{titlepage}

\setcounter{tocdepth}{2}
\tableofcontents
%\newpage

\section{Introduction and summary}

The six-dimensional $\cN{=}(2,0)$ theories are interacting non-gravitational theories whose basic massive excitations at a generic point of the vacuum moduli space are strings charged under two-form potentials whose field strengths are self-dual. Consistency requires that the charge lattice of the strings be simply-laced \cite{Henningson:2004dh}, and, indeed, the 6d (2,0) theories of type $\Type=A_n$, $D_n$, $E_{6,7,8}$ are realised as the low-energy limits of Type IIB strings on ALE spaces of the corresponding type \cite{Witten:1995zh}. The (2,0) theories of type $A_n$ and $D_n$ also arise as the low-energy limit of the worldvolume theory of M5-branes \cite{Strominger:1995ac,Dasgupta:1995zm,Witten:1995em}.

Although no satisfactory Lagrangian description of these theories is known, their mere existence sheds light on various non-perturbative properties of four-dimensional supersymmetric field theories, e.g.~the S-duality of  $\cN{=}4$ super Yang-Mills \cite{Witten:1995zh,Vafa:1997mh}, the physical realisation of the Seiberg-Witten curve of $\cN{=}2$ gauge theories \cite{Klemm:1996bj,Witten:1997sc}, and Argyres-Seiberg duality \cite{Argyres:2007cn,Argyres:2007tq,Gaiotto:2009we}.
In these constructions, the 4d theory is realised as the  compactification of the 6d theory on a Riemann surface $C$ with punctures; the punctures are the locations of half-BPS codimension-two defects of the 6d theory that are extended over 4d spacetime.  
In a particularly nice class of such defects, these are labeled by a homomorphism $\rho$ from $\su(2)$ to $\type$, or equivalently by a nilpotent orbit in $\type$.
For $\type=\su(N)$, the label becomes a Young diagram with $N$ boxes. 
We can also consider  twisted defects, around which we have an action of an outer-automorphism $o$ of $\Type$.
Then the defects are  labeled by homomorphisms $\rho$ from $\su(2)$ to $\fg$, where $\dual\fg$ is the subalgebra of $\type$ invariant under $o$, see Table~\ref{list}.
The study of these untwisted and twisted defects was initiated in \cite{Gaiotto:2009we}, and further developed in various papers including \cite{Gaiotto:2009gz,Tachikawa:2009rb,Benini:2009gi,Nanopoulos:2009xe,Gaiotto:2009hg,Nanopoulos:2009uw,Drukker:2010jp,Chacaltana:2010ks,Chacaltana:2011ze,Tachikawa:2010vg,Nishinaka}.

\begin{table}
\[
\begin{array}{|cccc|}
\hline
\Type & o & G^\vee & G \\
\hline
A_{2n-1} & \bZ_2 & C_n & B_n \\
D_{n} & \bZ_2 & B_{n-1} & C_{n-1} \\
\hline
\end{array}
\qquad
\begin{array}{|cccc|}
\hline
\Type & o & G^\vee & G \\
\hline
D_{4} & \bZ_3 & G_2 & G_2 \\
E_{6} & \bZ_2 & F_4 & F_4\\
\hline
\end{array}
\]
\caption{The relation among the type $\Type$, the twist $o$, the group $G$ and its dual $G^\vee$. 
When the outer-automorphism twist $o$ is trivial, $\Type=G=G^\vee$. \label{list}}
\end{table}

In these papers, the analysis was restricted to the defects of the 6d theories of type $A$ and $D$, where their brane realisations were fully used. The aim of this paper is to distill the results obtained by these methods and phrase them into a form independent of string theory constructions, which makes our framework applicable to twisted and untwisted defects in 6d theories of any type.
We will give algorithms to determine the following data for a homomorphism $\rho$:
\begin{itemize}
\item 
the dimension of the local Higgs branch\footnote{In general, it is not always possible to completely Higgs the gauge symmetry. By ``Higgs branch" we mean the branch where gauge symmetry is ``maximally Higgsed", \emph{i.e.}, where the number of massless abelian vector multiplets is the smallest. This maximally Higgsed branch was called the Kibble branch in \cite{Hanany:2010qu}.}, $\dim\Higgs(\rho)$, \eqref{dimHiggs}, 
\item 
the dimension of the local Coulomb branch \eqref{3dCoulomb}, $\dim\Coulomb(\rho)$, \eqref{twisted3dCoulomb},
\item 
the local contribution to the graded Coulomb branch, Sec~\ref{generalrecipe},
\item 
the flavour symmetry $\Flavour(\rho)$ and its level $k$, \eqref{4dk}, and\footnote{
When $\Flavour(\rho)$ has an $\mathfrak{sp}(n)$ factor, there can be the $\bZ_2$ global anomaly \cite{Witten:1982fp}. A half-hypermultiplet in a pseudoreal representation $R$, of $\Sp(n)$, contributes $k=l(R)$ to the level of the $\mathfrak{sp}(n)$. At the same time, the $\Sp(n)$ has a $\bZ_2$ global anomaly if and only if $l(R) = 1\pmod{2}$. 
Thus, when there is some S-duality frame in which the defect $\rho$ lies on a 3-punctured sphere consisting of free (half-)hypermultiplets, the level $k$ of $\mathfrak{sp}(n)$  is odd if and only if the corresponding $\Sp(n)$ factor has a $\bZ_2$ global anomaly. This is easily confirmed using quiver description when $\fg$ is of type $C$ or $D$. We do not have a definite derivation for other $\fg$. } 
\item 
the local contribution to the effective number of vector multiplets, $n_h(\rho)$,  \eqref{nh}, and hypermultiplets, $n_v(\rho)$, \eqref{nv}, which are linear combinations of the central charges $a$ and $c$.
\end{itemize}
The quantities listed above will be deduced using the two setups below:

\begin{figure}

\def\cigar{%
\vcenter{\hbox{\begin{tikzpicture}[scale=.5]
 \node  (A) at (-4, 2) {};
 \node  (B) at (3, 2) {};
 \node  (C) at (3, 0) {};
 \node  (D) at (-4, 0) {};
 \node  (P) at (1, 2) {};
 \node  (Q) at (1, 0) {};
 \node  (V) at (3.5, 1) {$\bullet$};
 \node  (E) at (-4, 1) {};
 \node  at (0, 1.5) {$S^1$};
 \draw (A.center) to (B.center);
 \draw [out=0,in=0,looseness=2] (B.center) to (C.center);
 \draw (C.center) to (D.center);
 \draw [out=0,in=0, looseness=1] (P.center) to (Q.center);
 \draw [out=180,in=180, looseness=1,style=dashed] (P.center) to (Q.center);
 \draw [style=dashed,color=red](V.center) to (E.center);
\end{tikzpicture}}}}
\def\redseg{%
\vcenter{\hbox{\begin{tikzpicture}[scale=.5]
 \node  (a) at (3.5, 0) {$|$};
 \node  (b) at (-4, 0) {};
 \draw [color=red](a.center) to (b.center);
\end{tikzpicture}}}}
\def\greenseg{%
\vcenter{\hbox{\begin{tikzpicture}[scale=.5]
 \node  (a) at (3.5, 0) {$|$};
 \node  (b) at (-4, 0) {};
 \draw [color=green!50!blue](a.center) to (b.center);
\end{tikzpicture}}}}
\begin{center}
\small
\textbf{1.}  6d $\cN{=}(2,0)$ theory of type $\Type$ on\[
\bR^{2,1}\times\cigar\times \tilde S^1
\]

\vspace{-5pt}

\begin{tabular}{ccccc}
& \qquad $\tilde S^1$  $\swarrow$ & &&  $\searrow$  $\  S^1$ \qquad \hbox{}\\[5pt]
\textbf{5.} & 5d $\cN{=}2$ SYM with group $\Type$ on & & \textbf{2.} & 5d $\cN{=}2$ SYM with group $G$ on \\
& $\bR^{2,1}\times\cigar$ & & & $\bR^{2,1}\times\greenseg\times\tilde S^1$ \\[10pt]
& $S^1$ \raisebox{-5pt}{\rotatebox{90}{$\longleftarrow$}} & &&\raisebox{-5pt}{\rotatebox{90}{$\longleftarrow$}} $\ \tilde S^1$\\[5pt]
\textbf{4.} & 4d $\cN{=}4$ SYM with group $\dual{G}$ on & & \textbf{3.} & 4d $\cN{=}4$ SYM with group $G$ on \\
& $\bR^{2,1}\times\redseg$ & $\Leftarrow$ S-dual $\Rightarrow$ & & $\bR^{2,1}\times\greenseg$ \qquad \hbox{} 
\end{tabular}
\end{center}

\caption{Chain of dualities used to study codimension-two defects of the 6d theories. \label{dualitychain}}
\end{figure}

\paragraph{Setup 1:} Consider the following configuration. (The steps are also shown in Fig.~\ref{dualitychain}.) Take a half-BPS codimension-two defect of the 6d theory of type $\Type$. 

\begin{enumerate}
\item  Consider 6d spacetime to be of the form $\bR^{2,1}\times$ (cigar)  $\times \tilde S^1$. Denote by $S^1$ the circle around the cigar. Let us then put the defect at the tip of the cigar. Let $o$ be the outer-automorphism monodromy that arises when we loop around the defect; this monodromy can be trivial.\footnote{We do not discuss the $\bZ_2$ outer-automorphism twists of the $A_{2n}$ theories, because they are subtle, and we do not understand them well enough. See, e.g., Sec.~4 and Sec.~6 of \cite{Tachikawa:2011ch} for the subtleties involved.} Let $\dual{G}$ be the part of $\Type$ invariant under $o$, and $G$ be the Langlands dual of $\dual{G}$.  If $o$ is trivial, then $G=\dual{G}=\Type$. See Table~\ref{list} for the other cases.
\item Reduce along $S^1$. We get 5d $\cN{=}2$ super Yang-Mills with gauge group $G$ on $\bR^{2,1}\times$ (a half-line) $\times \tilde S^1$. 
The codimension-two defect in 6d becomes a boundary condition at the end of the half-line, which for the defects we consider in this paper is given by\footnote{These are \emph{not} the only possible half-BPS boundary conditions of the 5d theory arising from half-BPS codimension-2 defects of the 6d theory. Many others are known. We concentrate on these boundary conditions because they can be uniformly described.}
\begin{equation}
\Phi_{1,2,3}(s)\sim \rho(\tau_{1,2,3})/s. \label{5dBC}
\end{equation} Here, $\Phi_i$ are the adjoint scalars of the gauge theory, $s$ is the distance to the boundary, $\tau_{1,2,3}$ are 1/2 the Pauli matrices, and $\rho$ is a homomorphism $\rho:\su(2)\to \fg$. 
We call $\rho$ the \emph{Nahm pole}.
\item Reduce further along $\tilde S^1$. Now we have a boundary condition for $\cN{=}4$ super Yang-Mills with gauge group $G$, given basically by \eqref{5dBC}, and which is among the class of boundary conditions studied in \cite{Gaiotto:2008sa}.
\item Now, invert the order of the reductions on $S^1$ and $\tilde{S}^1$, which amounts to S-duality of $\cN{=}4$ super Yang-Mills. In the S-dual description, we have different boundary condition for $\cN{=}4$ super Yang-Mills with gauge group $\dual{G}$, namely, a 3d $\cN{=}4$ superconformal theory, $T^\rho[\fg]$, living at the 3d boundary of the 4d bulk, and which has flavour symmetry $\dual{G}$, coupled to 4d super Yang-Mills \cite{Gaiotto:2008ak}.
\item Go back one step in the reduction. It is reasonable to assume that the codimension-2 defect of the 5d theory at this stage is given by coupling the 3d theory $T^\rho[\fg]$  to the defect. The presence of this 3d theory at the defect produces a pole in the adjoint scalar field, \begin{equation}
\Phi(z)\equiv \Phi_4(z)+i \Phi_5(z) \sim \tilde \rho(\sigma^+) /z,
\end{equation} where $z$ is a local complex coordinate for the cigar so that the tip is at $z=0$, and  $\tilde\rho$ is a new homomorphism, $\tilde\rho: \su(2)\to {}\dual{\fg}$, determined by the properties of $T^\rho[\fg]$. We call $\tilde{\rho}$ the \emph{Hitchin pole}.
\item Unfortunately, with our current understanding, we cannot go back one more step to say exactly what is at the defect of the 6d theory.
However, we can still study how the worldvolume fields $\phi_k(z)$ of dimension $k$ behave there, by studying the behavior of $p_k(\Phi(z))$ where $p_k$ is a degree-$k$ invariant polynomial of $\fg$.
\end{enumerate}

When $\fg=\su(N)$, a homomorphism $\rho:\su(2)\to \su(N)$, such as the Nahm pole, determines how the fundamental $N$-dimensional representation of $\su(N)$ decomposes under $\su(2)$ into the irreducibles $N=n_1+\cdots+n_k$, which determines a partition $[n_i]$ of $N$. On the other hand, the Hitchin pole $\tilde\rho$ is associated to a partition $[\tilde n_i]$, such that the Young diagram for $[\tilde n_i]$ is the transpose of the Young diagram for $[n_i]$.  We want to study the map
\begin{equation}
d: \{\text{$\rho:\su(2)\to \fg$ up to conjugacy}\} \to
\{\text{$\tilde\rho:\su(2)\to \dual{\fg}$ up to conjugacy}\}.\label{d-map}
\end{equation}
We will argue below that this $d$ is the \emph{Spaltenstein map}, well known in the nilpotent orbit literature.
This map $d$ fails to be a bijection (except for $\fg=\su(N)$), but it satisfies $d^3=d$. 
It is then possible, say, that defects corresponding to two distinct Nahm poles $\rho$ and $\rho'$ map to the same Hitchin pole, $d(\rho)=d(\rho')$.  
We will see that, when this happens, certain discrete groups $\Sommers(\rho)\neq \Sommers(\rho')$ introduced by Sommers and Achar \cite{Sommers,AcharSommers,Achar} enter in the description of the Hitchin system, and affect the properties of the Coulomb branch.

\begin{figure}
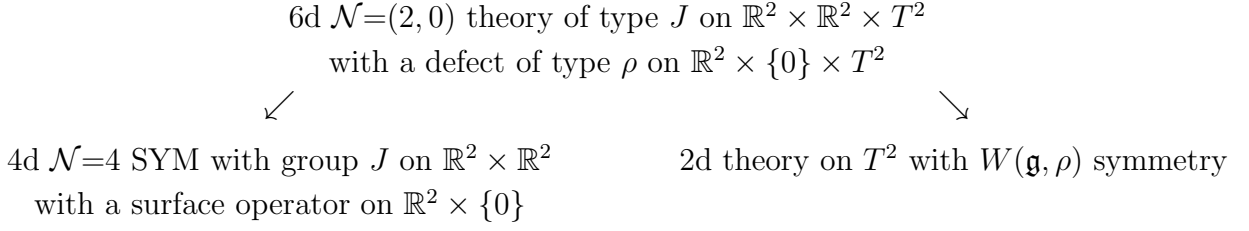

\begin{center}
6d $\cN{=}(2,0)$ theory of type $\Type$ on  $\bR^{2}\times \bR^2 \times T^2$ \\
 with a defect of type $\rho$ on $\bR^2\times \{0\}\times T^2$

\begin{tabular}{cccccc}
$\swarrow$ && &&  $\searrow$  \\[5pt]
4d $\cN{=}4$ SYM with group $\Type$ on $\bR^2\times\bR^2$ &&&& 2d theory on $T^2$ with $W(\fg,\rho)$ symmetry\\
 with a surface operator on $\bR^2\times\{0\}$
\end{tabular}
\end{center}
\caption{Surface operator of 4d SYM and a 2d theory \label{surfaceW} }
\end{figure}

\paragraph{Setup 2:} We also consider the following setup. See also Fig~\ref{surfaceW}. 
\begin{enumerate}
\item  We put the 6d theory of type $\Type$ on $\bR^2\times \bR^2\times T^2$, with complex coordinates $z_1,z_2,z_3$. We place the codimension-2 defect associated to $\rho:\su(2)\to \fg$ at $z_2=0$ so that it extends along $z_1$ and $z_3$.
\item When reduced along $T^2$, this becomes a half-BPS surface defect of $\cN{=}4$ SYM with gauge group $\Type$. This is very closely related to the surface operators studied by Gukov and Witten \cite{Gukov:2006jk,Gukov:2008sn}.
\item Instead, we can perform Nekrasov's deformation with parameter $\epsilon_{1,2}$ along the $z_{1,2}$-plane, respectively, which effectively confines excitations to be within the region $z_{1,2}\sim \epsilon_{1,2}$, which in turn gives us a 2d theory on $T^2$.
We assume that this theory has the W-symmetry $W(\fg,\rho)$ with parameter $b^2=\epsilon_2/\epsilon_1$, as suggested by recent works \cite{Wyllard:2010rp,Wyllard:2010vi,Tachikawa:2011dz,Kanno:2011fw}.
\end{enumerate}

This setup is useful because of the existence of formul\ae\ \cite{Bais:1987zk,deBoer:1993iz} for the 2d central charge and for the level of 2d current subalgebra of $W(\fg,\rho)$, which we use to constrain the form of the 4d central charges for the defect. Combining the data from Setup 1 and Setup 2 will enable us to write down algorithms for the local properties of the codimension-2 defect in terms of the label $\rho$.

The rest of the paper is organised as follows: 

In Sec.~\ref{3d}, we study the theories $T^\rho[\fg]$, since they live at the defect when the 6d theory is reduced on a circle, as we saw in Setup 1. In Sec.~\ref{TgHiggs}, we recall the Higgs branch of $T^\rho[\fg]$, and relate the possibility of Higgsing from $T^\rho[\fg]$ to $T^{\rho'}[\fg]$ with the inclusion of the corresponding nilpotent orbits $O_\rho\subset \bar O_{\rho'}$.
 In Sec.~\ref{TgCoulomb}, it will be argued that the Coulomb branch of $T^\rho[\fg]$ is given by the Spaltenstein dual orbit $d(O_\rho)$. In Sec.~\ref{induction}, we introduce the concept of induced nilpotent orbits, which we use to study the Coulomb branch of the mass-deformed version of $T^\rho[\fg]$ theory in Sec.~\ref{mass}. Then, in Sec.~\ref{term}, we illustrate various properties of nilpotent orbits, and contrast them with the physical properties of $T^\rho[\fg]$. 

In Sec.~\ref{4d}, we study the properties of the codimension-2 defects of the 6d $\cN{=}(2,0)$ theory using the analysis in Sec.~\ref{3d}.  In Sec.~\ref{dimensions}, the formul\ae\  for the dimensions of the local Higgs and the Coulomb branch will be given. In Sec.~\ref{surface}, we take a detour and study the relation between our defects and the surface operators studied by Gukov and Witten, as well as W-algebras of general type. Then, in Sec.~\ref{central}, we determine the local contributions to the central charges $n_v$ and $n_h$ from our defects. In Sec.~\ref{graded}, we review how the scaling dimensions of the local Coulomb branch operators were determined for theories of type $A$ and $D$, from which the general procedure will be physically induced. The procedure will be then tested in a few cases to show how it works.  In Sec.~\ref{cases}, we combine our methods to study various illustrative examples.  Among others, we find the properties of the $\bZ_3$-twisted defects of the $D_4$ theory, which are labeled by homomorphisms $\rho:\su(2)\to \fg_2$. We also study three-punctured spheres in 6d theories for various $\Type=A,D,E$ giving rise to free hypermultiplets. This provides an independent consistency check of our approach.

%We conclude in Sec.~\ref{summary} with a few discussions. 
We have three Appendices: 
In Appendix~\ref{conjecture}, we formulate our physical claim in Sec.~\ref{graded} in a mathematically precise way, which will hopefully make our results more readable to mathematicians.
In Appendix~\ref{explicitG2}, we collect the explicit formul\ae\   to embed $\fg_2$ inside $\so(8)$. Appendix~\ref{crash} contains tables of exceptional nilpotent orbits. 

Before proceeding, let us note that a fair amount of the theory of nilpotent orbits in Lie algebras will be used. The standard textbooks are \cite{Spaltenstein,Carter,CollingwoodMcGovern,McGovern}. On the other hand, some of our results heavily lean on more recent developments on nilpotent orbits. The necessary concepts and theorems are  introduced and explained along the way in the paper.

On notations: the Lie groups $G$ are taken to be compact and of adjoint type, while the Lie algebras $\fg=\fg_\bC$ are taken to be in their complexified version. By $G^\vee$ we mean a compact group of adjoint type, whose Lie algebra $\fg^\vee$ is the Langlands dual of the Lie algebra $\fg$ of $G$. We use the convention $\mathfrak{sp}(1)=\su(2)$.

\section{The 3d theory $T^\rho[\fg]$ and nilpotent orbits}\label{3d}
Let us study a class of 3d $\cN{=}4$ superconformal field theory introduced by Gaiotto and Witten in \cite{Gaiotto:2008ak}. Along the way, we introduce various concepts concerning nilpotent orbits.

\subsection{Higgs branch of $T^\rho[\fg]$ }\label{TgHiggs}
In the context of 4d $\cN{=}4$ super Yang-Mills with gauge group $G$ on a half-space $\bR^{2,1}\times\bR^{\geq 0}$,
Gaiotto and Witten considered a half-BPS boundary condition for the adjoint fields specified by a homomorphism $\rho: \su(2)\to \fg$.  The (adjoint) orbit for an element $e\in \fg$ is the set of elements in $\fg$ that are $G_\bC$-conjugate to $e$, i.e., are of the form, $\mathrm{ad}(g)\cdot e$ for some $g$ in $G_\bC$. We denote the orbit containing $e$ by $O^\fg_e = G_\bC\cdot e$. We often abbreviate $O^\fg_e$ as $O_e$.

The Jacobson-Morozov theorem states that the classification of the homomorphisms $\rho$, up to conjugacy, is equivalent to the classification of nilpotent elements $e$ in $\fg$, also up to conjugacy, through the correspondence $e=\rho(\sigma^+)$. The orbit $O_e$  is called a \emph{nilpotent orbit}.  
When $\fg=\su(n)$, a nilpotent orbit $O_e$ is specified by the size of its Jordan blocks, $n=n_1+\cdots+n_k$, or, equivalently, by a partition $p=[n_i]$ of $n$. 
We call $n_i$ the parts of the partition $p$.

When $\fg=\so(n)$, a nilpotent orbit $O_e$ is specified by a corresponding homomorphism $\rho:\su(2)\to \so(n)$, under which the $n$-dimensional vector representation is decomposed into irreducible representations of $\su(2)$. 
Hence, nilpotent orbits of $\so(n)$ again give rise to partitions $p=[n_i]$ of $n$, but with the requirement that any even part in $p=[n_i]$ must appear an even number of times. Such partitions are called \emph{B- or D-partitions}, when $n$ is odd or even, respectively. 
Given a partition $p=[n_i]$ satisfying this condition, there is a unique nilpotent orbit, except for the case when all the parts $n_i$ are even and each even integer appears even times. Such a partition, e.g.~$[4,4,2,2]$, is called a very even partition, and there are two distinct nilpotent orbits associated to it, exchanged by the outer automorphism of $\so(n)$. Two such orbits are distinguished by putting $I$ or $II$ as a superscript to a very even partition.

Similarly, for a nilpotent element $e$ in $\fg=\mathfrak{sp}(n)$, the corresponding $\rho:\su(2)\to\mathfrak{sp}(n)$ determines a partition $p=[n_i]$ of $2n$, with the condition that any odd part in $p=[n_i]$ appears an even number of times. Such a partition is called a \emph{C-partition.} 
In this case, each partition corresponds to a unique orbit.

The boundary condition for $\cN{=}4$ super Yang-Mills is given by a Neumann boundary condition involving $\rho$ for $\Phi_{1,2,3}$, and a Dirichlet boundary condition for $\Phi_{4,5,6}$: \begin{equation}
\Phi_{1,2,3}(s)\sim \rho(\tau_{1,2,3})/s, \qquad \Phi_{4,5,6}|_{s=0}=0. \label{4dBC}
\end{equation} 
Performing an S-duality on the 4d bulk, the gauge group of the bulk theory becomes $\dual{G}$, and at the boundary $s=0$ now lives a 3d superconformal field theory $T^\rho[\fg]$.  This 3d SCFT has $\dual{G}$ flavour symmetry on the Coulomb branch, which is coupled to the restriction of the bulk gauge fields to the boundary.

The simplest situation arises for $\rho=0$, in which case the theory $T^\rho[\fg]$ is often denoted just by $T[\fg]$. 
Its Coulomb branch is $\cN_{\dual{\fg}}$ and its Higgs branch is $\cN_{\fg}$, where $\cN_{\fg}$ denotes the \emph{nilpotent cone} of $\fg$ --- the subset of $\fg$ consisting of its nilpotent elements. The dimension of $\cN_\fg$ is given by \begin{equation}
\dim_\bC \cN_\fg = \dim G-\rank G.
\end{equation} This complex dimension is always even. In fact $\cN_{\fg}$ is a hyperk\"ahler cone, which is expected since the theory $T[\fg]$ has 3d $\cN{=}4$ superconformal symmetry. 

Now, for a homomorphism $\rho:\su(2)\to\fg$, we can give a Higgs vev $e=\rho(\sigma^+)$ to the theory $T[\fg]$.
Let us take the low-energy limit at this point. The moduli directions inside $\cN_{\fg}$ that are transverse to $O_e$ are in general singular, while the directions along $O_e$ are smooth. Thus, at low energies, the theory $T[\fg]$ becomes $(\dim_\bC O_e)/2$  free hypermultiplets plus the interacting 3d theory $T^\rho[\fg]$. The Higgs branch quaternionic dimension of $T^\rho[\fg]$ is then \begin{equation}
\dim_\bH \Higgs(T^\rho[\fg]) = \frac12(\dim G-\rank G - \dim_\bC O_e).
\end{equation}

For the classical Lie algebras, a nilpotent orbit $O_e$ is labeled by a partition $p=[n_i]$, and its dimension is given by\begin{equation}
\dim_\bC O_{[n_i]} = 
\begin{cases}
%\left\{
%\begin{array}{ll@{\quad \text{when}\ } l}
N^2 - {\displaystyle\sum_i} s_i^2 &\text{when } \fg=\su(N),\\
N(2N+1) - \frac12{\displaystyle\sum_i} s_i^2 + \frac12{\displaystyle\sum_{i\text{ odd}}} r_i &\text{when } \fg=\so(2N+1),\\
N(2N+1) - \frac12{\displaystyle\sum_i} s_i^2 - \frac12{\displaystyle\sum_{i\text{ odd}}} r_i &\text{when } \fg=\mathfrak{sp}(N),\\
N(2N-1) - \frac12{\displaystyle\sum_i} s_i^2 + \frac12{\displaystyle\sum_{i\text{ odd}}} r_i &\text{when } \fg=\so(2N),
%\end{array}\right.
\end{cases}
\end{equation} 
where $[s_i]$ is the transpose partition to $[n_i]$, and $r_k$ is the number of times the part $k$ appears in the partition $[n_i]$. 

When $\rho'$ can be chosen so that a nilpotent element $e'=\rho'(\sigma^+)$ can be found arbitrarily close to $e=\rho(\sigma^+)$,
we can change the Higgs vev of $T[\fg]$ from $e$ to $e'$, or equivalently, we can Higgs $T^\rho[\fg]$ to $T^{\rho'}[\fg]$. 
In other words, the set of limiting points of the orbit $O_{e'}$ contains $e$, or, equivalently, the closure $\bar O_{e'}$ of $O_{e'}$ contains $O_e$. 
This defines the standard partial ordering for nilpotent orbits, defined so that $O_{e'}\ge O_e$ if $\bar O_{e'}\supset O_e$.
So, partial ordering of nilpotent orbits in $\fg$ determines the partial ordering of the $T^\rho[\fg]$ theories via Higgsing. 
Under this partial ordering, the maximal (largest) nilpotent orbit is called the \emph{principal (or regular) nilpotent orbit} $O_\text{prin}$, which is the orbit through a generic nilpotent element.  

\begin{table}
\[
\begin{array}{c@{\qquad}c}
\su(6) & \so(8) \\[1em]
\xymatrix@=8pt{
&[6]\sp{d}\\
&[5,1]\sp{d}\\
&[4,2]\sp{dl}\sp{dr}\\
[4,1^2]\sp{dr}&&[3^2]\sp{dl}\\
&[3,2,1]\sp{dl}\sp{dr}\\
[3,1^3]\sp{dr}&&[2^3]\sp{dl}\\
&[2^2,1^2]\sp{d}\\
&[2,1^4]\sp{d}\\
&[1^6]
} &
\xymatrix@=8pt{
&[7,1]\sp{d}\\
&[5,3]\sp{d}\sp{dl}\sp{dr}\\
[4^2]^I \sp{dr} &  [5,1^3] \sp{d} & [4^2]^{II} \sp{dl} \\
& [3^2,1^2]\ns{d}\\
&[3,2^2,1]\sp{d}\sp{dl}\sp{dr}\\
[2^4]^I \sp{dr} &  [3,1^5] \sp{d} & [2^4]^{II} \sp{dl} \\
& [2^2,1^4]\sp{d}\\
& [1^8]
} 
\end{array}
\]
\caption{Partial ordering of the $\su(6)$ and $\so(8)$ nilpotent orbits.\label{susohasse} }
\end{table}

Partial ordering for classical $\fg$ is given by the ordering of the corresponding partitions, defined so that $p=[n_i]\ge p'=[n'_i]$ if and only if $\sum_{i=1}^k n_i \ge \sum_{i=1}^k n'_i$ for all $k$. Two very-even orbits of $\so(4N)$ that have the same partition type cannot be compared. See the examples in Table~\ref{susohasse}.
For $\fg=\su(N)$, it is particularly clear that partial ordering of nilpotent orbits is consistent with physics of the $T^\rho(\fg)$ theories, since the 3d mirror of $T^\rho[\fg]$ is given by the 3d quiver gauge theory of the form\begin{equation}
[\SU(N)]-\SU(\ell_1)-\SU(\ell_2)-\cdots 
\end{equation} where $\ell_i= N-\sum_{i=1}^k n_i$. The condition $p\ge p'$ implies that $\ell_i\le \ell_i'$.
Then, it is easy to see that we can go to the Coulomb branch of $\SU(\ell_i')$  where the unbroken theory is $\SU(\ell_i)$, i.e. we can Higgs from the theory corresponding to $p'$ to the one corresponding to $p$.  This point was already made in, e.g., ~\cite{Nanopoulos:2009uw} in the context of the corresponding 4d theory.

Let us denote by $\Flavour(\rho)$ the subgroup of $G$ commuting with the image of $\rho$.
Since $T^\rho[\fg]$ is obtained by giving the Higgs vev $e=\rho(\sigma^+)$ to $T[\fg]$, 
the theory $T^\rho[\fg]$ has Higgs flavour symmetry $\Flavour(\rho)$. Therefore, we can give $T^\rho[\fg]$ mass deformations associated to the lowest component of the 3d $\cN{=}4$ current multiplet. 
In the language of the S-dual boundary condition, this corresponds to deforming the condition \eqref{4dBC} to
\begin{equation}
\Phi_{1,2,3}(s)\sim \rho(\tau_{1,2,3})/s, \qquad \Phi_{4,5,6}|_{s=0}=m_{4,5,6}. \label{massive4dBC}
\end{equation} where $m_{4,5,6}$ are elements of $\fg$ that commute among themselves as well as with $\rho$.

For classical $\fg$, the Lie algebra $\flavour(\rho)$ for the flavour symmetry group $\Flavour(\rho)$ for $\rho$ of partition type $[n_i]$ is given by the following formul\ae\ : \begin{equation}
\begin{array}{l@{\quad\text{when}\ }l}
\mathfrak{s}[ \op{i} \mathfrak{u}(r_i)  ] & \fg=\su(N) ,\\
 \op{i\, \text{odd}}\so(r_i) \oplus \op{i\, \text{even}} \mathfrak{sp}(r_i/2)  & \fg=\so(2N+1)\, \text{or}\, \so(2N), \\
\op{i\, \text{odd}}\mathfrak{sp}(r_i/2) \oplus \op{i\, \text{even}} \so(r_i)  & \fg=\mathfrak{sp}(N). \\
\end{array}\label{classicalflavour}
\end{equation}

\subsection{The Coulomb branch of $T^\rho[\fg]$ and the Spaltenstein map}\label{TgCoulomb}
Next, let us consider the Coulomb branch of the $T^\rho[\fg]$ theory. This is a subset of the Coulomb branch of $T[\fg]$, which is the nilpotent cone $\cN_{\dual{\fg}}$.  Hence, it is natural to expect that the Coulomb branch of $T^\rho[\fg]$ be a union of closures of certain nilpotent orbits in $\dual{\fg}$; we will see that this is indeed the case.
When $\fg$ is of classical type, the theories $T^\rho[\fg]$ can be constructed via an arrangement of branes. 
Let $p$ be the partition corresponding to the nilpotent element $e=\rho(\sigma^+)$.
When $\fg$ is of type $A$ \cite{Gaiotto:2008ak}, or type $C$ and $D$ \cite{Benini:2010uu},
the Coulomb branch is in fact the closure of a single nilpotent orbit $O_{\tilde e}$ where $\tilde e$ is a nilpotent element of $\dual{\fg}$, whose partition type is given by $p^t$ when $\fg$ is of type $A$, $p^+{}^t{}_B$ when $\fg$ is of type $C$, and $p^t{}_D$ when $\fg$ is of type $D$. Here, for a  partition $p=[n_1,\ldots,n_k]$, $p^t$ is the transpose partition to $p$, $p^+$ is the augmented partition $[n_1,\ldots,n_k,1]$, $p^-$ is the reduced partition $[n_1,\ldots,n_{k}-1]$,
and $p_{B,C,D}$ stand for the $B$-, $C$-, and $D$-collapses, respectively, of $p$, and are defined to be the unique maximal $B$-, $C$-, $D$-partition $q$ satisfying $p\geq q$.\footnote{Algorithmically, the $C$-collapse is obtained by the following procedure: we pick the largest odd integer $n$ appearing odd number of times in $p$.  Pick the largest integer $m$ which is smaller than $n-1$ in the parts of $p$. Then change the parts $(n,m)\to (n-1,m+1)$. We repeat this process until it stops. 
Similarly, the $B$- or $D$-collapse is obtained by picking the largest even integer $n$ appearing an odd number of times in $p$, picking the largest integer $m$ which is smaller than $n-1$ in the parts of $p$, and then changing the parts $(n,m)\to (n-1,m+1)$, and repeating until it stops.}

It turns out that this combinatorial operation agrees with a map $d:\cN_\fg/G \to \cN_{\dual{\fg}}/\dual{G}$ defined for any simple Lie algebra $\fg$, known as  the \emph{Spaltenstein map}.\footnote{Originally, Spaltenstein \cite{Spaltenstein} defined $d_{LS}:\cN_\fg/G\to \cN_\fg/G$, i.e., a map that takes nilpotent orbits in $\fg$ to nilpotent orbits in $\fg$. This map is called the Lusztig-Spaltenstein map. Later, Barbasch and Vogan \cite{BarbaschVogan} found that this map can be described in a more natural way if one considers instead a map
$d_{BV}:\cN_\fg/G\to \cN_{\dual{\fg}}/\dual{G}$, which takes nilpotent orbits in $\fg$ to nilpotent orbits in the Langlands dual $\dual\fg$. For simply-laced $\fg$, $d_{BV}=d_{LS}$. We will find it convenient to use $d_{BV}$ in this paper, and will refer to it as the Spaltenstein map. We denote it simply by $d$. } 
The statements in the previous paragraph then amount to $d(O_p)=O_q$, where
\begin{equation}\label{Ddual}
\begin{aligned}
q&={p^t} &  \text{for}\ &\fg=\su(N),%\label{Adual}
&
q&={(p^t)^-{}_C} &  \text{for}\ & \fg=\so(2N+1),\\%\label{Bdual}
q&={(p^+)^t{}_B} &  \text{for}\ & \fg=\mathfrak{sp}(N),%\label{Cdual}
&
q&={p^t{}_D} &  \text{for}\ & \fg=\so(2N).%\label{Ddual}
\end{aligned}\end{equation} Here we added the formula for the Lie algebras of type $B$.

In other words, our proposal is that \emph{the Coulomb branch of $T^\rho[\fg]$ is given by the closure of the Spaltenstein dual orbit $d( O_e)$ to the orbit $O_e$, where  $e=\rho(\sigma^+)$.}
Notice that if $O_e$ is a nilpotent orbit in $\fg$, then $d(O_e)$ is a nilpotent orbit in the Langlands dual, $\dual{\fg}$.

Let us discuss first one piece of evidence for this proposal. The map $d$ is known to reverse the partial ordering of nilpotent orbits: $d(O)\le d(O')$ if $O\ge O'$. 
Physically, this means that the Coulomb branch of $T^{\rho}[\fg]$ is a subset of the Coulomb branch of $T^{\rho'}[\fg]$ if $T^{\rho'}[\fg]$ can be Higgsed to $T^\rho[\fg]$. We can state this result in terms of the mixed branches of $T[\fg]$ theory, too. Its maximal Coulomb branch is $\cN_{\dual{\fg}}$ and the maximal Higgs branch is $\cN_{\fg}$. What are the mixed branches? We propose that the total vacuum moduli space of $T[\fg]$ is given by \begin{equation}
\{(x,y)\in \cN_{\dual{\fg}} \times \cN_{\fg} \ | \ d(O_x) \ge O_y  \, \text{and}\,  d(O_y) \ge O_x \}.
\end{equation} It then follows that if we give a vev $e\in \cN_{\fg}$ to the Higgs branch, the Coulomb branch is restricted to the closure of $d(O_e)$. 

Next, we want to understand how the theory behaves under mass deformations. To do this, we first need to learn the concept of induction of orbits. 

\subsection{Orbit induction}\label{induction}
So far we have considered only nilpotent orbits, i.e., conjugacy classes of nilpotent elements of $\fg$. Let us now redirect our attention to semisimple (i.e.~diagonalisable) elements of $\fg$. Take a semisimple element $m\in \fg$. The subgroup $L\subset G$ which leaves $m$ invariant under conjugation is called a \emph{Levi subgroup}; in physical terms, $L$ is the subgroup of $G$ unbroken by a vev in the adjoint field.
The orbit $O_m$ through $m$ has dimensions \begin{equation}
\dim_\bC O_m = \dim G - \dim L.
\end{equation} 
For $\fg=\su(N)$, if we choose $m$ to be a semisimple element of the form \begin{equation}
m=\diag(\underbrace{m_1,\ldots,m_1}_{\ell_1},\ldots,\underbrace{m_k,\ldots,m_k}_{\ell_k})\label{typical}
\end{equation} with the $m_i$ all distinct (and $\sum l_i m_i = 0$), then the corresponding Levi subgroup is
\begin{equation}
L=\mathrm{S}(\U(\ell_1)\times \cdots \U(\ell_k)).\label{levi}
\end{equation}
Let us denote the centre of $L$ by $Z$, which contains of exponentials of elements of the form \eqref{typical}, 
but possibly with some $m_i$ equal to each other. % and maybe some discrete semisimple elements.

We would like to know what happens to the semisimple orbit $O_m$ if we take the limit $m_i\to 0$ while keeping $L$ fixed. For example, for $\fg=\su(2)$,
we can take the limit \begin{equation}
\begin{pmatrix}
m & 1 \\
0&-m 
\end{pmatrix} \to 
e=\begin{pmatrix}
0&1\\
0&0
\end{pmatrix}.\label{foo}
\end{equation} The left hand side is diagonalisable as long as $m\ne 0$, and $L=\U(1)$, and its $m\to 0$ limit is the nilpotent element $e$. In terms of orbits, we have $O_m\to O_e$.  

We can also consider a generic element $x\in \fg$, not necessarily semisimple or nilpotent.  Any such $x$ has a decomposition $x=x_s+x_n$ where $x_s$ is semisimple, $x_n$ is nilpotent and $[x_n,x_s]=0$.
Let $L$ be the Levi subgroup that stabilises $x_s$. Then $x_n\in\fl$, where $\fl$ is the Lie algebra of $L$.  The pair $(\fl, O^\fl_{x_n})$, where $O^\fl_{x_n}$ is the nilpotent orbit in $\fl$ that contains $x_n$, is called the $\fg$-\emph{Jordan class} of $x$, which is a natural generalisation of the Jordan normal form for $N\times N$ matrices. 
The dimension of $O_x$ is given by \begin{equation}
\dim_\bC O_x = \dim G- \dim L + \dim_\bC O^\fl_{x_n}. \label{aaa}
\end{equation}
Now, for a fixed Jordan class $(\fl,O^\fl_e)$, consider the orbit $O_{m+e}$ where  $m$ is a generic element  in the centre $\fz$ of  $\fl$ .  The orbit $O_{m+e}$ in the limit $m\to 0$ tends to another nilpotent orbit $O_{e'}$, which we denote by $\Ind^\fg_\fl O^\fl_e$: \begin{equation}
\lim_{m\to 0} O_{m+e} \to O_{e'} \equiv\Ind^\fg_\fl O^\fl_e.\label{bbb}
\end{equation} This process is called \emph{induction}. 
As an example, let us take $\fg=\su(N)$ and the Levi subalgebra $\fl$ to be the Lie algebra of \eqref{levi}. Then $O_e=\Ind^\fg_\fl O^\fl_0$ is a nilpotent orbit of $\su(N)$, whose partition $[n_i]$ is the transpose partition to $[\ell_i]$. The first example \eqref{foo} was for $[\ell_i]=[1,1]$ and $[n_i]=[2]$. The algorithm to calculate, for classical $\fg$,  the induced orbits given the partition can be found in \cite{CollingwoodMcGovern}. For exceptional $\fg$, the table of the induction is given in \cite{GraafElashvili}.

One property of induction is\begin{equation}
\Ind^\fg_{\fl'} \Ind^{\fl'}_\fl O^\fl_e=\Ind^\fg_{\fl}  O^\fl_e, \label{bosh}
\end{equation}
where $\fl\subset \fl'\subset \fg$.
Also, the induction and the Spaltenstein map are related through: \begin{equation}
d(O^\fg_e) = \Ind^{\dual{\fg}}_{\dual{\fl}} d(O^\fl_e).\label{important}
\end{equation}
More generally, we can consider \begin{equation}
\lim_{m\to z} O_{m+e} \to O_{x} \equiv\Ind^\fg_\fl (z+O^\fl_e)
\end{equation} where $z$ is an element of the centre $\fz$ of the Levi subalgebra $\fl$.
When $z$ is a generic element of $\fz$, $O_x$ is just the orbit $O_{z+e}$. However, for nongeneric $z$ such that the Levi algebra $\fl'$ which commutes with $z$ is bigger, $\fl'\supsetneq \fl$,  $O_x$ is the orbit through $z+ \Ind^{\fl'}_\fl O^\fl_e$. The property \eqref{bosh} then guarantees that the orbit $\Ind^\fg_\fl (z+O^\fl_e)$, as subsets of $\fg$, changes continuously as we change $z$.
Finally, induction preserves codimension:% which follows from \eqref{aaa} and \eqref{bbb}: 
\begin{equation}
\dim G-\dim_\bC \Ind^\fg_\fl O^\fl_e = \dim L-\dim_\bC O^\fl_e.\label{codim}
\end{equation}

Note that inductions from two different Levi subalgebras can lead to the same orbit. 
Let $\fg=\so(8)$, and consider the Levi subalgebra $\fl=\u(2)\times \so(4)$ which is a commutant of an element of the form \begin{equation}
\diag(m,m,-m,-m,0,0,0,0),
\end{equation} and another Levi subalgebra of the form $\fl'=\u(3)\times \u(1)$ which is a commutant of an element of the form \begin{equation}
\diag(m,m,m,m',-m,-m,-m,-m').
\end{equation}
The principal elements in $\fl$ and $\fl'$ have the partition type $[3,2,2,1]$ and $[3,3,1,1]$ in $\so(8)$ respectively. Therefore, we have
\begin{equation}
\Ind^\fg_{\fl} O^\fl_0 = \Ind^\fg_{\fl} d(O^\fl_\text{prin} ) = d(O_{[3,2,2,1]}) = O_{[3,3,1,1]},
\end{equation} and  \begin{equation}
\Ind^\fg_{\fl'} O^{\fl'}_0 = \Ind^\fg_{\fl'} d(O^{\fl'}_\text{prin} ) = d(O_{[3,3,1,1]}) = O_{[3,3,1,1]},
\end{equation}
which are the same.\footnote{The limit is as subsets of $\fg$. In \cite{Vogan1,Vogan2}, Vogan showed that the limit can be a finite cover of $O_x$ if the limit is taken in another way as follows. Namely, consider the variety $X_m=G_\bC \times_{P_\bC} (m+O^{\fl}_e+\mathfrak{n})$, where $P_\bC$ is a parabolic subgroup containing the complexified Levi subgroup $L_\bC$, and $\mathfrak{n}$ is the nilradical of the Lie algebra of $P_\bC$. It was shown that $X_m$ is generically the orbit $\Ind^\fg_\fl (m+O^\fl_e)$, but when $m=0$  it is a partial resolution of singularities of a \emph{finite cover of} the orbit $O_x=\Ind^\fg_\fl (O^\fl_e)$. When $O^\fl_e$ is the zero orbit, $X_0$ is just $T^*(G_\bC/P_\bC)$ and is smooth, which can then be blown down to the closure of a finite cover of $O_x$.  See also \cite{Fu,Namikawa}.  
With this procedure, the induction from $\fl$ gives $O_{[3,3,1,1]}$, while  the induction from $\fl'$ gives the double cover of $O_{[3,3,1,1]}$. Vogan conjectured that when two inductions give the same nilpotent orbit $O_e$, the finite covers of $O_e$ one obtains in this construction are always different. It was announced in \cite{covers1,covers2} that the conjecture was proved. \label{voganfootnote} }
For both lines, the first equality is the duality between the zero orbit and the principal orbit; the second is the relation of $d$ and $\Ind$ in \eqref{important}, and the third is the formula \eqref{Ddual}.
This means that the theories $T^{[3,2,2,1]}[\so(8)]$ and $T^{[3,3,1,1]}[\so(8)]$ have the same Coulomb branch. However, the these theories are different 3d theories, with different Higgs branches. 

\subsection{Mass deformations of $T^\rho[\fg]$}\label{mass}
The $T^\rho[\fg]$ theory has the Higgs flavour symmetry $G_\rho$, so can be deformed by a mass parameter $m_{4,5,6}$ which is in the Cartan of $\fg_\rho$. We call the deformed version $T^{\rho,m}[\fg]$.
The mass deformation should modify the Coulomb branch to be an orbit whose semisimple part is $m=m_4+im_5$, because that is what gives the boundary value of $\Phi_{4,5,6}$, in particular of $\Phi=\Phi_4+i\Phi_5$ in \eqref{massive4dBC}. This should be a deformation of the Coulomb branch of $T^\rho[\fg]$, which we proposed to be $d(O_e)$ where $e=\rho(\sigma^+)$. 

These required properties can be nicely realised if we assume that \emph{the Coulomb branch of $T^{\rho,m}[\fg]$, as a complex manifold, is given by $
\Ind^{\dual{\fg}}_{\dual{\fl}} (m+d(O^\fl_e))$ where $\fl$ is the smallest Levi subalgebra containing $e$.}\footnote{The mass term $m_6$ changes the K\"ahler structure of the nilpotent orbit, and partially resolves its singularities. This should be given by the construction of Vogan, see footnote \ref{voganfootnote}. In particular, when $\rho(\sigma^+)$ is a principal element in a Levi subalgebra $\fl$, the smooth resolution is $T^*(G_\bC/P_\bC)$. \label{voganfootnote0.5}}
This proposal passes a few consistency checks: \begin{itemize}
\item First, $T^{\rho,m}[\su(N)]$ can be  realised as a quiver gauge theory, and its Coulomb branch when $m$ is generic is just the orbit $O_m$ \cite{Gaiotto:2008ak}.  Let us confirm that this agrees the proposal above. Suppose $e=\rho(\sigma^+)$ has the partition type $[\ell_i]$, or equivalently $e$ has Jordan blocks of size $\ell_i$. The smallest Levi subalgebra $\fl$, containing $e$, is just \eqref{levi}. Hence, $e$ is a principal orbit in $\fl$, and $d(O^\fl_e)=O^\fl_0$. Then $\Ind^{\dual{\fg}}_{\dual{\fl}} (m+O_0) = O_m$, because $m$ is already the generic element in the centre of $\fl$.
\item Second, it behaves as required when $m\to 0$, because 
\begin{equation}
\lim_{m\to0}\Ind^{\dual{\fg}}_{\dual{\fl}} (m+d(O^\fl_e))=\Ind^{\dual{\fg}}_{\dual{\fl}} (d(O^\fl_e)) = d(O^\fg_e)
\end{equation}
thanks to the fundamental relation between $d$ and $\Ind$, \eqref{important}.\footnote{If the limit is taken as in footnote \ref{voganfootnote}, or in other words by giving a nonzero real mass term and then taking the zero mass limit  as in footnote \ref{voganfootnote0.5}, it can result in a finite cover of the orbit $d(O_e^\fg)$. This suggests that the Coulomb branch of the massless $T^\rho[\fg]$ could, itself, be a finite cover.\label{voganfootnote1}}
\item Third, it behaves nicely under the spontaneous breaking of the bulk 4d $\cN{=}4$ theory and the S-duality. In the massive boundary condition \eqref{massive4dBC}, let us make the mass parameters $m$ very big. Then the bulk gauge symmetry $G$ is broken to $L'$, the Levi subgroup commuting with $m$. In this limit, the boundary condition \eqref{massive4dBC} can be considered as a massless boundary condition \eqref{4dBC} for the 4d $\cN{=}4$ theory with gauge group $L'$, with the same Nahm pole $\rho$.
Let us study this from the S-dual point of view. The boundary condition is now the 4d $\cN{=}4$ theory with gauge group $\dual{G}$, coupled to $T^{\rho,m}[\fg]$. This has the Coulomb branch $O_{m+e'}$ where $e'$ is a nilpotent element in $O_{e'}=\Ind^{\dual{\fl}'}_{\dual{\fl}} d(O_e^\fl) = d(O_e^{\fl'})$. Now couple the bulk $\dual{G}$ gauge group to this orbit $O_{m+e'}$. The semi-simple part $m$ breaks the gauge group to $\dual{L'{}}$, to which the orbit $O_{e'}$ couples. Indeed, $O_{e'}=d(O_e^{\fl'})$ is the Coulomb branch of $T^{\rho}[\fl']$ theory.
\end{itemize}

\subsection{Distinguished, rigid and special}\label{term}
In this section, we will explain the properties of nilpotent orbits in more detailed terms; a reader who is mainly interested in the properties of defects of the 6d theory  can skip it.  

\subsubsection{Distinguished orbits and the Bala-Carter classification}
Let us summarise the general feature of nilpotent orbits in $\fg$.
One way to characterise a given nilpotent orbit $O_e$ through $e\in \fg$ is to consider the smallest Levi subalgebra $\fl$ containing $e$, and specify $e$ inside $\fl$.  The orbit $O^\fl_e$ in $\fl$ has the property that the smallest Levi subalgebra containing $e$ is $\fl$ itself.  Such an orbit is called \emph{distinguished} in $\fl$. The theorem of Bala and Carter states that the nilpotent orbit of $\fg$ is in bijection with the pair $(\fl,O^\fl_e)$ where $\fl$ is a Levi subalgebra and $O^\fl_e$ is a \emph{distinguished} nilpotent orbit in it.  Note that $O_e$ is distinguished in $\fg$ if and only if its flavour symmetry $\Flavour(O)$ is a discrete group.

For $\fg=\su(N)$, the only distinguished orbit is the principal orbit, i.e.~the one with the largest Jordan block, with the partition $[N]$.
Therefore, in the pair $(\fl,O^\fl_e)$, the distinguished orbit in each simple factor of type $A$ of $\fl$ is uniquely fixed. 
For other classical $\fg$, an orbit is distinguished if the corresponding partition $[n_i]$ has no repeated parts.
For example, a nilpotent element $e$ of $D_8$ whose partition is $[5,4,4,3]$ is in a Levi subalgebra $\U(4)\times \SO(8)$.
The element $e$ is a sum of nilpotent elements $e'\in \su(4)$ and $e''\in \so(8)$, of partition types $[4]$ and $[5,3]$, respectively. They are both distinguished; $[4]$ is principal as it should be, and $[5,3]$ is a non-principal distinguished orbit of $\so(8)$.

An orbit is called \emph{rigid} if it is not induced from any nilpotent orbit in any proper Levi subalgebra. 
Zero orbit is always rigid, and is the only rigid orbit for $\fg=\su(N)$. 
For any other $\fg$, the minimal nilpotent orbit (which is the orbit of a generator whose root vector is long) is known to be always rigid. But the concept of rigid orbits does not play any role in our paper.

\subsubsection{Special orbits and the Spaltenstein map}
Let us come back to the point that the Spaltenstein map \begin{equation}
d:\{\text{nilpotent orbits of }\fg\}\to \{\text{nilpotent orbits of }{}\dual{\fg}\}
\end{equation} is not quite an involution for $\fg\ne \su(N)$; they satisfy $d^3=d$.
If a nilpotent orbit $O_e$ is an image of $d$, i.e. $O_e=d(O_{e'})$, the orbit $O_e$ is called \emph{special}. 
Then the Spaltenstein map is an involution $d^2=1$ on the set of special orbits. 
In our physical setup, the Spaltenstein map $d$ connects the type of $T^\rho[\fg]$ theory, which is specified by the orbit $O_e$ through $e=\rho(\sigma^+)$, and its Coulomb branch, $d(O_e)$. As the appearance of the orbit $O_e$ and $d(O_e)$ in our setup is not symmetric, the failure of $d$ to be an involution should not surprise us. 

For a special $O_e$, the set of orbits $O$ such that $d^2(O)=O_e$ is called the special piece of $O_e$; for such non-special $O$, $O_e$ is the unique smallest special orbit larger than $O$.  Among nilpotent orbits of a fixed $\fg$, there are usually more special orbits than non-special orbits. The Spaltenstein map is order-reversing, $d(O)\ge d(O')$ if $O\le O'$, and therefore $d$ is an order-reversing involution on the special nilpotent orbits, as is illustrated in Tables~\ref{F4}, \ref{E6}, \ref{E7} and \ref{E8} of exceptional nilpotent orbits.
There, the dotted line connects two orbits in the same special piece. 

So far, we introduced in this section three adjectives for the nilpotent orbits, \emph{distinguished}, \emph{rigid}, and \emph{special}.
Intuitively, a distinguished orbit $O$ is `rather large', because the only Levi subalgebra containing a nilpotent element $e\in O$ is $\fg$ itself. Conversely, a rigid orbit $O$ is  `rather small', because it is not induced from any smaller Levi subalgebra.  No distinguished orbit is rigid. This is a consequence of a more general fact that a distinguished orbit $O$ is always induced from a zero orbit from a proper Levi subalgebra $\fl$, i.e. $O=\Ind^\fg_\fl O^\fl_0$. This in turn means that a distinguished $O$ is always special, because $O=\Ind^\fg_\fl O^\fl_0 = \Ind^\fg_\fl d(O^\fl_\text{prin})= d(O_e)$, where $e\in O^\fl_\text{prin}\subset \fl\subset \fg$. Rigid orbits can either be special or non-special.

For  general $\fg$, $O_0$ is always the smallest.
The principal orbit $O_\text{prin}$ is the maximal orbit and is the dual of $O_0$, therefore the dimension is \begin{equation}
\dim_\bC O_\text{prin} = \dim_\bC \Ind^\fg_{\fh} O^{\fh}_0= \dim G - \rank G \label{prindim}
\end{equation} where $\fh$ is the Cartan subalgebra; here we used \eqref{codim}. Its closure is the whole nilpotent cone, $\bar O_\text{prin}=\cN_{\fg}$.
The orbit of a generator $E_\alpha$ for a long root $\alpha$  is always the minimal nilpotent orbit $O_\text{min}$ (which is, strictly speaking, next-to-minimal in the ordering). 
Let us consider $\su(2)$ generated by $E_{\plusminus\alpha}$, and consider the corresponding Levi subalgebra  $\fh'=\su(2)\times \u(1)^{\rank G-1}$. 
The next-to-maximal orbit is called the subregular orbit $O_\text{subreg}$, and equals $d(O_\text{min})$. This is given by \begin{equation}
O_\text{subreg}=d(O_\text{min})=\Ind^\fg_{\fh'} d(O^{\fh'}_\text{prin}) =\Ind^\fg_{\fh'} O^{\fh'}_0,
\end{equation} which  therefore has dimension \begin{equation}
\dim_\bC O_\text{subreg}=\dim G-\rank G -2.\label{subregdim}
\end{equation}
We see that $O_\text{subreg}$ is complex codimension-2 inside $\bar O_\text{prin}$. 
It is a classic result by Brieskorn and Slodowy \cite{Brieskorn,Slodowy} that the transverse space to $O_\text{subreg}$ inside $\bar O_\text{prin}$ 
is an ALE space of type $\Type$, with a discrete action of group $S$,
such that $\dual{\fg}$ is the subalgebra of the simply-laced Lie algebra of type $\Type$ under the outer-automorphism $S$.

%\newpage

\section{Properties of codimension-2 defects}\label{4d}

Let us now return to the problem of computing the local properties of our class of codimension-2 defects of the 6d theories, in the context of the 4d $\cN{=}2$ theories obtained by compactifying the 6d theories on a Riemann surface.

Pick a 6d $\cN{=}(2,0)$ theory of type $\Type$, and compactify it on a Riemann surface $C$ of genus $g$
so that it preserves 4d $\cN{=}2$ supersymmetry.
We pick points $p_i$ $(i=1,2,\ldots)$ on $C$, and put codimension-2 defects in the class we are studying at the $p_i$. 
Each $p_i$ has an associated outer-automorphism $o_i$ of the Lie algebra $\type$. 
Let us denote by $\dual{\fg_i}$ the Lie subalgebra of $\type$ that is invariant under $o_i$, 
and $\fg_i$ the Langlands dual of $\dual{\fg_i}$. Then, the defect at $p_i$ is labeled by a homomorphism $\rho_i:\su(2)\to\fg_i$, 
or, equivalently, by a nilpotent orbit $O_{\rho_i(\sigma^+)}$ of $\fg_i$. In what follows, when we refer to a specific defect on $C$, we will omit the index $i$ to simplify our notation. When the punctures on $C$ are too few, it may not be possible to take a naive limit to get a 4d superconformal theory \cite{Gaiotto:2011xs}; in this paper we always work in the generic case, where such limit can be taken.
When this simplifying assumption is met, the introduction of a puncture of type $\rho$ will: \begin{itemize}
\item add a flavour symmetry group factor $\Flavour(\rho)$,
\item increase the dimensions of the Higgs and the Coulomb branches by, respectively, $\dim_\bH \Higgs(\rho)$ and $\dim_\bC \Coulomb_{4d}(\rho)$, and 
\item increase $n_h$ and $n_v$ by $n_h(\rho)$ and $n_v(\rho)$. Accordingly, the central charges $a$ and $c$ also get a contribution.
\end{itemize}
All these contributions are local, in the sense that they do not depend on the genus of $C$ or on the nature of the other punctures. Moreover, $\Coulomb_{4d}(\rho)$ will always be a linear space with an action of the scaling transformation. We will describe below how to find  these quantities, and we will make various related observations along the way.  
Admittedly, our discussions will not be watertight by themselves. 
We provide in Sec.~\ref{cases} ample examples where our methods can be tested in various ways.

As a comment on the terminology, let us note that a puncture with zero Nahm pole is called the \emph{full} puncture, and an untwisted puncture with subregular Nahm pole is called the \emph{simple} puncture in the previous literature. An untwisted puncture with the principal Nahm pole is the same as having no puncture at all. In the twisted case, a puncture with the principal Nahm pole was called a twisted simple puncture.

\subsection{Branch dimensions and local solutions to the Hitchin system}\label{dimensions}
First, let us consider the local contributions of a defect to the dimensions of the Coulomb branch and the Higgs branch.
Consider 5d $\cN{=}2$ super Yang-Mills with gauge group $\Type$ on $\bR^{2,1}\times C$, coupled to the 3d theory $T^{\rho_i}[\fg]$, which wraps $\bR^{2,1}$ and so lives at a point on $C$. In terms of our Setup 1, in Fig.~\ref{dualitychain}, we are going from Step 1 directly to Step 5 by compactifying on $\tilde{S}^1$. This system was analyzed in \cite{Benini:2010uu} for $\Type$ of type $A$ or type $D$, but the arguments are valid for general $\Type$. Under this compactification on $\tilde S^1$,
the dimension of the Coulomb branch doubles, while the dimension of the Higgs branch is -preserved. 
The local contribution to the Higgs branch quaternionic dimension is \begin{equation}
\dim_\bH \Higgs(\rho) = \dim_\bH \Higgs(T^\rho[\fg])= \frac12(\dim G-\rank G -\dim_\bC O_{\rho(\sigma^+)}).\label{dimHiggs}
\end{equation}
The total quaternionic dimension of the Higgs branch is then \cite{Benini:2010uu} \begin{equation}
\dim_\bH\Higgs=\sum_i \dim_\bH\Higgs(\rho_i) + \rank \fg',\label{totalHiggs}
\end{equation} where $\fg'$ is the subalgebra of $\type$ preserved by the outer-automorphism monodromy on the Riemann surface $C$  (noting that there can be monodromy both around punctures and around handles of $C$). 

The Coulomb branch of the 3d theory is given by the moduli space of the Hitchin system of gauge group $\Type$, with an outer-automorphism twist $o_i$ around the $i$-th puncture $p_i$, coupled to the Coulomb branch $d(O_{\rho_i})$ of the theory $T^{\rho_i}[\fg]$.

So, to compute the local contribution of a defect to the Coulomb branch, we first need to understand the local boundary condition for the Hitchin system near the puncture on $C$. Before doing this, let us see note that the outer automorphism $o$ introduces a grading for the Lie algebra $\type$. The outer automorphism $o$ can be trivial, of order 2 (for $\type=A_{2N-1},D_{N}$), or of order 3 (for $\type=D_4$). The Lie algebra $\type$ splits into a direct sum of eigenspaces under the action of $o$:
\begin{equation}
\begin{array}{ll}
\type =\type_1 + \type_{-1} & \text{for $o$ of order 2},\\
\type =\type_1 + \type_{\omega}+\type_{\omega^2} & \text{for $o$ of order 3}.
\label{gradeddecomposition}
\end{array}
\end{equation}
Here, the lower indices denote the eigenvalues under the action of $o$, e.g., $o(\type_{\omega^2})=\omega^2\type_{\omega^2}$. Also, by definition, $\type_1=\dual{\fg}$. The grading means that, e.g., $[\type_{\omega},\type_{\omega^2}]\subset \type_{1}$. %Also, of course, $o_3$ exists only for $\type=D_4$. 
The specific embedding of $\type_1=G_2$ in $\type=D_4$ can be found in Appendix~\ref{explicitG2}.

Now, choose a defect, and let us set the origin of a local coordinate $z$ to be the position of the defect on $C$. If the outer automorphism $o$ associated to the defect is trivial, we call the defect \emph{untwisted}, and it is labeled by a homomorphism $\rho:\su(2)\to\type$ (since, for trivial $o$, $\type=\fg=\dual{\fg}$), or, equivalently, by a nilpotent orbit $O_{\rho}$ in $\type$. Notice that the Spaltenstein-dual nilpotent orbit $d(O_{\rho})$ is also in $\type$. The Higgs field $\Phi$ in the Hitchin system then behaves as\begin{equation}
\Phi(z) = \left[\frac{\Phi_{-1}}{z} + \Phi_0+ \cdots \right]dz,\label{hitchinpole}
\end{equation}where $\Phi_{-1}$ is an element in $d(O_{\rho_i})$, and $\Phi_0$ is a generic element in $\type$. 
We see that the introduction of an untwisted defect of type $\rho$ increases the dimension of the Coulomb branch by \begin{equation}
\dim_\bC \Coulomb_{4d}(\rho) =\dim_\bH \Coulomb_{3d}(\rho) = \frac12\dim_\bC d(O_\rho).\label{3dCoulomb}
\end{equation} In the following, we just call them $\dim\Coulomb(\rho)$.

When $o$ is nontrivial, we call our defect \emph{twisted}, and we need to impose  
\begin{equation}
\Phi(e^{2\pi i}z) = g[ o( \Phi(z)) ] g^{-1}
\end{equation}
where $g$ parameterises the coset $\Type/{}\dual{G}$.
Specifically, when $o$ is of order 2, our twisted defect is labeled by a nilpotent orbit in $\fg$, while the Spaltenstein dual orbit lives in $\dual{\fg}=\type_1$. The boundary condition for the Higgs field in this case is
\begin{equation}
\Phi(z) \sim \left[\frac{\Phi_{-1}}{z} + \frac{\Phi_{-1/2}}{z^{1/2}} + \Phi_0+  \cdots \right] dz
\label{hitchinpole2}
\end{equation}
where $\Phi_{-1}$ is an element of $d(O_{\rho_i})$, $\Phi_{-1/2}$ is a generic element in $\type_{-1}$, and $\Phi_{0}$ is a generic element in $\type_{1}$.

On the other hand, when $o$ is of order 3, the defect is again labeled by a nilpotent orbit in $\fg$, and the Spaltenstein dual orbit lives in $\dual{\fg}=\type_1$, but now the boundary condition for the Higgs field is\begin{equation}
\Phi(z) \sim \left[\frac{\Phi_{-1}}{z} + \frac{\Phi_{-2/3}}{z^{2/3}}+ \frac{\Phi_{-1/3}}{z^{1/3}} + \Phi_0+  \cdots\right] dz. 
\label{hitchinpole3}
\end{equation}
Here, $\Phi_{-1}$ is an element of $d(O_{\rho_i})$, $\Phi_{-1/3}$ is a generic element in $\type_{\omega}$, and $\Phi_{-2/3}$ is a generic element in $\type_{\omega^2}$.

Altogether, we see that the introduction of a twisted defect of type $\rho$ increases the dimension of the Coulomb branch by \begin{equation}
\dim \Coulomb(\rho) = \frac12\dim_\bC d(O_\rho) + \frac12 \dim \Type/\dual{G}.\label{twisted3dCoulomb}
\end{equation}
The second term in \eqref{twisted3dCoulomb} may be a half-integer when $o$ is of order 2.  This is not a problem because twisted punctures must always come in pairs.

So, finally, for a theory on a surface of genus $g$, the total Coulomb branch dimension is \begin{equation}
\dim \Coulomb=\sum_i \dim \Coulomb(\rho_i) +(g-1) \dim G. \label{total3dCoulomb}
\end{equation}

\subsection{Surface operators of 4d $\cN{=}4$ super Yang-Mills and our defects}\label{surface}

As another way to study the defect, we use Setup 2, given in the introduction; see Fig.~\ref{surfaceW}. There, we compactified the 6d theory of type $\type$ on a torus $T^2$, which a defect of type $\rho$ fully wraps. Then, at low energies, we get $\cN{=}4$ super Yang-Mills with gauge group $\Type$ with a surface operator characterised by $\rho$. 
This surface operator is closely related to those  studied by Gukov and Witten \cite{Gukov:2006jk,Gukov:2008sn}.
In our case, we have a singularity in the adjoint field of the form \eqref{hitchinpole}, where the residue was in the Spaltenstein dual orbit $d(O_\rho)$.  By construction, this surface operator is S-duality invariant, because the 6d defect wraps the whole $T^2$. In \cite{Gukov:2008sn}, it was argued that the residue of such S-duality-invariant defect should be in a special nilpotent orbit, and indeed $d(O_\rho)$ is special. 

We can turn on the mass parameter $m$ corresponding to the flavour symmetry $\Flavour(\rho)$ of the defect of type $\rho$. 
This changes the residue to lie in the orbit $O=\Ind^{\dual{\fg}}_{\dual{\fl}} (m+d(O^\fl_\rho))$, where $\fl$ is the smallest Levi subalgebra containing $\rho(\sigma^+)$ as discussed in Sec.~\ref{mass}. 
The orbit $O$ is semisimple when $m$ is a generic element in the centre of $\fl$ and $d(O^\fl_\rho)=O^\fl_0$, or equivalently, when $\rho$ is a principal orbit in the Levi subalgebra $\fl\subset \fg$. 
Then the Coulomb branch of $T^{\rho,m}[\fg]$, which is $O$, is smooth, and there is no problem to treat it as a non-linear sigma model with the target $O$. 
This is then really just the surface operator in \cite{Gukov:2006jk}, where this was called a surface operator of Levi type $\fl$. In other words, \emph{the surface operator of Levi type $\fl$ in \cite{Gukov:2006jk} comes from a generic mass deformation of 6d defect of type $\rho$, corresponding to the principal nilpotent element in $\fl$.}

For other choice of $\rho$ and $m$, the orbit $O$ in which the Hitchin pole resides, or equivalently the target of the nonlinear sigma model living on the surface operator, is singular. 
Therefore we need to specify what physically happens at these singularities. 
The construction from 6d naturally provides one: we have the dimensional reduction on $S^1$ of the 3d theory $T^{\rho}[\fg]$ at the surface operator. 
Note that more than one $\rho$ can give rise to the same Hitchin pole in $d(O_\rho)$; nonetheless, they give different physics. 

Let us consider putting the 6d theory and the defect on a more general Riemann surface $C$,
and perform Nekrasov's deformation on the 4d side, with parameters $\epsilon_{1,2}$.
This will lead to a 2d theory on $C$.
When $\fg=\su(N)$ ,  it was suggested in \cite{Wyllard:2010rp} that
this theory has the W-symmetry $W(\su(N),\rho)$ at the parameter $b^2=\epsilon_2/\epsilon_1$,
obtained by the quantum Drinfeld-Sokolov reduction. 
This proposal has been checked to some extent by recent works \cite{Wyllard:2010vi,Tachikawa:2011dz,Kanno:2011fw}.
It seems natural to propose then that \emph{this setup, with general $\fg$ and $\rho$, give rise to a 2d system with the symmetry $W(\fg,\rho)$.} This is in accord with the conjecture made when $\rho$ is principal in a Levi subalgebra  in \cite{Braverman:2010ef} for the finite W-algebra.

The 2d central charge was calculated in \cite{Bais:1987zk,deBoer:1993iz}\ %\footnote{Note that the preprint version of \cite{deBoer:1993iz} has a typo exactly at this equation.} 
and has the value \begin{equation}
c_\text{2d}=\dim \fg_0 -\frac12\dim \fg_{1/2}+\frac{24}{b^2}\rho_\fg\cdot\rho_\fg + 12\rho_\fg\cdot \frac{h}{2}+24b^2 \frac{h}2\cdot\frac{h}2\label{2dc}
\end{equation} where $\rho_\fg$ is the Weyl vector of $\fg$, $h=\rho(\sigma^3)$, and we decomposed $\fg$ into \begin{equation}
\fg=\bigoplus_{j\in \frac12\bZ} \fg_j
\end{equation} where $j$ is the eigenvalue of the action of $h/2$. Note that when $\rho$ is  principal,
the corresponding $h/2$ is the  Weyl vector $\rho_{\dual{\fg}}$ of the dual algebra $\dual{\fg}$, and the formula \eqref{2dc} becomes 
the central charge of the standard W-algebra $W(\fg)$
\begin{equation}
c_\text{2d}(\fg,\rho)=\rank G + 24\left|\tfrac{1}b \rho_\fg + b \rho_{\dual{\fg}} \right|^2, 
\end{equation} which shows the symmetry exchanging $(1/b,\fg)\longleftrightarrow (b,{}\dual{\fg})$.

The defect of type $\rho$ has flavour symmetry $\Flavour(\rho)$, and the corresponding W-algebra $W(\fg,\rho)$ has the affine Lie subalgebra $\hat \flavour(\rho)$. The current algebra level of $\hat \flavour(\rho)$ was also calculated in \cite{deBoer:1993iz}. Recall that $\Flavour(\rho)$ commutes with $\rho(\su(2))$. Therefore, the adjoint representation $\fg$ can be decomposed as \begin{equation}
\fg=\bigoplus_{j\in \frac12\bZ} R_j \otimes V_j\label{RV}
\end{equation}  where $V_j$ is the irreducible representation of $\su(2)$ of spin $j$, and $R_j$ is a representation of $\flavour(\rho)$.
Choose generators $T^{a,b}$ of a simple subalgebra $\flavour' \subset \flavour(\rho)$ so that $\tr_{\flavour'} T^aT^b=\coxeter(\flavour')\delta^{ab}$, where $\coxeter$ is the dual Coxeter number.
Denote by $f$ the natural embedding $f:\flavour(\rho)\to \fg$. Then the level of $\hat \flavour'$ is given by \begin{equation}
k_{2d}(\flavour') \delta^{ab}= k_{2d} \frac{1}{\coxeter(\fg)} \tr_{\fg}f(T^a)f(T^b) + \sum_j 2j\, \tr_{R_j} T^a T^b\label{2dk}
\end{equation} where $k_{2d}=-\coxeter(\fg)+1/b^2$ is the level of the affine $\fg$ algebra before the Drinfeld-Sokolov reduction.

\subsection{Central charges}\label{central}
Four-dimensional conformal field theories have two Weyl anomaly coefficients, $a$ and $c$. For 4d $\cN{=}2$ theories, it is more convenient to use $n_v=4(2a-c)$ and $n_h=4(5c-4a)$, normalised so that $n_v$ counts the number of free vector multiplets and $n_h$ counts the number of free hypermultiplets, for completely free theories.  Adding a defect of type $\rho$ increases these central charges by $n_v(\rho)$ and $n_h(\rho)$. The total $n_v$ and $n_h$ of a 4d $\cN{=}2$ SCFT are
\begin{equation}\label{totalnvnh}
\begin{split}
n_v&=\sum_i n_v(\rho_i) + (g-1) (\tfrac43  \coxeter(\Type) \dim\Type+ \rank\Type),\\
n_h&=\sum_i n_h(\rho_i) + (g-1) (\tfrac43 \coxeter(\Type)\dim\Type  ).
\end{split}\end{equation} The global terms, which are proportional to $(g-1)$, were calculated in \cite{Benini:2009mz,Alday:2009qq} using the anomaly polynomials of the 6d $\cN=(2,0)$ theories determined in \cite{Harvey:1998bx,Intriligator:2000eq,Yi:2001bz}.

Let us consider how to obtain $n_{v,h}(\rho)$ for general $\fg$, for which descriptions as quiver gauge theories are not necessarily available.
Note  that not only $n_{v,h}(\rho)$, but also the 2d central charge $c_{2d}$ \eqref{2dc} should come from the anomaly polynomial of the defect of type $\rho$.  On flat space, a codimension-2 defect has its $a(\rho)$ and $c(\rho)$; the $\SO(2)$ rotation of the transverse space to the defect is a flavour symmetry, and it has a flavour symmetry central charge $k_T(\rho)$. The $n_v(\rho)$ and $n_h(\rho)$ are a certain linear combination of these three fundamental quantities, determined by the R-symmetry twist needed to preserve 4d $\cN{=}2$ supersymmetry when the 6d theory is put on a curved Riemann surface. 

The central charge of the W-algebra $W(\fg,\rho)$ should have the same origin. Note that the standard W-algebra $W(\type)=W(\type,\rho_\text{prin})$ corresponds to the absence of the defect when $\fg$ is simply-laced; therefore, the contribution from the presence of the defect of type $\rho$ is:\footnote{We normalise the length of roots so that 
 $\rho_{\fg}\cdot\rho_{\fg}=\rho_{\type}\cdot\rho_{\type}$ so that the term proportional to $1/b^2$ goes away,
since  the equivariant integral of the anomaly over the worldvolume of the defect produces terms only of the form $b^2$ or $1$ in \eqref{deltac}, as discussed in \cite{Tachikawa:2011dz}. } \begin{multline}
\delta c_{2d}(\fg,\rho)=c_{2d}(\fg,\rho)-c_{2d}(\type,\rho_\text{prin}) \\
= (\dim \fg_0-\rank \type)+\frac12\dim \fg_{1/2} + 12(\rho_\fg\cdot\frac h2-\rho_\type\cdot\rho_{\type}) + 24 b^2 (\frac{h}2\cdot\frac{h}2 - \rho_{\type}\cdot\rho_{\type}).\label{deltac}
\end{multline}
This suggests that $a(\rho)$, $c(\rho)$ and $k_T(\rho)$, and hence also $n_h(\rho)$ and $n_v(\rho)$, could be expressed as linear combinations of \begin{equation}
\dim \fg_0-\rank \type,  \quad
\dim \fg_{1/2} , \quad
\rho_{\fg}\cdot\frac{h}2-\rho_{\type}\cdot\rho_\type , \quad 
\frac{h}2\cdot\frac{h}2 - \rho_{\type}\cdot\rho_{\type}.
\end{equation}  
Since we already know the quantities $n_h(\rho)$ and $n_v(\rho)$ for type $\Type=A$, $C$ and $D$ from the analysis of quiver gauge theories, we can readily check if the suggestion above works. It indeed does, and we find that $n_h(\rho)$ and $n_v(\rho)$ are given by  \begin{subequations}\label{nvnh}\begin{align}
n_h(\rho)&=8(\rho_{\type}\cdot\rho_\type-\rho_{\fg}\cdot\frac{h}2)+\frac12\dim \fg_{1/2},\label{nh}\\
n_v(\rho)&=8(\rho_{\type}\cdot\rho_\type-\rho_{\fg}\cdot\frac{h}2)+\frac12(\rank \type-\dim \fg_{0}).\label{nv}
\end{align}\end{subequations} Note that  $\rho_\type\cdot\rho_\type=\frac{1}{12}\coxeter(\Type)\dim\Type$.

The relation between the level of the 2d current subalgebra $\hat \flavour(\rho)$ of $W(\fg,\rho)$ and the level of the 4d flavour symmetry $\flavour(\rho)$  was already studied in \cite{Tachikawa:2011dz} for $\fg=\su(N)$. The analysis there showed that $k_{4d}(\flavour')=2k_{2d}(\flavour')|_{1/b^2=0}$. Using \eqref{2dk}, we find \begin{equation}
k_{4d}(\flavour') \delta^{ab}=  -2\tr_{\fg} f(T^a)f(T^b) + 2\sum_j 2j\, \tr_{R_j} T^a T^b,
=2\sum_j \tr_{R_j} T^a T^b.\label{4dk}
\end{equation}  where $R_j$ and $f$ were already defined in \eqref{RV}. As a check, let $\rho=0$, $\flavour(\rho)=\fg$. Then, $R_0=\fg$, $R_{i>0}=0$, and we have $k_{4d}(\fg)=2\coxeter(\fg)$, as expected.

\subsection{Contribution to the 4d Coulomb branch}\label{graded}
\subsubsection{Preliminaries}
The Coulomb branch of any 3d $\cN{=}4$ theory is a hyperk\"ahler manifold, and is in particular a holomorphic symplectic manifold. If it arises as an $S^1$ compactification of a 4d $\cN{=}2$ theory, this holomorphic symplectic manifold is completely integrable \cite{Donagi:1995cf,Seiberg:1996nz}, i.e.~it is a lagrangian fibration over a K\"ahler base $B$, such that the generic fibre is an abelian variety.
The base, $B$, is the Coulomb branch of corresponding 4d theory.
In our case, the Coulomb branch of the 3d theory is the moduli space of the Hitchin system. We computed the local contribution by a defect of type $\rho$ to the quaternionic dimension of the 3d Coulomb branch, as thus also to the complex dimension of the local 4d Coulomb branch, in \eqref{3dCoulomb}, \eqref{twisted3dCoulomb}.

The 4d $\cN{=}2$ theory is superconformal, so its Coulomb branch actually has a finer structure. 
The scaling symmetry  sends $\Phi(z)\to t\Phi(z)$, which preserves the form of the singularities because nilpotent orbits are cones. 
The scaling symmetry also makes $B$ into a cone. 
For all known cases, $B$ is in fact just a graded vector space.
Assuming this statement to be always true, one can choose the generators $u_i$ of the chiral ring of the 4d Coulomb branch which form a basis for $B$ unambiguously. 
Let $n_k$ be the total number of $u_i$ whose scaling dimension is $k$.
Then  we should have
\begin{equation}\label{refined}
\dim_\bC(B) = \sum_k n_k.
\end{equation}
The $n_k$ receive a local contribution $n_k(\rho)$ from a defect of type $\rho$. The local contribution $n_k(\rho)$ and the local contribution $n_v(\rho)$ to the effective number of vector multiplets $n_v$ are related by
\begin{equation}\label{oldnv}
n_v(\rho) = \sum_k (2k-1) n_k(\rho).
\end{equation} This expression was proven in \cite{Argyres:2007tq,Shapere:2008zf}.
Now we would like to compute the local contributions $n_k(\rho)$.

Let $P^{(d_a)}(\Phi)$ $(a=1,\ldots,\rank \Type)$ be the degree-$d_a$ symmetric invariant polynomial of $\type$, so that $P^{(d_a)}$ generates all the invariant polynomials. Let $\phi^{(d_a)}(z)$ be the invariant polynomials constructed from the Hitchin field $\Phi(z)$, i.e.~$\phi^{(d_a)}(z)\equiv P^{(d_a)}(\Phi(z))$.
Then, $\phi^{(d_a)}(z_1)$ and $\phi^{(d_b)}(z_2)$, for any $d_a,d_b,z_1,z_2$, Poisson-commute by construction. We expect them to provide a set of complete integrals of motion. 
We  assign to $\phi^{(d_a)}(z)$ a scaling dimension $d_a$. 
 
Let us introduce punctures of type $\rho_i$ at $z=z_i$. Then, the singularities \eqref{hitchinpole}, \eqref{hitchinpole2}, \eqref{hitchinpole3} give rise to a pole of order at most $p_{d_a}(\rho)$ in the $\phi^{d_a}(z)$ at $z=z_i$; $p_{d_a}$ may be fractional when $\rho$ is a twisted puncture.  The number of degrees of freedom in the meromorphic $d_a$-differential $\phi^{(d_a)}(z)$ is \begin{equation}
\sum_i p_{d_a} (\rho_i)  + (1-g)(2d_a-1).\label{totalCoulomb}
\end{equation} 
Considering the second term above to be the contribution from the bulk of the Riemann surface, we see that a puncture of type $\rho$, inserted say at $z=0$, effectively adds $p_{d_a}(\rho)$ Coulomb branch operators of scaling dimension $d_a$.\footnote{It is not a problem if $p_{d_a}$ is fractional, as is the case for twisted punctures, because the total number of Coulomb branch operators of scaling dimension $d_a$ is always an integer.} More concretely, we can identify these operators with the coefficients $\phi^{(d_a)}_{k}$ of the poles of order $z^{-k}$ in $\phi^{(d_a)}(z)$, where $0<k\le p_{d_a}$.   
However, these coefficients $\phi^{(d_a)}_k$ are not always the most elementary Coulomb branch operators. Rather, they are polynomials in the true generators of the Coulomb branch operators introduced by $\rho$. Indeed, the coefficients $\phi^{(d_a)}_k$ usually satisfy rather intricate constraints.\footnote{Gukov and Witten \cite{Gukov:2008sn} call these constraints the fingerprint of a Hitchin pole, and pointed out their relation to the Kazhdan-Lusztig map. % \cite{KLmap,KLmapSpaltenstein1,KLmapSpaltenstein2,KLmapSpaltenstein3}. 
In principle, this map should also determine the information we need, but we have not been able to deduce the algorithm connecting the two.}

To see how this is possible, let us consider the untwisted boundary condition \eqref{hitchinpole}; the twisted case can be treated similarly.  We consider \begin{equation}
\Phi(z)=e\frac{dz}z + \Phi_0dz + \cdots \label{phiz}
\end{equation} where $e$ is a fixed element in the Spaltenstein-dual orbit $d(O_\rho)$.\footnote{Note that in fixing the residue to be $e$, we break the naive scale invariance $\Phi(z)\to t\Phi(z)$. However, the scaling $e\to te$ can be undone by an action of $\fg$, which we can combine with the original scaling symmetry to define a new scaling action.} 
The allowed continuous gauge transformations are of the form
\begin{equation}
g(z)=g_0 + g_1 z + \cdots, \qquad g_0\in \fg_e \equiv \{x \ |\  [e,x ]=0\},\  g_{i>0}\in \fg. \label{allowedgauge}
\end{equation}   
We want to find functions of $\Phi_0$ that are invariant under \eqref{allowedgauge}. 
The coefficients $\phi^{(d_a)}_k$ introduced above are indeed invariants, but they are not necessarily the most basic ones.  Let us study the cases $\fg=\su(N)$ and $\fg=\so(2N)$ first, to get intuition about the mechanism at work, and then we will state our general proposal in Sec.~\ref{generalrecipe}.

\subsubsection{The case $\fg=\su(N)$}

Let us define $\phi^{(d)}(z)$, $d=2,\ldots,N$, through $\det (x+\Phi(z))= x^N+ \sum_d \phi^{(d)}(z) x^{N-d}$. 
For a puncture $\rho$, the $\{p_{d}(\rho)\}$ was called the \emph{pole structure} in \cite{Gaiotto:2009we}, and the algorithm to compute it was given. The $\phi^{(d)}_k$, $2\le d\le N$, $1\le k\le p_d(\rho)$ are actually independent in this case. So, they directly form a basis of generators for the Coulomb branch operators. 
As an example, take $N=7$ and $\rho=[3,2,1,1]$.
The pole structure for $\rho$ is  $\{p_d\}_{d=2,\ldots,7}=\{1,2,3,3,4,4\}$. 
So, this defect adds 17 operators to the Coulomb branch. 
The Hitchin orbit $d(O_\rho)$ has partition $[4,2,1]$, so, using \eqref{3dCoulomb}, we find $\dim_\bC d(O_\rho)=34$, which agrees with $2\sum p_d$. Also, using \eqref{nv}, we can compute $n_v(\rho)$, and it agrees with $\sum (2d-1)p_d$. This case was further analyzed in \cite{Gaiotto:2009gz,Nanopoulos:2009uw}.

\subsubsection{The case $\fg=\so(2N)$}\label{sopoles}

Let us define $\phi^{(d)}(z)$, $d=2,4\ldots,2N$, through $\det (x+\Phi(z))= x^N+ \sum_d \phi^{(d)}(z) x^{N-d}$.  The term with $d=2N$ is actually a square, $\phi^{(2N)}(z)=\left(\tilde\phi^{(N)}(z)\right)^2$. The Pfaffian $\tilde\phi^{(N)}(z)$ has a pole of order up to $\tilde{p}_N$. So, the pole structure for an $\so(2N)$ defect is of the form $\{p_2,p_4,p_6,\dots,p_{2N-2};\tilde{p}_N\}$. The structure of the $\phi^{(d)}_k$ and the constraints they satisfy was studied in detail in \cite{Chacaltana:2011ze}. 
Here we summarise the results, and make relevant observations along the way. 
We first consider $\fg=\so(8)$.

\paragraph{First example.}
Take $\rho=[5,3]$.
Then, the Hitchin pole is $d(O_\rho)=O_{[2^2,1^4]}$, which has complex dimension 10. Then, $\dim \Coulomb(\rho)=5$ and $n_v(\rho)=39$. 
From the Hitchin pole, we find a pole structure $\{p_2,p_4,p_6;\tilde p_4\}=\{1,2,2;1\}$.
There is just one constraint, $\phi_2^{(4)}=(\phi_1^{(2)})^2/4$, and so the defect contributes to the total number of operators $n_d$ of scaling dimension $d$ with $(n_2(\rho),n_4(\rho),n_6(\rho))=(1,2,2)$, which does reproduce $\dim \Coulomb(\rho)$ and $n_v(\rho)$.

\paragraph{Second example.}
Take $\rho=[3,3,1,1]$.
Then $d(O_\rho)=O_{[3,3,1,1]}$, with $\dim_\bC O_{[3,3,1,1]}=18$. $n_v(\rho)=69$. From the Hitchin pole, we find $\{p_2,p_4,p_6;\tilde p_4\}=\{1,2,4;2\}$. One finds that $\phi_4^{(6)}$, seen as a polynomial in the components of $\Phi_0$ in \eqref{phiz}, is always a square, $\phi_4^{(6)}=(a^{(3)}_2)^2$. However, unlike the previous example, $a ^{(3)}_2$ is not a pole coefficient $\phi^{(k)}_l$, but a new invariant of dimension 3, which is a more basic Coulomb branch operator than $\phi_4^{(6)}$. So, it is $a^{(3)}_2$, not $\phi_4^{(6)}$, that should be added to the set of generators of the Coulomb branch. Let us try to understand where this $a^{(3)}_2$ parameter comes from.

For a fixed element $e$ in the Hitchin pole, $O_\rho$, there exists an $\so(6)$ Lie subalgebra of $\so(8)$ that contains $e$. We can restrict the Higgs field $\Phi(z)$ to such $\so(6)$, and the corresponding $\so(6)$ Pfaffian is of the form
\begin{equation}
\Pfaff\left[\Phi(z)\Bigm|_{\so(6)}\right]=\frac{a^{(3)}_2}{z^2}+\dots
\end{equation}
This is equivalent to restricting the $\so(8)$-nilpotent orbit with partition $\rho=[3,3,1,1]$ to the $\so(6)$-nilpotent orbit with partition $[3,3]$.

So, $a^{(3)}_2$ arises as the coefficient of the leading pole in this $\so(6)$-restricted Pfaffian. Now, $a^{(3)}_2$ is a good Coulomb branch operator in the $\so(8)$ theory because 
both the magnitude and its chosen sign are invariant under continuous transformations \eqref{allowedgauge}.
We find $(n_2(\rho),n_3(\rho),n_4(\rho),n_6(\rho))=(1,1,4,3)$, which reproduce $\dim_\bC O$ and $n_v(\rho)$.

\paragraph{Third example.}
Take $\rho'=[3,2,2,1]$. Then $d(O_{\rho'})=O_{[3,3,1,1]}$. This is exactly the same Hitchin pole as in the previous example, which was for $\rho=[3,3,1,1]$.  Using the terminology introduced in Sec.~\ref{term}, $O_{\rho'}$  is a \emph{non-special} nilpotent orbit, while $O_\rho$ is special, and, in this case, $d(O_{\rho'})=d(O_\rho)=O_\rho$.

Naively, the Hitchin systems associated to $\rho$ and $\rho'$ are the same. 
However, $n_v(\rho')=75$ is different from $n_v(\rho)=69$. This can be reproduced if we consider $\phi_4^{(6)}$, rather than $a^{(3)}_2$, to be a Coulomb branch operator. In that case, we get $(n_2(\rho'),n_4(\rho'),n_6(\rho'))=(1,4,4)$, which reproduces the desired $n_v(\rho')=75$. But then, how we can prevent $\phi^{(6)}_4$ from being a square?

Consider the subgroup of $\SO(8)$ unbroken by a fixed element of the Hitchin pole, $O_{\rho}$, which is $(\OO(2)\times \OO(2))/\bZ_2$. This group has 2 connected components.  A (semisimple) element of $(\OO(2)\times \OO(2))/\bZ_2$ not connected to the identity is\begin{equation}
s: \bR^8 = \bR^3\oplus \bR^3\oplus \bR^1\oplus \bR^1 \ni a\oplus b\oplus c\oplus d \mapsto  a \oplus (-b) \oplus c \oplus (-d).
\end{equation} 
Now, $s$, of course, belongs to $\SO(8)$, but its restriction to the $\bR^3\oplus \bR^3$ above does \emph{not} belong to $\SO(6)$. This means that $s$ will flip the sign of the $\so(6)$-restricted Pfaffian, i.e., the sign of $a_2^{(3)}$.

Hence, our rule is that for the special Nahm pole $O_\rho$ we impose invariance only under continuous gauge transformations \eqref{allowedgauge}, whereas for the non-special Nahm pole $O_{\rho'}$ we impose invariance both under such continuous gauge transformations as well as under the transformation $s$ in $(\OO(2)\times \OO(2))/\bZ_2\subset \so(8)$. In both cases the Hitchin pole is $d(O_\rho)=d(O_\rho')=O_\rho$, but we impose different invariance conditions for the Higgs field $\Phi$.

In terms of the corresponding quiver gauge theories, the difference arises as follows. The puncture $\rho$, together with several twisted punctures of type $[6]$, gives rise to a 4d quiver theory with gauge group \begin{equation}
\Sp(1)-\SO(6)-\Sp(3)-\SO(8) - \cdots,
\end{equation} whereas $\rho'$ corresponds to a quiver theory with gauge group
\begin{equation}
\Sp(1)-\SO(7)-\Sp(3)-\SO(8) - \cdots .
\end{equation}  Therefore, a Coulomb branch operator with scaling dimension 3 (belonging to $\SO(6)$) in the quiver theory for $\rho$ is converted to a Coulomb branch operator with scaling dimension 6 (belonging to $\SO(7)$) in the quiver theory for $\rho'$. 

\paragraph{A more intricate example.}
Let $\fg=\so(40)$, and \begin{equation}\label{rho4}\begin{aligned}
\rho_1&=[9,9,7,7,3,3,1,1],\quad& 
\rho_2&=[9,8,8,7,3,3,1,1],\quad\\
\rho_3&=[9,9,7,7,3,2,2,1],\quad&
\rho_4&=[9,8,8,7,3,2,2,1].
\end{aligned}\end{equation} In this case, $d(O_{\rho_1})=d(O_{\rho_2})=d(O_{\rho_3})=d(O_{\rho_4})=O_{\rho}$, where $\rho=[7^2,5^2,4^2,3^2,1^2]$.
Here, $O_{\rho_1}$ is special, while $O_{\rho_{2,3,4}}$ are non-special. The four nilpotent orbits $O_{\rho_{1,2,3,4}}$ constitute a special piece. Plus, $d(O_{\rho})=O_{\rho_1}$. 

Computing the $\phi^{(d)}_k$ for the Hitchin pole $O_\rho$, one finds that $\phi^{(14)}_{12}$ and $\phi^{(38)}_{36}$  are squares, \begin{equation}
\phi^{(14)}_{12}=(a^{(7)}_{6})^2,\qquad
\phi^{(38)}_{36}=(a^{(19)}_{18})^2,
\end{equation} where \begin{equation}
\Pfaff \left[\Phi(z) \Bigm|_{\so(2n)}\right]=\frac{a^{(n)}_{n-1}}{z^{n-1}}+\dots
\end{equation}
where $\so(14)$ and $\so(38)$  are certain subalgebras of $\so(40)$, and the corresponding restricted partitions are $[7^2]$ and $[7^2,5^2,4^2,3^2]$.

As such, $a^{(7)}_{6}$ flips sign under an element $s_{[7,5]}\in \SO(40)$, while $a^{(19)}_{18}$ does so under $s_{[3,1]}\in \SO(40)$. Here, $s_{[n_1,n_2]}$ is defined as the element \begin{equation}
s_{[n_1,n_2]}: \bR^{40} = \bR^{n_1}\oplus \bR^{n_2}\oplus \bR^{40-n_1-n_2}
\ni a\oplus b \oplus c \mapsto (-a)\oplus (-b) \oplus c\label{fubar}
\end{equation} where the decomposition of $\bR^{40}$ to subspaces is done according to the partition $p$. 

The 4d quivers corresponding to $\rho_i$ are given respectively by \begin{equation}\begin{aligned}
\rho_1:& \cdots -\SO(14) - \cdots - \SO(38) - \cdots  ,&
\rho_2:& \cdots -\SO(14) - \cdots - \SO(39) - \cdots  ,\\
\rho_3:& \cdots -\SO(15) - \cdots - \SO(38) - \cdots  ,&
\rho_4:& \cdots -\SO(15) - \cdots - \SO(39) - \cdots  .
\end{aligned}\end{equation} So, the change in the local Coulomb branch can be accounted for if we impose invariance of operators under discrete elements 
\begin{equation}\label{AAA}
\begin{aligned}
\rho_1:& \emptyset, & \rho_2: & s_{[3,1]},\\
\rho_3:& s_{[7,5]}, & \rho_4: & s_{[3,1]}, s_{[7,5]},
\end{aligned}\end{equation} in addition to the invariance under the continuous part \eqref{allowedgauge}.
Note that the Hasse diagram of the partial order among $\rho_i$ is given by 
\begin{equation}
\vcenter{\hbox{\xymatrix@=8pt{
& \rho_1 \ns{dl}\ns{dr} \\
\rho_2 \ns{dr} && \rho_3 \ns{dl} \\
& \rho_4
}}},
\end{equation} which is the reverse of the inclusion of the discrete groups imposed in \eqref{AAA}.

\subsubsection{General recipe}\label{generalrecipe}
With the experience gained from the analysis of defects for $\fg=\su(N)$ and $\fg=\so(2N)$, let us propose the general procedure to determine the number and the scaling dimensions of the Coulomb branch operators associated to a defect labeled by $\rho:\su(2)\to\fg$.  In this section $G$ is taken to be in the adjoint form. 
We define $A(O)$ for an orbit $O$ as follows: pick $e\in O$, and let $\Flavour(O)$ be the subgroup of $G$ commuting with $e$, and $\Flavour(O)^\circ$ the connected component of $\Flavour(O)$ that contains the identity. Then $A(O)=\Flavour(O)/\Flavour(O)^\circ$ is the group of components of $\Flavour(O)$. $A(O)$ is trivial when $\fg=A_n$, is $(\bZ_2)^k$ for some $k$ when $\fg=B_n$, $C_n$ or $D_n$, and is $S_k$ for some $k$ when $\fg$ is exceptional. In particular, $A(O)$ is a Coxeter group.

Then we claim:
\emph{The Hitchin pole is given by $e\in d(O_\rho)$, and the Hitchin field $\Phi(z)$ has the form \eqref{phiz}. We first find all the functions of $\Phi(z)$ invariant under the connected part of the gauge group, generated by \eqref{allowedgauge}. Then we further impose invariance under a subgroup $\Sommers(\rho) \subset A(d(O_\rho))$. The resulting invariant functions are the local Coulomb branch operators.}

The remaining problem is the identification of the subgroup $\Sommers(\rho)$. Here, relatively recent work by Sommers and Achar \cite{Sommers,AcharSommers,Achar} on the theory of nilpotent orbits comes to the rescue.
They first constructed a map $f:O \mapsto (d(O),C(O))$ assigning to a nilpotent orbit in $\fg$
a pair, consisting of the Spaltenstein dual orbit $d(O)$ in $\dual{\fg}$ and a conjugacy class $C(O)$ in $\bar A(d(O))$, 
which is a certain quotient of $A(O)$ introduced by Lusztig. 
These maps have the following properties \cite{Sommers,Achar}
\begin{itemize}
\item $f(O_1)=f(O_2)$ if and only if $O_1=O_2$.
\item $f(O)=(d(O),1)$ if and only if $O$ is special.
\end{itemize}

Just as $A(d(O))$,  $\bar A(d(O))$ is also a Coxeter group.
Using this fact, they then assigned  a subset of simple reflections $\bar r_1,\ldots, \bar r_\ell \in \bar A(d(O))$  whose product lies in $C(O)$, see \cite{AcharSommers}. 
Let $r_1,\ldots, r_\ell$ be the corresponding simple reflections in $A(d(O))$, and let $\Sommers(O)$ be the subgroup of $A(d(O))$ generated by them.  In particular, $\Sommers(O)$ is trivial if $O$ is special.

For $\fg=\so(2N)$, the group $\Sommers(O_{\rho})$ obtained in this way agrees with 
the subgroup $\Sommers(O_\rho)$  identified physically in \cite{Chacaltana:2011ze} and reviewed in the last subsection.
Thus, we propose that \emph{the Coulomb branch operators for the defect of type $\rho$ are invariant under the subgroup $\Sommers(O_\rho)$}. 

\subsubsection{$\Sommers(O_\rho)$ for classical $\fg$}\label{subsec:sommers}
In this section we discuss $\Sommers(O_\rho)$ for $\fg=B_n,C_n,D_n$, and see how it can be used to determine the Coulomb branch operators. 
Let us first describe $A(O)$ and $\bar A(O)$, as in \cite{Sommers,AcharSommers,Achar}.
Let $\fg$ be of type $B$, ($C$, $D$), and take a nilpotent orbit $O$ whose partition type is $p=[p_1,p_2,\ldots]$.
Let $G$ be the corresponding compact adjoint group.
Then  $\Flavour(O)$, which is the commutant of $e\in O$ in $G$, is given by the following formula \cite{CollingwoodMcGovern}:
\begin{equation}\label{classicalflavourgroups}
\begin{aligned}
&\mathrm{S}\Bigl[ \prod_i (\mathrm{U}(r_i))^i_\Delta\Bigr]/\{\text{scalar matrices in }\mathrm{SL}(N)\}
&&\quad\text{when } G=\mathrm{SU}(N) ,\\
&\mathrm{S}\Bigl[ \prod_{i\, \text{odd}}(\mathrm{O}(r_i))^i_\Delta \times \prod_{i\, \text{even}} (\mathrm{Sp}(r_i/2))^i_\Delta\Bigr]
&&\quad\text{when } G=\mathrm{SO}(2N+1),\\
&\mathrm{S}\Bigl[ \prod_{i\, \text{odd}}(\mathrm{O}(r_i))^i_\Delta \times \prod_{i\, \text{even}} (\mathrm{Sp}(r_i/2))^i_\Delta\Bigr]/ \bZ_2
&&\quad\text{when } G=\mathrm{SO}(2N), \\
&\Bigl[\prod_{i\, \text{odd}}(\mathrm{Sp}(r_i/2))^i_\Delta \times \prod_{i\, \text{even}} (\mathrm{O}(r_i))^i_\Delta\Bigr]/\bZ_2
&&\quad\text{when } G=\mathrm{Sp}(N),
\end{aligned}
\end{equation}
where $H^i_\Delta$ stands for the diagonal subgroup of the direct product of $i$ copies of the group $H$. 
Therefore, $F(O)$ 
has a subgroup $\mathrm{O}(r_j)$ for each odd (even, odd) part $j$ in $p$ where $r_j=\#\{i| p_i=j\}$. Denote by $x_j$ a parity transformation in $\mathrm{O}(d_j)$. 
An integer  $j$ is called markable when there is an odd (even, even) $i$ such that $p_i=j$. 
Then $A(O)$ is generated by $x_j$ for type $C$, and $x_jx_{j'}$ for types $B$ and $D$.
For type $C$, the simple reflections of $\bar A(O)$ are $x_j$ for markable $j$.
For type $B$ and $D$, the simple reflections of $\bar A(O)$ are $x_jx_{j'}$ for 
markable $j$ and $j'$ such that there is no markable $j''$ such that $j<j''<j'$.

Now we can state how to obtain $\Sommers(O)$. 
Let $\fg$ be of type $B$, $C$, or $D$. 
Let $p$, $q$ be the partition type of $O_\rho$, $d(O_\rho)$ respectively. 
Let $q=[q_1,q_2,\ldots]$.
The conjugacy class $C(O_\rho)$ is given as follows.  
It is possible to split the parts of $p^t$ to $\mu$ and $\nu$, i.e.~$p^t=\mu\cup \nu$, so that  $\nu$ is a distinguished partition of type $C$, $D$, $D$, and $\mu$ is a partition of type $B$, $C$, $C$, respectively when $\fg=B_n,$ $C_n$, $D_n$. The partition $\nu=[\nu_1,\ldots,\nu_y]$ is guaranteed to be the subparts of $q$, and all parts of $\nu$ are markable.  Then the semisimple element  
in the class $C(O_\rho)$  is given by \begin{equation}
s=x_{\nu_{1}} x_{\nu_{2}} \cdots x_{\nu_y}.
\end{equation}   

For example, take $\rho_4$ in \eqref{rho4} of the $D_{20}$ theory. The partition is $p=[9,8^2,7,3,2^2,1]$ and the dual orbit has the partition type $q=[7^2,5^2,4^2,3^2,1^2]$. As such, $A(O_q)$ is generated by the parity transformation of 
$(\mathrm{O}(2)\times \mathrm{O}(2)\times \mathrm{O}(2)\times \mathrm{O}(2))/\bZ_2$, i.e.~it is $(\bZ_2)^3$.
The dual partition to $p$ is $p^t=[8,7,5,4^4,3,1]$.  Then $\nu=[7,5,3,1]$ and $\mu=[8,4^4]$; indeed $\nu$ is a subset of $q$. The element $s=x_1x_3x_5x_7$ is the product of simple reflections $x_1x_3$ and $x_5x_7$, which were denoted by  $s_{[3,1]}$ and  $s_{[7,5]}$ in \eqref{fubar}.
Then the discrete subgroup $\Sommers(O_{\rho_4})$ is generated by $s_{[3,1]}$ and $s_{[7,5]}$.

As another example, let us consider twisted defects of $A_3$ theory, which are labeled by nilpotent orbits of $B_2$. Take defects with Nahm poles $\rho_1=[3,1,1]$ and $\rho_2=[2,2,1]$.
The Hitchin poles are both $q=[2^2]$. Then $A(O_q)=\bZ_2$ is the parity transformation of $\mathrm{O}(2)$. 
For $\rho_1$, $\nu=\emptyset$, and thus $C(O_{\rho_1})$ and $\Sommers(O_{\rho_1})$ are both trivial. 
For $\rho_2$, $\nu=[2]$, and thus $x_2\in C(O_{\rho_2})$ is the nontrivial element of $A(O_q)$, and   $\Sommers(O_{\rho_2})=A(O_q)$. 
The study of the invariant polynomials $\phi^{(2,3,4)}$ shows that the degrees of poles of $\phi^{(2)}(z)$, $\phi^{(3)}(z)$ and $\phi^{(4)}(z)$ are  $1,3/2,3$ respectively, and there is one constraint of the form $\phi^{(4)}_3=(a^{(2)}_{3/2})^2$, where $a^{(2)}_{3/2}$ is a component of $\Phi_{-1/2}$. 
Therefore, we identify that the nontrivial element of $A(O_q)=\bZ_2$ acts on  $a^{(2)}_{3/2}$ by the multiplication by $-1$.
Let us denote by $n_d$ the number of Coulomb branch operators with scaling dimension $d$.
For $\rho_1$, we find $(n_2,n_3,n_4)=(2,3/2,2)$ and $n_v(\rho_1)=55/2$;
For $\rho_2$, we find  $(n_2,n_3,n_4)=(1,3/2,3)$ and $n_v(\rho_2)=63/2$.
These $n_v$ agree with what we obtain from the general formula \eqref{nv}. 
We can also check the number of Coulomb branch operators to the gauge theory analysis in \cite{Tachikawa:2009rb} and we find agreement. 

The conjugacy class $C(O)$ for any exceptional nilpotent orbit was determined and tabulated in \cite{Sommers}. The discrete groups $\Sommers(O)\subset A(O)$ can then be easily determined. 
We will show how they can be used  in Sec.~\ref{defects-g2} and in Sec.~\ref{defects-e8}.

\subsection{Case studies}\label{cases}
In this subsection, we will study various examples, to see how our algorithms work, and also to check the outcome against known results in the literature. Each subsections can be read separately. 

\subsubsection{$\fg=A_n$ and $D_n$}
Let us first perform more checks of the formul\ae\ \eqref{nvnh} for $n_v(\rho)$ and $n_h(\rho)$ by comparing them to the known results in the literature. These numbers  were calculated for $\fg$ of type $A$, type $C$ and  type $D$ in \cite{Nanopoulos:2009uw,Chacaltana:2010ks,Chacaltana:2011ze}. Their procedure relied on computing the contributions to graded dimensions of the Coulomb branch, $n_k(\rho)$.
Given the $n_k(\rho)$, the relation \eqref{oldnv} gives $n_v(\rho)$.

For the classical groups, a formula for the difference $n_h(\rho)-n_v(\rho)$ was derived from the quiver description \cite{Gaiotto:2009we,Tachikawa:2009rb}. 
Let $p$ be the partition corresponding to $\rho$. Define the transpose partition, $p^t=[s_i]$ and let $r_i= s_{i}-s_{i+1}$ be the number of times that ``$i$" appears in $p$, as before. Then we have
\begin{equation}\label{oldnhmnv}
n_h(\rho)-n_v(\rho) = \begin{cases}
\tfrac{1}{2}\Bigl(-N +{\displaystyle \sum_i} s_i^2\Bigr)&\text{if}\ \fg=A_{N-1},\\
\tfrac{1}{4}{\displaystyle \sum_i} s_i^2 - \tfrac{1}{2} {\displaystyle \sum_{i\, \text{odd}}} s_i &\text{if}\ \fg= D_N.
\end{cases}
\end{equation}
Combining this with \eqref{oldnv} yields a formula for $n_h(\rho)$.

In similar fashion, the levels of the simple factors in the flavour symmetry algebra, $\mathfrak{f}(\rho)$, given in \eqref{classicalflavour} were determined to be
\begin{equation}
 k_i = 2\sum_{j\leq i} s_j\label{anlevels}
\end{equation}
for each $\su(r_i)$  factor in  $A_{N-1}$. For $D_N$, each $\so(r_i)$ factor had level
\begin{equation}
k_i = \begin{cases}
2\Bigl({\displaystyle \sum_{j\leq i}} s_j\Bigr)-4&\text{if}\  r_i\geq 4\\
4\Bigl({\displaystyle \sum_{j\leq i}} s_j\Bigr)-8&\text{if}\  r_i=3
 \end{cases}
\label{dnlevels}
\end{equation}
and each $\mathfrak{sp}(r_i/2)$ had level
\begin{equation}
k_i = \sum_{j\leq i} s_j. \label{dnlevelsx}
\end{equation}

\begin{table}
\hspace*{-1em}\scalebox{1}{$\begin{array}{c||c|c|c|r@{}r@{}r@{}r@{}r@{}r}
\rho&\mathfrak{f}(\rho)&n_h&n_v & (n_2,&n_3,&n_4,&\dots,&n_{N{-}1},&n_N)\\
\hline
\hline
 [1^N]&\mathfrak{su}(N)_{2N}&\tfrac23(N{+}1)N(N{-}1)&\tfrac16N(N{-}1)(4N{+}1)&(1,&2,&3,&\dots,&N{-}2,&N{-}1)  \rule[-7pt]{0pt}{25pt}\\
 \hline
 \relax
 [2,1^{N{-}2}]&{\textstyle\mathfrak{su}(N{{-}}1)_{2N{-}2}\atop\textstyle \oplus\mathfrak{u}(1)}
&\tfrac23(N^3{-}11N{+}6)&\tfrac16(4N^2{+}5N{-}3)(N{-}2)&(1,&2,&3,&\dots,&N{-}2,&N{-}2)  \rule[-12pt]{0pt}{35pt}\\
\hline
 \relax 
 [N{-}1,1]&\mathfrak{u}(1)&N^2&(N{+}1)(N{-}1)&(1,&1,&1,&\dots,&1,&1)
\end{array}$}
\caption{The punctures for $A_{N-1}$ with zero, minimal and subregular Nahm poles. \label{anpuncts}}
\end{table}

\begin{table}
\hspace*{-3em}\scalebox{.9}{$\begin{array}{c||c|c|c|r@{}r@{}r@{}r@{}r@{}r@{}r@{}r}
\rho&\mathfrak{f}(\rho)&n_h&n_v&(n_2,&n_4,&n_6,&\dots,&n_{2N{-}4},&n_{2N{-}2}; &\tilde{n}_N)\\
\hline
\hline
[1^{2N}]&\mathfrak{so}(N)_{2N{+}4}&\tfrac43N(N{-}1)(2N{-}1)&\tfrac13N(N{-}1)(8N{-}7)&(1,&3,&5,&\dots,&2N{-}5,&2N{-}3;&N{-}1)   \rule[-7pt]{0pt}{25pt}\\
\hline \relax
[2^2,1^{2N{-}4}]&{\textstyle\mathfrak{su}(2)_{2N}\oplus\atop\textstyle\mathfrak{so}(2N{{-}}4)_{4N{{-}}8}}&\tfrac23(4N^3{-}6N^2{-}7N{+}12)&\tfrac13(8N^3{-}15N^2{-}5N{+}15)&(1,&3,&5,&\dots,&2N{-}5,&2N{-}4;&N{-}1)  \rule[-12pt]{0pt}{35pt}\\
\hline \relax
[2N{-}3,3]&{-}&4N^2{-}4N{-}10&4N^2{-}4N{-}9&(1,&1,&2,&\dots,&2,&2;&1)
\end{array}$}
\caption{The punctures for $D_N$ with zero, minimal and subregular Nahm poles. \label{dnpuncts}}
\end{table}

We can then compare the new formul\ae\ \eqref{nvnh}, \eqref{4dk} against
the ones calculated by old independent methods. 
The results for $\rho$ the zero, minimal and sub-regular orbits are 
 in Tables~\ref{anpuncts},\ref{dnpuncts}, showing perfect agreement.

\subsubsection{The full puncture for general simply-laced $\fg$}\label{subsec:full}
Next, let us consider the case where $\rho$ is the zero nilpotent orbit of simply-laced $\fg=\type$, i.e., the maximal (or full) untwisted puncture. This is a special puncture, and there is no complication from $\Sommers(\rho)$.
The flavour symmetry $\Flavour(\rho)$ is just the whole of $\Type$, and the level is obviously $2\coxeter(\Type)$. Therefore, two such punctures can be connected to an $\cN{=}2$ vector multiplet of gauge group $\Type$ with zero beta function. $n_v(\rho)$ and $n_h(\rho)$ can be calculated from \eqref{nvnh} and we find\begin{align}
n_v(\rho)&=\frac23 \coxeter(\Type)\dim\Type  -\frac12(\dim \Type-\rank \Type), &
n_h(\rho)&=\frac23 \coxeter(\Type)\dim\Type.
\end{align}
The Hitchin pole for this puncture is the principal orbit $O_\text{prin}$, with $\dim_\bC O_\text{prin}=\dim \Type-\rank \Type$.
The invariant polynomials $\phi^{(d_i)}(z)$ have poles of order $d_i-1$ at most, where the $d_i$ are the degrees of the Casimir invariants of $\type$. (For $\type=D_N$, one of these invariant polynomials will be the Pfaffian, $\tilde{\phi}^{(N)}$.) Since $\sum_i (d_i-1)= \frac12 \dim_\bC O_\text{prin}$, the coefficients $\phi^{(d_i)}_{k}$ for $1\le k \le d_i-1$ are all independent. We can also verify that\begin{equation}
n_v(\rho)=\sum_i (2d_i-1)(d_i-1).
\end{equation} We conclude that the local Coulomb branch operators for the defect of type $\rho=0$ consist of $(d_i-1)$ operators of scaling dimension $d_i$. 
\footnote{
A comment is in order here. Recall that our operators $\phi^{(d_i)}_k$ is the coefficient of $z^{-k}$ of $P^{(d_i)}({\Phi_{-1}}/z + \Phi_0)$ where $\Phi_{-1}\in O_\text{prin}$ and $\Phi_0$  is a generic element. We have fixed $\Phi_{-1}$ to a particular principal nilpotent element above; but instead, we can put $\Phi_0$ to a diagonalisable element $a$ and consider \begin{equation}
\phi^{(d_i)}_k= \text{coefficient of $t^{d_i-k}$ of}\   P^{(d_i)}(x + at)\label{shift}
\end{equation} as functions on $x\in O_\text{prin}$.  The principal nilpotent orbit $O_\text{prin}$ is a holomorphic symplectic manifold with the standard Kostant-Kirillov symplectic form, and these $\phi^{(d_i)}_k$ are known to form complete integrals of motion. The holomorphic symplectic form of the Hitchin system on a sphere comes from the symplectic forms on the nilpotent orbits giving the residues, the analysis of a single nilpotent orbit can be thought of as a local version of our problem.
This integrable system, for the principal nilpotent orbit, is called the Gelfand-Zeitlin (or Cetlin) system \cite{GuilleminSternberg}, and the way to generate Poisson-commuting functions by shifting the invariant polynomial in a shift in a constant direction $at$ as in \eqref{shift} is called the shift-of-argument method of Mishchenko--Fomenko \cite{MishchenkoFomenko}. The restriction of these functions on a smaller nilpotent orbit $O$ is known to give complete integrals of motion if Elashvili's condition is satisfied \cite{Bolsinov}, and Elashvili's condition is now known to be always satisfied \cite{CharbonnelMoreau}.   Therefore, all nilpotent orbits are now known to be completely integrable.
%\footnote{
The authors thank the posters at \href{http://mathoverflow.net/questions/85467/}{the entry 85467 of MathOverflow} for discussions, in particular A. Chervov.
%} 
In a sense we need a stronger version of this statement, that the ring of integrals of motion is a polynomial ring, whose generators are our local Coulomb branch operators. 
See Appendix~\ref{conjecture} for a more precise mathematical formulation.
}

\subsubsection{A family of interacting three-punctured spheres}\label{R2n}

Consider the 6d $A_{2N-1}$ theory on the three-punctured sphere with defects:
\begin{itemize}
\item the untwisted defect with subregular Nahm pole $[2N-1,1]$ of $\su(2N)$.
\item two copies of the twisted defect whose Nahm pole is the zero orbit $[1^{2N+1}]$ of $\so(2N+1)$.
\end{itemize}
From \eqref{nvnh}, the twisted defect $[1^{2N+1}]$ contributes
\begin{equation}
\begin{split}
n_h([1^{2N+1}])&=\tfrac{4}{3}N(4N^2-1)\\
n_v([1^{2N+1}])&=\tfrac{1}{6}(32N^3-6N^2-5N-3)
\end{split}
\end{equation}
Its contribution to the graded dimension of the Coulomb branch is
\begin{equation}
(n_2,n_3,n_4,\dots,n_{2N-1},n_{2N}) = \bigl(1,\tfrac{5}{2},3,\dots,\tfrac{4N-3}{2},2N-1\bigr)
\end{equation}
which agrees with the prediction of \eqref{twisted3dCoulomb} for the contribution to the total Coulomb branch dimension,
\begin{equation}
\dim \Coulomb([1^{2N+1}])= \tfrac{1}{2}(4N^2-N-1)
\end{equation}
The contribution \eqref{dimHiggs} to the Higgs branch dimension is $\dim\Higgs([1^{2N+1}])=N^2$ and the flavour symmetry algebra and its level are $\mathfrak{f}([1^{2N+1}])= \so(2N+1)_{4N-2}$.

Putting these results together with ones for the untwisted sector puncture, collected in Table~\ref{anpuncts}, we find that this three-punctured sphere has
\begin{equation}
\begin{split}
n_h&=4N^2\\
n_v&=(2N+1)(N-1)\\
(n_2,n_3,\dots,n_{2N-1},n_{2N})&=(0,1,0,\dots,1,0)\\
\dim\Higgs&=2N^2+N+1\\
\mathfrak{f}&\supset \so(2N+1)_{4N-2}\oplus\so(2N+1)_{4N-2}\oplus\mathfrak{u}(1)
\end{split}
\end{equation}
where, in the last line, we have allowed for an enhancement of the flavour symmetry algebra over the na\"\i ve one associated to the punctures. In fact, if we allow for the enhancement to $\mathfrak{f}=\so(4N+2)_{4N-2}\oplus\mathfrak{u}(1)$ (and further enhanced to $(\mathfrak{e}_6)_6$ when $N=3$), these are precisely the invariants of the family of interacting SCFTs that we called $R_{2,2N-1}$ in \cite{Chacaltana:2010ks}. It is quite common to find different fixtures which nonetheless realize the same SCFT. In \cite{Chacaltana:2010ks}, $R_{2,2N-1}$  was realized by the compactification of the $A_{2N-2}$ theory on a three-punctured sphere with untwisted sector Nahm poles $[(N-1)^2,1],[(N-1)^2,1],[1^{2N-1}]$; on the evidence, we believe that these two series of three-punctured spheres give rise to the same SCFTs in the 4d limit.  In particular, $R_{2,3}$ is the $(E_6)_6$ SCFT of Minahan and Nemeschansky.

\subsubsection{The torus with twist lines}
Let us illustrate the formul\ae\ \eqref{dimHiggs}, \eqref{twisted3dCoulomb} of the dimensions of the Coulomb and the Higgs branches  with some simple examples.

\begin{itemize}

\item As a first example, consider  the $A_{3}$ theory, compactified on $T^2$, with the insertion of two simple (principal) punctures, $[5]$, from the $\mathbb{Z}_2$ twisted sector. From \eqref{twisted3dCoulomb}, each puncture contributes
\begin{equation}
\dim\Coulomb([5]) = \tfrac{1}{2}\dim (\su(4)/\so(5)) = \tfrac{5}{2}
\end{equation}
to the Coulomb branch dimension, but nothing to the Higgs branch dimension. Since we have twisted sector punctures, $\fg'=\so(5)$. Hence
\begin{equation}
 \dim\Coulomb = 5,\qquad \dim\Higgs=2
\end{equation}
The physical interpretation of this theory is an $\SU(4)\times \Sp(2)$ gauge theory, with matter in the $(6,4)$, see \cite{Tachikawa:2009rb}. The Coulomb branch is, as expected, 5-dimensional. At a generic point on the Higgs branch, the gauge symmetry is broken to $\U(1)^3$, which, also as expected, yields a 2-dimensional Higgs branch. Moreover, $n_v=25$, $n_h=24$, as calculated from \eqref{totalnvnh}, agree with the physical interpretation.

\item
As a second example, consider the $A_{2N-1}$ theory again, compactified on a $T^2$, with one insertion of the simple (subregular) puncture, $[2N-1,1]$. $d(O_{\text{subreg}})=O_\text{min}$, so the Coulomb branch has dimension
\begin{equation}\label{CoulombExample1}
    \dim \Coulomb = \frac{1}{2} \dim(O_\text{min}) = 2N-1
\end{equation}
The dimension of the Higgs branch depends on whether we insert an outer-automorphism twist line on the torus. The subregular puncture contributes 1 to $\dim\Higgs$. \emph{Without} the twist line, $\fg'=\fg=su(2N)$, which yields
\begin{subequations}
\begin{equation}\label{HiggsExample1}
    \dim\Higgs = 1 + (2N-1) = 2N\quad.
\end{equation}
\emph{With} the insertion of an outer-automorphism twist line, $\fg'=\so(2N+1)$, and 
\begin{equation}\label{HiggsExample1twisted}
    \dim\Higgs = N+1\quad.
\end{equation}
\end{subequations}

These are in accord with the physical interpretation of these compactifications of the $A_{2N-1}$ (2,0) theory.
\begin{itemize}
 \item \emph{Without} the outer automorphism twist, we have $\SU(2N)$ gauge theory with one free hypermultiplet and one hypermultiplet in the adjoint. That clearly agrees with \eqref{CoulombExample1},\eqref{HiggsExample1}.
 \item \emph{With} the outer automorphism twist, we have two possibilities, related by S-duality.
\begin{itemize}
\item An $\SU(2N)$ gauge theory, with matter in the anti-symmetric tensor of dimension $N(2N-1)$ plus the symmetric tensor of dimension $N(2N+1)$. On the Higgs branch, the gauge symmetry is broken down to the $\mathrm{U}(1)^N$ which is the Cartan torus of the $\Sp(N)$ subgroup which preserves the VEV of the antisymmetric. The Higgs branch is thus $(N+1)$-dimensional, as predicted by \eqref{HiggsExample1twisted}.
\item A $\Spin(2N+1)$ gauge theory, coupled to the SCFT we just discussed in \S\ref{R2n}. All the other numerical invariants match this physical description. The only one we cannot check directly for general $N$ is \eqref{HiggsExample1twisted}, because the precise nature of the Higgs branch is not currently known. For $N=2$, however, we can be quite precise. $R_{2,3}$ is the $(E_6)_6$ SCFT of Minahan and Nemeschansky.
Its Higgs branch is known \cite{Gaiotto:2008nz} to be the minimal nilpotent orbit of $E_6$, which is 11-dimensional. Upon the $\Spin(5)$ gauging, the $\Spin(5)$ is broken, at a generic point on the Higgs branch, to the Cartan torus. The Higgs branch is $11-8=3$ dimensional, again as predicted by \eqref{HiggsExample1twisted}.
\end{itemize}
\end{itemize}
\end{itemize}

\subsubsection{Defects of $G_2$}\label{defects-g2}
Let us treat a case with an outer-automorphism twist in detail: $\fg=G_2$. This comes from the $\bZ_3$ outer-automorphism $o$ of $D_4$.  This example will also serve as a model for studying exceptional nilpotent orbits.
\paragraph{Nilpotent orbits of $G_2$}

The nilpotent orbits of $G_2$ are tabulated in Table~\ref{G2}. 
Let $E_{\gamma}$ be the generator of $\fg$ corresponding to the root vector $\gamma$.
\begin{itemize}
\item Consider the zero orbit. The smallest Levi subalgebra containing it is $\U(1)^2\subset G_2$. The Bala-Carter classification assigns a label to this orbit, called the \emph{Bala-Carter label}, essentially the non-Abelian part of the Levi subalgebra. In this case the Bala-Carter label is just $0$. 
\item Next, consider the orbit containing $E_\alpha$, where $\alpha$ is the long simple root of $G_2$. The smallest Levi subalgebra containing this orbit is $\U(1)\times A_1 \subset G_2$, and the $E_{\alpha}$ orbit is distinguished in it, because it is principal. This orbit has Bala-Carter label $A_1$.
\item Similarly, consider the orbit containing $E_\beta$, where $\beta$ is the short simple root of $G_2$. The smallest Levi subalgebra containing this orbit is again $\U(1)\times A_1\subset G_2$ by construction, but this subalgebra is \emph{not} conjugate to the one for $E_\alpha$. To specify that this $A_1$ involves the short root, rather than the long one, we we write its Bala-Carter label as $\tilde A_1$.
\item Let us next consider the principal orbit, which contains generic nilpotent elements in $G_2$. The smallest Levi subalgebra that contains it is $G_2$ itself, and so its Bala-Carter label is, accordingly, $G_2$. 
\item Finally, consider the orbit containing $E_\gamma+E_\alpha$, where $\gamma$ is the lowest root.  
Recall that the extended Dynkin diagram for $G_2$ is $\circ - \circ \Rrightarrow \circ$, so here $\gamma$ and $\alpha$ correspond, respectively, to the leftmost and middle node. 
Therefore $E_\gamma+E_\alpha$ is distinguished in a subalgebra $A_2$ of $G_2$. However, this is not a Levi subalgebra.
There is an order-3 semisimple group element $s$ of $G_2$ such that the subalgebra of $G_2$ commuting with it is this $A_2$;
such an algebra is called \emph{pseudo-Levi subalgebra}. The actual smallest Levi subalgebra containing $E_\gamma+E_\alpha$ is $G_2$ itself, and this orbit is distinguished in $G_2$. This orbit has Bala-Carter label $G_2(a_1)$. Let us explain this funny label below.
\end{itemize}

\begin{table}
\[
\rowcolors{0}{}{gray!10}
\begin{array}{c|c|c|c|c|l}
\text{B-C label} & \circ \Rrightarrow\circ & \dim_\bC O& A(O) & \flavour(O) &\text{Orbit properties} \\
\hline
\hline
0 & 00 & 0 & 1 & G_2 & \text{rigid} \\
A_1 & 10 & 6 & 1 & A_1 & \text{rigid, non-special} \\
\tilde A_1 & 01 & 8 & 1 & A_1 & \text{rigid, non-special} \\
G_2(a_1) & 20 & 10 & S_3  & 1 &  \text{distinguished}\\
G_2 & 22 & 12 & 1 &1 &  \text{distinguished}
\end{array}
\quad
\vcenter{\hbox{\xymatrix@=8pt{
G_2 \sp{d}\\
G_2(a_1) \ns{d}\\
\relax\rigid{\tilde A_1} \ns{d}\\
\relax\rigid{A_1} \sp{d}\\
\relax\rigid{0}
}
\quad
\xymatrix@=8pt{
G_2 \sp{d}\\
G_2(a_1) \sp{d}\\
\relax\rigid{0}
}}}
\]
\caption{Nilpotent orbits of $G_2$ and their partial ordering under closure.\label{G2}}
\end{table}

For a nilpotent element $e\in\fg$, pick a map $\rho:\su(2)\to \fg$ such that $e=\rho(\sigma^+)$. Let $h=\rho(\sigma^3)$,
and choose simple roots $\alpha_i$ so that the products $\alpha_i\cdot h$ are non-negative; one can show that these products can only be 0, 1 or 2.
We mark the nodes of the Dynkin diagram by the corresponding integers $\alpha_i\cdot h$, and the resulting marked Dynkin diagram is called \emph{the weighted Dynkin diagram} for the nilpotent orbit containing $e$.
For instance, the principal orbit is known to have $\alpha_i\cdot h=2$ for all $i$. For other distinguished orbits of $\fg=X_N$, the Bala-Carter label is $X_N(a_i)$, where $i$ is the number of zeros in the weighted Dynkin diagram.
If there is more than one  distinguished orbit with the same number of zeroes in the weighted Dynkin diagram, 
we label them $X_N(a_i)$, $X_N(b_i)$, \ldots, corresponding to the distinguished orbits in decreasing order in their orbit dimensions.
Coming back to the orbit for $E_\gamma+E_\alpha$,  the weighted Dynkin diagram is $2\Rrightarrow 0$, and so the Bala-Carter label for this orbit is $G_2(a_1)$.

Among the five orbits, $0$, $A_1$ and $\tilde A_1$ are rigid, while $G_2(a_1)$ and $G_2$ are distinguished. The Spaltenstein map exchanges $0$ and $G_2$, and maps all three of $A_1$, $\tilde A_1$ and $G_2(a_1)$ to $G_2(a_1)$. Thus, $A_1$ and $\tilde A_1$ are non-special. 
The transverse space $X$ to the subregular orbit $G_2(a_1)$ inside the closure of the principal orbit $G_2$ is complex two-dimensional and has an action of $S_3=A(G_2(a_1))$, which is the symmetric group on three letters. $X$ is in fact an ALE space of type $D_4$, and the presence of $S_3$ is related to the fact that $\dual{G_2}=G_2$ is the subalgebra of $D_4$ invariant under $S_3$.

\paragraph{Properties of defects}
Now that we reviewed the nilpotent orbits of $G_2$, the dimension of the Coulomb branch $\dim_\bC\Coulomb_{4d}(\rho)$ 
and the Higgs branch $\dim_\bH\Higgs(\rho)$ can be determined easily using \eqref{twisted3dCoulomb} and \eqref{dimHiggs}.
$n_v(\rho)$ and $n_h(\rho)$ follow from \eqref{nvnh}.

The Hitchin pole has the structure that we found in \eqref{hitchinpole3}: \begin{equation}
\Phi(z)= \left[\frac{e}z + \frac{\Phi_{-2/3}}{z^{2/3}}+\frac{\Phi_{-1/3}}{z^{1/3}} + \Phi_0 \right]dz
\label{g2hitchin}
\end{equation} where $\Phi_{-2/3}$, $\Phi_{-1/3}$, $\Phi_0$ are generic elements of $D_4$ whose eigenvalue under $o$ is, respectively, $\omega^2=e^{4\pi i/3}$, $\omega=e^{2\pi i/3}$ and 1.

As for gauge-invariant fields, we start from $\phi^{(2,4,6)}(z)$ and $\tilde\phi^{(4)}$, and redefine \begin{equation}
\phi^{[4]}=\phi^{(4)}-\frac14(\phi^{(2)}){}^2,\quad
\phi^{[6]}=\phi^{(6)}-\frac16\phi^{(2)}\phi^{[4]}.
\end{equation}  Then $\phi^{(2)}$ and $\phi^{[6]}$ are invariant under $o$, whereas \begin{equation}
\phi^{(4)}_\omega=\phi^{[4]}+2\sqrt{3} i \tilde\phi^{(4)},\qquad
\phi^{(4)}_{\omega^2}=\phi^{[4]}-2\sqrt{3} i \tilde\phi^{(4)}
\end{equation} become the eigenstates for $\omega$, $\omega^2$, respectively. So,  $\phi_2, \phi^{(4)}_\omega, \phi^{(4)}_{\omega^2}, \phi^{[6]}$ form a well-defined basis of differentials to describe the Coulomb branch; we will denote their respective pole orders by $\{p_2,p_{4,\omega},p_{4,\omega^2},p_6\}$.

The special orbits are $0$, $G_2(a_1)$ and $G_2$, with dimensions $0$, $10$ and $12$ respectively; since $\dim \SO(8)/G_2=14$, we expect the number of Coulomb branch operators to be $7$, $12$ and $13$. 
We can study the pole structures and the constraints satisfied by the pole coefficients for these Hitchin poles using the explicit embedding of $G_2$ in $D_4$ in Appendix~\ref{explicitG2}. We find
\begin{itemize}
\item For the Hitchin pole $O_{G_2}$,  $\{p_2,p_{4,\omega},p_{4,\omega^2},p_6\}=\{1,\frac{10}{3},\frac{11}{3},5\}$. There are no constraints.
\item For the Hitchin pole $O_{G_2(a_1)}$,  $\{p_2,p_{4,\omega},p_{4,\omega^2},p_6\}=\{1,\frac{10}{3},\frac{8}{3},5\}$. The constraints are of the form \begin{align}
\phi^{[4]}_{10/3}&=-6\left(\left(a^{(2)}_{5/3}\right)^2+\left(a'^{(2)}_{5/3}\right)^2\right),& 
\phi^{[6]}_5&=8a^{(2)}_{5/3}\left(\left(a^{(2)}_{5/3}\right)^2-3\left(a'^{(2)}_{5/3}\right)^2\right)\label{g2a1-constraint}
\end{align} where $a^{(2)}_{5/3}$ and $a'{}^{(2)}_{5/3}$ are particular components of $\Phi_{-1/3}$.
\item For the Hitchin pole $O_0$,  $\{p_2,p_{4,\omega},p_{4,\omega^2},p_6\}=\{1,\frac{7}{3},\frac{8}{3},4\}$. We find the constraints of the form
\begin{equation}\label{g2-0-const}
\begin{aligned}
\phi^{[4]}_{8/3}&=-6\left(a^{(2)}_{4/3}\right)^2,&
\phi^{[4]}_{7/3}&=\phi^{(2)}_{1} a^{(2)}_{4/3} ,\\ 
\phi^{[6]}_{4}&=-8\left(a^{(2)}_{4/3}\right)^3,& 
\phi^{[6]}_{3}&=\tfrac{1}{54}\left(\phi^{(2)}_1\right)^3+2a^{(2)}_{4/3}\phi^{(4)}_{5/3}
\end{aligned}\end{equation} where $a^{(2)}_{4/3}$ is a component of $\Phi_{-1/3}$.
\end{itemize}

\begin{table}
\[
\rowcolors{0}{}{gray!10}
\begin{array}{c||c|c|c|c|c|c|c|c|cc}
O_\rho  & d(O_\rho) & \Sommers(\rho)  & \flavour(\rho) & \dimH   & \dimC  & n_h(\rho) & n_v(\rho) &k_{4d}&  (n_2,n_4,n_6) \\
\hline\hline
G_2 &   0  &   1   &   & 0 & 7 & 48 & 49 && (2,3,2) \\
G_2(a_1) &   G_2(a_1)  &  1&  & 1 & 12 & 88 & 88 && (3,5,4) \\
\tilde A_1 &   G_2(a_1)  &  S_2 &  A_1 &  2& 12 & 93 & 92 &5& (2,6,4) \\
 A_1 &   G_2(a_1)  &  S_3 &\tilde A_1 &   3& 12 & 102 & 100 &14& (1,6,5) \\
 0 &   G_2  & 1  &  G_2&  6 & 12   & 112 & 107&8 & (1,7,5) 
\end{array}
\]
\caption{Properties of the $G_2$ defects.\label{G2detaileddata}}
\end{table}

Using these data, we find the properties of the defects as follows: \begin{itemize}
\item For the Nahm pole with Bala-Carter label $G_2$, the Hitchin pole is the zero orbit, $0$.  The constraint structure  \eqref{g2-0-const} means that we have two dimension-2 operators $\phi^{(2)}_1$ and $a^{(2)}_{4/3}$ , \emph{three}\footnote{This is not a typo. Since a $\bZ_3$ puncture cannot be introduced alone, we need to introduce, say, a pair. Then the two punctures combined will have $5/3+4/3+5/3+4/3=6$ Coulomb branch operators of dimension 4, from \eqref{totalCoulomb}. Then we need to assign three Coulomb branch operators to each of the two punctures. } dimension-4 operators 
$\phi^{[4]}_{5/3}$, $\phi^{[4]}_{2/3}$,  $\phi^{[4]}_{4/3}$,  $\phi^{[4]}_{1/3}$, and two dimension-6 operators $\phi^{[6]}_2$, $\phi^{[6]}_1$. The total $n_v=2\cdot 3+3\cdot 7+2\cdot 11=49$ agrees with what we get from \eqref{nv}.
\item For the Nahm pole with label $G_2(a_1)$, which is special, the Hitchin pole is again $G_2(a_1)$. 
The discrete group $A(O)$  is $S_3$.
The constraint structure \eqref{g2a1-constraint} is exactly what one expects for there is an action of $S_3$ on the plane spanned by $a^{(2)}_{5/3}$ and $a'^{(2)}_{5/3}$. Then there are three operators of dimension 2, five operators of dimension 4, and four operators of dimension 6. The total $n_v$ is $88$, which agrees with that from \eqref{nv}.
\item For the Nahm pole with label $\tilde A_1$, which is non-special, the Hitchin pole is again $G_2(a_1)$. Sommers \cite{Sommers} assigns a subgroup $S_2\subset S_3$ to this Nahm pole. Accordingly, we  pick $a^{(2)}_{5/3}$ and $(a'^{(2)}_{5/3})^2$ to be the generators of the Coulomb branch operators.
Then we have  two operators of dimension 2, six operators of dimension 4, and four operators of dimension 6. The total $n_v$ is $92$, which agrees with that from \eqref{nv}.  The flavour symmetry is $A_1$. Under $\tilde A_1\times A_1\subset G_2$, $\fg_2$ decomposes as $(\mathbf3,\mathbf1)+(\mathbf1,\mathbf3)+(\mathbf2,\mathbf4)$. Using \eqref{4dk}, we find the flavour central charge to be $k_{4d}(A_1)=5$. 
\item For the Nahm pole with label $A_1$, which is also non-special, the Hitchin pole is again $G_2(a_1)$. Sommers assigns to this defect the group $S_3$ itself. Accordingly, 
we pick $\phi^{[4]}_{10/3}$ and $\phi^{[6]}_5$ to be the generators of the Coulomb branch operators. We have then one operator of dimension 2, six operators of dimension 4, and five operators of dimension 6. The total $n_v$ is $49$, which agrees that from \eqref{nv}. The flavour symmetry is $\tilde A_1$. Using the decomposition under $A_1\times \tilde A_1$ given in the previous example, together with the formula \eqref{4dk}, we find the flavour central charge to be $k_{4d}(\tilde A_1)=14$.
\item Finally, for the Nahm pole with label $0$, i.e., the zero orbit, the Hitchin pole is $G_2$. There are no constraints, and thus we directly have 
one operator of dimension 2, seven operators of dimension 4, and five operators of dimension 6. The total $n_v$ is $107$, which agrees with that from \eqref{nv}.
\end{itemize}
The information we just obtained is summarised in Table~\ref{G2detaileddata}.

\subsubsection{The special piece of $E_8(a_7)$ of  $\fg=E_8$.}\label{defects-e8}

\begin{table}
\[
\raisebox{4em}{\small\xymatrix@=7pt@C=-20pt{
&E_8(a_7)\ns{d}\\
&E_7(a_5)\ns{dl}\ns{dr}\\
E_6(a_3)+A_1 \ns{d}\ns{drr} && D_6(a_2)\ns{d}\ns{dll} \\
A_5+A_1 \ns{dr} &&D_5(a_1)+A_2 \ns{dl}\\
&A_4+A_3
}}
\rowcolors{2}{}{gray!10}
\begin{array}{c|c|c|c|c}
\rho & \Sommers(\rho) & n_v(\rho) &   \text{\small known ops} & \text{\small known $n_v$} \\
\hline
\hline
E_8(a_7) & \emptyset &4064&\pz6,\pz6,\pz6,\pz6 & 44 \\
E_7(a_5) &(12) & 4076 &\pz6,\pz6,\pz6,12  & 56\\
D_6(a_2) &(12),(34) & 4088 &\pz6,\pz6,12,12 & 68\\
E_6(a_3)+A_1 & (12),(23) &4100&\pz6,\pz6,12,18 & 80\\
A_5+A_1&  (12),(23),(45) &4112&\pz6,12,12,18 & 92\\
D_5(a_1)+A_2& (12),(23),(34)& 4136&\pz6,12,18,24 & 116\\
A_4+A_3& (12),(23),(34),(45)& 4184   &12,18,24,30  & 164
\end{array}
%\xymatrix@=8pt@C=-15pt{
%&\emptyset\ns{d}\\
%&(12)\ns{dl}\ns{dr}\\
%(12),(23) \ns{d}\ns{drr} && (12),(34)\ns{d}\ns{dll} \\
%(12),(23),(45) \ns{dr} &&(12),(23),(34) \ns{dl}\\
%&(12),(23),(34),(45)
%}
\]
\caption{A special piece in the set of nilpotent orbits of $E_8$,  the corresponding subgroups of $S_5=A(E_8(a_7))$, $n_v$ and the scaling dimensions of operators governed by subgroups of $S_5$. The  fourth column shows the scaling dimensions of 4 Coulomb branch operators out of the 104 that each of the 7 defects in the list has. The remaining 100 operators have the same scaling dimensions. The fifth column shows the contribution to $n_v$ just from the known 4 operators.\label{E8a7}}
\end{table}

For another check of our proposal about how the discrete groups $\Sommers(\rho)$ affect the generators of the Coulomb branch, let us take a look at the nilpotent orbits in $E_8$ that map via Spaltenstein to the nilpotent orbit $E_8(a_7)$. There are 7 such orbits, shown in Table~\ref{E8a7}.
%Among these, the special nilpotent orbit is $E_8(a_7)$ itself, which is self-dual under Spaltenstein.
$A(E_8(a_7))$ is $S_5$, and the subgroup of $S_5$ assigned to each of the 7 nilpotent orbits by Sommers \cite{Sommers} is also shown in the table, in terms of the generating reflections $(i,i+1)$, which act on the set $\{1,2,3,4,5\}$.
Looking up $h$ in Table~\ref{tableE8-1} and using \eqref{nv}, one can compute $n_v(\rho)$ for each nilpotent orbit.  

Since $\dim_\bC E_8(a_7)=208$, the special defect should add 104 operators to the Coulomb branch. The scaling dimensions of four of them can be determined as follows. 
Since $A(E_8(a_7))$ is $S_5$, for the special nilpotent orbit $\rho=E_8(a_7)$ there will be four operators of dimension $d$ on which $S_5$ would act as the Weyl group of $A_4$. Then, the Coulomb branch for the orbit $\rho=A_4+A_3$ must contain four operators of scaling dimensions $\{2d,3d,4d,5d\}$. These scaling dimensions should be contained in the set of degrees of Casimir invariants of $E_8$, $\{2,8,12,14,18,20,24,30\}$. The only possibility is $d=6$. Then, for each of the 7 defects, $\Sommers(\rho)$ determines the scaling dimensions of these four operators, which are listed in the fourth column of Table~\ref{E8a7}, while the contribution to $n_v$ from just these four operators is listed in the fifth column. The remaining 104-4=100 Coulomb branch operators have unknown scaling dimensions (though still constrained by the total $n_v(\rho)$), but whichever they are, they should be completely the same for the 7 defects. As a consistency check, the difference between the full local contribution $n_v(\rho)$ (from the 104 operators) and the contribution to $n_v$ from just the known 4 operators should be a constant. This is indeed so. The difference between entries on the same row in the third and fifth columns of Table~\ref{E8a7} is always 4020.

\subsubsection{Free-field fixtures}

\begin{table}
\[
%\rowcolors{0}{}{gray!10}
\begin{array}{c||c|c|c|c|c|c|c|cc}
\fg & O_\rho   & \dual{\fg} & d(O_\rho)  & \dimH  & \dimC & n_h & n_v & (n_2,n_3,n_4,n_5,n_6) \\
\hline\hline
\su(6) & [1^6] & \su(6) & [6] & 15 & \pm15 & \pm140 & \pm125 & \pm(1,2,3,4,\pz5) \\
\so(7) & [7] & \mathfrak{sp}(3) & [1^6] & \pz0 & \pm\pz7 & \pm\pz52 & \pm\pz53  &  \pm(1,\tfrac32,1,\tfrac32,\pz2) \\
\so(7) & [3^2,1] & \mathfrak{sp}(3) & [2^3] & \pz2 & \pm13 & \pm108 & \pm107  &  \pm(1,\tfrac32,3,\tfrac72,\pz4)\\
\hline
\text{bulk} &&&& \pz3 & -35 &  -280 & -285 & -(3,5,7,9,11) \\
\hline
\text{total} &&&& 20 & \pm\pz0 &  \pm\pz20 & \pm\pz\pz0 &  \pm(0,0,0,0, \pz0) 
\end{array}
\]
\caption{Three defects on a sphere giving a hypermultiplet in the $\mathbf{20}$ of $\SU(6)$. \label{20}}
\end{table}

Our final group of examples will be 3-punctured spheres known to correspond to free hypermultiplets by methods different from the ones in this paper. 
Our formul\ae\ should reproduce their properties.
\begin{itemize}
\item First, compactify the 6d theory of type $A_5$ on a sphere with 3 punctures, where the first is the full untwisted defect with Nahm pole $[1^6]$, the second is a twisted defect with principal Nahm pole $[7]$ of $\so(7)$, and the third is a twisted defect with Nahm pole $[3^2,1]$ of $\so(7)$. In \cite{Tachikawa:2011yr}, it was found that this sphere describes a hypermultiplet in the 3-index antisymmetric tensor representation of $\SU(6)$, by 
determining the Hitchin poles from the Seiberg-Witten curve; see Sec.~3.2 of that paper.
The properties of defects, calculated using the methods in this paper, are in Table~\ref{20}. The individual contributions from the defects are to a contribution from the bulk, using \eqref{totalHiggs}, \eqref{total3dCoulomb}, \eqref{totalnvnh} and \eqref{totalCoulomb}. We see that, indeed, this sphere describes 20 free hypermultiplets. The flavour symmetries for the defects are $\flavour([1^6])=\su(6)$, $\flavour([7])=1$ and $\flavour([3^2,1])=\so(2)$. 

\begin{table}
\[
%\rowcolors{0}{}{gray!10}
\begin{array}{c||c|c|c|c|c|c|c|r@{\ }r@{\ }r}
\fg & O_\rho   & \dual{\fg} & d(O_\rho)  & \dimH  & \dimC & n_h & n_v & (n_2,&n_4,&n_6) \\
\hline\hline
\fg_2 & 0 & \fg_2 & G_2 & \pz6 & \pm13 & \pm112 & \pm107 & (1,&7,&5) \\
\mathfrak{sp}(3) & [6] & \so(7) & [1^7] &\pz0 & \pm\pz\tfrac72 & \pm\pz24 & \pm\pz\tfrac{49}2  & (1,&\tfrac32,&1) \\
\mathfrak{sp}(3) & [2,1^4] & \so(7) & [5,1^2] & \pz6 & \pm\pz\tfrac{23}2 & \pm102 & \pm\tfrac{193}2  & (1,&\tfrac{11}2,&5) \\
\hline
\text{bulk} &&&& \pz2 & -28 &  -224 & -228 & -(3,&14,&11) \\
\hline
\text{total} &&&& 14 & \pm\pz0 &  \pm\pz14 & \pm\pz\pz 0 & (0,&0,&0) 
\end{array}
\]
\caption{Three defects on a sphere giving 2 hypermultiplets in the $\mathbf{7}$ of $G_2$.  Notice that $[2,1^4]$ of $\mathfrak{sp}(3)$ is non-special. \label{7}}
\end{table}

\item Second, compactify the 6d theory of type $D_4$ on a sphere with 3 punctures, where the first is the full $\bZ_3$-twisted defect with Nahm pole $0$ of $\fg_2$, the second is a $\bZ_2$-twisted defect with principal Nahm pole $[6]$ of $\mathfrak{sp}(3)$, and the third is a $\bZ_2$-twisted defect with Nahm pole $[2,1^4]$ of $\mathfrak{sp}(3)$. In \cite{Tachikawa:2010vg}, this was shown to describe 2 hypermultiplets in the $\mathbf{7}$ of $G_2$, using S-duality arguments.
The properties of the defects, calculated by the methods in this paper, are in Table~\ref{7}. We see that this sphere describes 14 free hypermultiplets, as expected. The flavour symmetries of the defects are $\flavour(O_0)=\fg_2$, $\flavour([6])=1$ and $\flavour([2,1^4])=\mathfrak{sp}(2)$. Eq. \eqref{4dk} tells us that the level of the last flavour symmetry is $k=7$, which agrees with the fact that 2 hypermultiplets in the $\mathbf{7}$ of $G_2$  count as 7 half-hypermultiplets in the fundamental of $\mathfrak{sp}(2)$.

\item Third, compactify the 6d theory of type $E_6$ on a sphere with 3 untwisted punctures, with Nahm poles $0$, $E_6(a_1)$ and $A_2+2A_1$. 
From the analysis of the Seiberg-Witten geometry in  \cite{Tachikawa:2011yr}, by extracting the Hitchin poles, one finds that this sphere describes 2 hypermultiplets in the $\mathbf{27}$ of $E_6$. The other properties of these defects, calculated using the methods in this paper, are in Table~\ref{27}.   Adding the bulk contribution as before, we see that this sphere indeed describes 54 free hypermultiplets. The flavour symmetries of the defects are $\flavour(0)=\mathfrak{e}_6$, $\flavour(E_6(a_1))=1$, and $\flavour(A_2+2A_1)=\su(2)+\u(1)$.

\begin{table}
\[
%\rowcolors{0}{}{gray!10}
\begin{array}{c||c|c|c|c|c|r@{\ }r@{\ }r@{\ }r@{\ }r}
 O_\rho    & d(O_\rho)  & \dimH  & \dimC& n_h & n_v & (n_2,n_5,&n_6,&n_8,&n_9,&n_{12}) \\
\hline\hline
0 & E_6 & 36 & \pm36& \pm\pz624 & \pm\pz588& (1,4,&5,&7,&8,&11) \\
E_6(a_1) & A_1 & \pz1 & \pm11 & \pm\pz168 & \pm\pz167 & (1,1,&2,&2,&2,&3) \\
A_2+2A_1 & A_4+A_1 & 11 & \pm31 & \pm\pz510 & \pm\pz499 & (1,4,&4,&6,&7,&9) \\
\hline
\text{bulk} && \pz6 & -78 &  -1248 & -1254 & -(3,9,&11,&15,&17,&23) \\
\hline
\text{total} && 54 & \pm\pz0 &  \pm\pz\pz54 &\pm\pz\pz\pz0 & (0,0,&0,&0,&0,&0) 
\end{array}
\]
\caption{Three defects on a sphere giving 2 hypermultiplets in the $\mathbf{27}$ of $E_6$.  \label{27}}
\end{table}

\item Fourth, take the 6d theory of type $E_7$ on a sphere with 3 untwisted defects, with Nahm poles $0$, $E_7(a_1)$ and $A_3+A_2+A_1$. 
From the analysis of the Seiberg-Witten geometry in \cite{Tachikawa:2011yr}, as in the previous example, one finds that this sphere describes 3 half-hypermultiplets in the $\mathbf{56}$ of $E_7$; the Hitchin poles were discussed in Sec.~3.3 of that paper.
The other properties of these defects, calculated using the methods in this paper, are in Table~\ref{56}.   So, this sphere indeed describes 84 free hypermultiplets.

\begin{table}
\[
%\rowcolors{0}{}{gray!10}
\scalebox{.95}{$
\begin{array}{c||c|c|c|c|c|r@{\ }r@{\ }r@{\ }r@{\ }r@{\ }r@{\ }r}
 O_\rho    & d(O_\rho)  & \dimH  & \dimC & n_h & n_v & (n_2,&n_6,&n_8,&n_{10},&n_{12},&n_{14},&n_{18}) \\
\hline\hline
0 & E_7 & 63 & \pm\pz63& \pm1596 & \pm1533 & (1,&5,&7,&9,&11,&13,&17) \\
E_7(a_1) & A_1 & \pz1 & \pm\pz17 & \pm\pz384 & \pm\pz383 & (1,&2,&2,&2,&3,&3,&4) \\
A_3+A_2+A_1 & A_4+A_2 & 13 & \pm\pz53 & \pm1296 & \pm1283 & (1,&4,&6,&8,&9,&11,&14) \\
\hline
\text{bulk} && \pz7 & -133 &  -3192 & -3199 & -(3,&11,&15,&19,&23,&27,&35) \\
\hline
\text{total} && 84 & \pm\pz\pz0 &  \pm\pz\pz84 & \pm\pz\pz\pz0 & (0,&0,&0,&0,&0,&0,&0) 
\end{array}
$}
\]
\caption{Three defects on a sphere giving 3 half-hypermultiplets in the $\mathbf{56}$ of $E_7$.  \label{56}}
\end{table}

 The flavour symmetries are $\flavour(0)=\mathfrak{e}_7$, $\flavour(E_7(a_1))=1$ and $\flavour(A_3+A_2+A_1)=\su(2)\simeq \so(3)$. Indeed, 3 half-hypermultiplets have $\SO(3)$ flavour symmetry. Let us use \eqref{4dk} to compute the flavour central charge $k$ for this $SO(3)$. From Table~\ref{tableE7}, the weighted Dynkin diagram of the defect $A_3+A_2+A_1$ is $\overset{\scriptstyle\phantom{000}0\phantom{00}}{\scriptstyle 002000}$, which can be Weyl-reflected to $\overset{\scriptstyle\phantom{0,0,0-}2\phantom{,0,0}}{\scriptstyle 2,2,2{-}6,2,2}$. This corresponds to the obvious embedding of $A_3+A_2+A_1$ in $E_7$, according to the Dynkin sub-diagram. Let $\rho:\su(2)\to \su(4)+\su(3)+\su(2)\to \mathfrak{e}_7$ be such embedding.  A $\u(1)$ subalgebra of the Cartan in the direction $\overset{\scriptstyle\phantom{000}0\phantom{00}}{\scriptstyle 000100}$ clearly commutes with this $\rho(\su(2))$ subalgebra.  This $\u(1)$ subalgebra should be the Cartan subalgebra of the flavour $\so(3)_\text{flavour}$. The normalisation can be fixed by demanding that $\mathfrak{e}_7$ correctly decomposes into representations of $\rho(\su(2))\times \so(3)_\text{flavour}$. One finds that \begin{equation}
 \mathfrak{e}_7=(\mathbf{7},\mathbf{5})+
 (\mathbf{5},\mathbf{7})+
 (\mathbf{5},\mathbf{3})+
 (\mathbf{3},\mathbf{9})+
 (\mathbf{3},\mathbf{5})+
 (\mathbf{3},\mathbf{1})+
 (\mathbf{1},\mathbf{3}),
\end{equation} where $(\mathbf{n},\mathbf{m})$ stands for the tensor product of the irreducible representation of dimension $n$ of $\rho(\su(2))$ and that of dimension $m$ of $\so(3)_\text{flavour}$. Using \eqref{4dk}, we find $k_{4d}(\so(3)_\text{flavour})=224$.
Now, 3 half-hypermultiplets in the $\mathbf{56}$ count as 28 copies of the $\mathbf{3}$ of $\so(3)$, and so yield $k_{4d}(\so(3))=28\cdot 8=224$, which agrees with our computation.

\end{itemize}

\section*{Acknowledgements}
O.~C.~would like to thank the warm hospitality of the ICTP in Trieste, Italy, where part of this work was done. His research is supported in part by the INCT-Matem\'atica and the
ICTP-SAIFR in Brazil through a Capes postdoctoral fellowship.
J.~D.~would like to thank the hospitality of the organizers of the Workshop \emph{\href{http://nils.carqueville.net/newgauge/Home.html}{``New perspectives on supersymmetric gauge theories"}} at LMU Munich, while this work was being completed. The research of J.~D.~is based on work supported by the National Science Foundation under Grant No. PHY-0969020, and by the United States-Israel Binational Science Foundation under Grant \#2006157.
Y.~T.~would like to thank Yukawa Institute for Theoretical Physics at Kyoto University, where he stayed while this work was being completed.  The  work of Y.~T.~is  supported in part by World Premier International Research Center Initiative
(WPI Initiative),  MEXT, Japan through the Institute for the Physics and Mathematics
of the Universe, the University of Tokyo.

\appendix

\section{Mathematical conjecture}\label{conjecture}
Here, we reformulate our claim in Sec.~\ref{graded}, which was based on various physical consideration, as a mathematical conjecture. 
We claim that for a nilpotent element $e\in\fg$, there is a natural polynomial subalgebra $R_+$ of $\bC[\fg]$ constructed using the data of $e'\in\fg$ which is in the Spaltenstein dual orbit of $O_e$, such that the sum of the degrees of the generators of $R_+$ is given by a formula involving the weighted Dynkin diagram of $e$. 

Let $\fg$ be a complex simple, simply-laced Lie algebra.\footnote{When $\fg$ is not simply-laced, the following conjecture can be similarly formulated, but it requires the use of a simply-laced Lie algebra $\Type$ and an outer-automorphism $o$ such that $\fg^\vee=\type^o$. }
Let $G$ be the corresponding simple adjoint group.
Pick a nilpotent element $e\in\fg$, and fix an $\sl(2)$ triple $(e,h,f)$  containing it.
Let $e'\in d(O_e)$ be a nilpotent element in the dual orbit of $O_e$ in the sense of Spaltenstein,
and fix an $\sl(2)$ triple $(e',h',f')$ containing it. 

Let $X_e$ be a subgroup of the stabiliser $G_{e'}$ defined as follows.
If $O_e$ is special, $X_e=G_{e'}{}^\circ$ is the component connected to the identity. 
If $O_e$ is not special, let $C$ be the conjugacy class of $\bar A(O_{e'})$ assigned to $O_e$ by Sommers \cite{Sommers} and Achar \cite{Achar}. As $\bar A(O_{e'})$ is a Coxeter group, we can find a set of simple reflections $r_i$ such that $r_1r_2\cdots r_k \in C$. Let $\cC=\langle r_1,r_2,\ldots,r_k\rangle$ be the subgroup of $\bar A(O_{e'})$ generated by those simple reflections, as in \cite{AcharSommers}.
We let $X_e$ be the preimage of $\cC$ under the natural surjection $G_{e'}\twoheadrightarrow \bar A(O_{e'})$.

We decompose $\fg^*=\bigoplus_k \fg^*_k$, where $\fg^*_k$ is the eigenspace of the adjoint action of $h'$ of eigenvalue $k$. 
This grading carries over as  the grading of $R=\bC[\fg]^{X_e}=\bigoplus_k R_k$. For an element $x\in R_k$, we write $\grade(x)=k$. Also, for a homogeneous element $x\in \bC[\fg]$, we write $\deg(x)$ for the degree of $x$ as a polynomial.

\paragraph{Conjecture 1.} Let $R_+=\bigoplus_{k>0} R_k$. 
Then $R_+$ is a polynomial algebra with $(\dim O_{e'})/2$ homogeneous generators. 

\paragraph{Conjecture 2.} Denote the generators by  $u_i$,  where $i=1,\ldots,(\dim O_{e'})/2$. Then \begin{equation}
\sum_i \left[2\grade(u_i)+2\deg(u_i)-1\right] = 8 \rho\cdot(\rho-\frac{h}2) +\frac12(\rank \fg - \dim \fg_{h}),\label{relation}
\end{equation} where we picked a Cartan subalgebra $\fh$ containing $h$, chose the simple roots $\alpha_i$ in it so that $\alpha_i\cdot h$ is all nonnegative, and let $\rho$ the Weyl vector; the dot uses the standard normalisation that the roots have squared lengths two. $\fg_h$ is the stabiliser of $h$.

\paragraph{Example.} Let $(e,h,f)=(0,0,0)$. Then the dual $(e',h',f')$ is the principal orbit, and in particular $h'=2\rho$.
Pick an invariant polynomial $P_d\in \bC[\fg]^G$ of degree $d$, and define an element $P_{d,k}\in \bC[\fg]$ for $k=1,\ldots,d-1$ as a polynomial function $P_{d,k}: \fg \to \bC$ as follows: \begin{equation}
P_{d,k}(x)=\text{the coefficient of $z^{-k}$ of } P_d(e'/z+x).
\end{equation} It is easily seen that $P_{d,k}\in R$, $\grade(P_{d,k})=k$, $\deg(P_{d,k})=d-k$. We claim $P_{d,k}$ are the polynomial generators of $R_+$. Then, the number of generators is \begin{equation}
\sum_{a=1}^{\rank g} (d_a-1) = \frac12{\dim \fg -\rank \fg}=\frac12 \dim O_{e'}.
\end{equation} Here, $d_a$ is one plus the $a$-th exponent of $\fg$. 
The relation \eqref{relation} becomes \begin{equation}
\sum_{a=1}^{\rank g} (d_a-1)(2d_a-1) = 8\rho\cdot\rho + \frac12(\rank \fg-\dim \fg).
\end{equation} This equality can be easily checked, using the known formul\ae\ for $\sum d_a$ and $\sum d_a{}^2$. Both sides are just $r\coxeter(4\coxeter+1)/6$, where $r$, $\coxeter$ are the rank and the dual Coxeter number of $\fg$.

\section{Embedding the nilpotent orbits of $G_2$ in $D_4$}\label{explicitG2}
Here we construct an explicit embedding of $\fg_2$ in $\so(8)$, and compute the $\sl(2)$ triples $(e,h,f)$ for all nilpotent orbits.
First, we describe the action of triality $S_3$ on $\so(8)$. Note that $\mathrm{Spin}(8)$ contains an $({\SU(2)}^4)/\mathbb{Z}_2$ subgroup, under which the adjoint decomposes as
\begin{equation}
\mathbf{28} = (\mathbf3,\mathbf1,\mathbf1,\mathbf1)+(\mathbf1,\mathbf3,\mathbf1,\mathbf1)+(\mathbf1,\mathbf1,\mathbf3,\mathbf1)+(\mathbf1,\mathbf1,\mathbf1,\mathbf3)+(\mathbf2,\mathbf2,\mathbf2,\mathbf2)
\end{equation}
Under this decomposition, the action of triality fixes one of the $\mathfrak{sl}(2)$ subalgebras, and permutes the other three.
The invariant subalgebra is $\mathfrak{g}_2\subset \mathfrak{so}(8)$, under which
\begin{equation}
\mathbf{28} = \mathbf{14} + \mathbf7 \otimes V
\label{D4gens}
\end{equation}
where $V$ is the 2-dimensional irreducible representation of $S_3$. In terms of our previous decomposition,
\begin{equation}
G_2 \supset (\SU(2)\times {\SU(2)}_D)/\mathbb{Z}_2\label{g2g2}
\end{equation}
where the first $\SU(2)$ is the one you kept fixed, and ${\SU(2)}_D$ is the diagonal $\SU(2)$ of the three which are permuted by triality. Under this embedding,
\begin{align}
\mathbf{14} &= (\mathbf3,\mathbf1)+(\mathbf1,\mathbf3)+(\mathbf2,\mathbf4),&
\mathbf7 &= (\mathbf1,\mathbf3) + (\mathbf2,\mathbf2).
\label{g2repsdecomp}
\end{align}

An explicit basis of anti-symmetric $8\times 8$ matrices for this $\mathfrak{g}_2$ subalgebra is as follows. 
First, we realize $\sl(2)_{L+}\oplus\sl(2)_{L-}\oplus\sl(2)_{R+}\oplus\sl(2)_{R-}\subset\so(8)$ as 
\begin{equation}
\begin{aligned}
H_{L,\plusminus} &=  P_\plusminus\otimes      \sigma_2         \otimes   \mathbbm{1}, &
H_{R,\plusminus} &=      P_\plusminus\otimes \mathbbm{1}     \otimes     \sigma_2,\\
E_{L,\plusminus} &= P_\plusminus\otimes\tfrac{1}{2} (\sigma_3+i\sigma_1) \otimes \sigma_2,&
E_{R,\plusminus} &=P_\plusminus\otimes \tfrac{1}{2} \sigma_2 \otimes (\sigma_3+i\sigma_1),\\
F_{L,\plusminus} &= P_\plusminus\otimes\tfrac{1}{2} (\sigma_3-i\sigma_1) \otimes \sigma_2,&
F_{R,\plusminus} &= P_\plusminus\otimes\tfrac{1}{2} \sigma_2 \otimes (\sigma_3-i\sigma_1)
\end{aligned}\end{equation} where $P_\plusminus=\tfrac12(1\plusminus\sigma_3)$.

Now, we  let $S_3$ leave $\sl(2)_{L+}$ invariant and permute $\sl(2)_{L-},\sl(2)_{R\plusminus}$. Thus,
we identify the first $\sl(2)$ factor in \eqref{g2g2} with $\sl(2)_{L+}$,\begin{equation}
H_1=H_{L+},\quad
E_1=E_{L+},\quad
F_1=F_{L+},
\label{sl2invt}
\end{equation} and the diagonal $\sl(2)_D$ in \eqref{g2g2} with the combination
\begin{equation}\label{sl2diag}
H_2=H_{L-}+H_{R+}+H_{R-},\quad
E_2=E_{L-}+E_{R+}+E_{R-},\quad
F_2=F_{L-}+F_{R+}+F_{R-},
\end{equation}
which is clearly invariant under $S_3$ permutations. So, \eqref{sl2invt} and \eqref{sl2diag} are, respectively, the 
$(\mathbf3,\mathbf1)$ and the $(\mathbf1,\mathbf3)$ in the decomposition of the $\mathbf{14}$ in \eqref{g2repsdecomp}.

On the other hand, the highest weight of the $(\mathbf2,\mathbf4)$ in the $\mathbf{14}$ is
\begin{equation}
S_{1,3}  = \tfrac{1}{4} \sigma_2\otimes (\sigma_3+i\sigma_1)\otimes (\sigma_3+i\sigma_1).
\end{equation}
The remaining weights in the $(\mathbf2,\mathbf4)$ can be found by using $F_1$ and $F_2$ as lowering operators,
\begin{align}
[F_1,S_{i,j}]&=S_{i-2,j},&
[F_2,S_{i,j}]&=S_{i,j-2}.
\label{otherweights}
\end{align}
With this choice of Cartan, the simple roots of $\fg_2$ correspond to $E_2$ (short root) and $S_{1,-3}$ (long root). 

Now, the explicit $\rho:\su(2)\to\fg_2$ homomorphisms corresponding to the 5 nilpotent orbits in $\fg_2$ are shown in Table~\ref{g2triples}. These can be found from the weighted Dynkin diagrams for the $\fg_2$ nilpotent orbits \cite{CollingwoodMcGovern} and our explicit expressions for the $\mathbf{14}$.

\begin{table}\[
\begin{array}{c|ccc}
\text{B-C} & \rho(H)& \rho(\sigma^+) & \rho(\sigma^-) \\
\hline
0& 0 &0&0 \\
 A_1 & H_1 & E_1& F_1 \\
 \tilde{A}_1 & \tfrac{1}{2}(3H_1+H_2) &  S_{1,1}&  S_{-1,-1} \\
 G_2(a_1) & 2H_1&  S_{1,3}+S_{1,-1} & S_{-1,-3}+S_{-1,1} \\
 G_2 & 5H_1+H_2 & \sqrt{10} S_{1,-3} +\sqrt{6} E_2& \sqrt{10} S_{-1,3} +\sqrt{6} F_2
\end{array}
\]
\caption{Explicit distinguished triples, for nilpotent orbits of $\fg_2$, as embedded in $\so(8)$.\label{g2triples}}
\end{table}

To construct generic elements in $\so(8)$, as in \eqref{g2hitchin}, we also need explicit expressions for the 
$\mathbf7 \otimes V$ in \eqref{D4gens}, which is the part of the $\mathbf{28}$ that is \emph{not} invariant under $S_3$.
In \eqref{g2hitchin}, we use only the $\bZ_3$ subgroup of $S_3$, so we decompose the 2-dimensional
representation $V$ of $S_3$ into irreducible representations of $\bZ_3$, namely, the two 1-dimensional
representations transforming as $\omega$ and $\omega^2$, where $\omega\neq 1$ is a cube root of 1.

Under such decomposition, we have $\mathbf{7}\times V=\mathbf{7}^{(\omega)} +\mathbf{7}^{(\omega^2)}$,
where both $\mathbf7^{(\omega)}$ and $\mathbf7^{(\omega^2)}$ decompose as \eqref{g2repsdecomp}, e.g., $\mathbf7^{(\omega)}=(\mathbf1,\mathbf3)^{(\omega)}+(\mathbf2,\mathbf2)^{(\omega)}$.

The explicit expressions for $(\mathbf1,\mathbf3)^{(\omega)}$ and $(\mathbf1,\mathbf3)^{(\omega^2)}$, which transform as $\sl(2)_2$-triplets, are
\begin{equation}\begin{split}
H^{(\omega)} &=H_{L-}+\omega H_{R+}+\omega^2 H_{R-},\\
H^{(\omega^2)} &=H_{L-}+\omega^2 H_{R+}+\omega H_{R-}
\end{split}\end{equation}
and corresponding expressions for $E^{(\omega)},F^{(\omega)}$ and $E^{(\omega^2)},F^{(\omega^2)}$. The highest weights for the $(\mathbf2,\mathbf2)^{(\omega)}$ and $(\mathbf2,\mathbf2)^{(\omega^2)}$ are
\begin{equation}
\begin{split}
T^{(\omega)}_{1,1} &= \tfrac{1}{2}(\sigma_2\otimes 1 - i \sigma_1\otimes\sigma_2)\otimes(\sigma_3+i\sigma_1)
-\tfrac{1}{2}\sigma_2\otimes(\sigma_3+i\sigma_1)\otimes 1\\ &\quad -\tfrac{i}{2}(\omega-\omega^2)\sigma_1\otimes (\sigma_3+i\sigma_1)\otimes\sigma_2\\
T^{(\omega^2)}_{1,1} &= \tfrac{1}{2}(\sigma_2\otimes 1 - i \sigma_1\otimes\sigma_2)\otimes(\sigma_3+i\sigma_1)
-\tfrac{1}{2}\sigma_2\otimes(\sigma_3+i\sigma_1)\otimes 1\\ &\quad -\tfrac{i}{2}(\omega^2-\omega)\sigma_1\otimes (\sigma_3+i\sigma_1)\otimes\sigma_2
\end{split}
\end{equation}
The rest of the weights for $(\mathbf2,\mathbf2)^{(\omega)}$ and $(\mathbf2,\mathbf2)^{(\omega^2)}$ can be found by using $F_1$ and $F_2$ as lowering operators, as in \eqref{otherweights}.

\newpage

\section{Tables of exceptional nilpotent orbits}\label{crash}
We reproduce the tables of properties of nilpotent orbits in the exceptional groups below;
the data are taken from \cite{Spaltenstein,Carter,CollingwoodMcGovern,McGovern}.
They can also be obtained online  at \href{http://lie.math.okstate.edu/UMRK/UMRK.html}{\texttt{http://lie.\allowbreak math.\allowbreak okstate.\allowbreak edu/UMRK/UMRK.html}}, or at \href{http://www.liegroups.org/tables/unipotentOrbits/}{\texttt{http://www.liegroups.org/}}.

The data for $G_2$ is in Table~\ref{G2}, in the body of the paper.
The data for $F_4$ is in Table~\ref{F4}, the data for $E_6$ is in Table~\ref{E6},
the data for $E_7$ is in Table~\ref{tableE7} and ~\ref{E7}, and
the data for $E_8$ is in Table~\ref{tableE8-1}, \ref{tableE8-2} and ~\ref{E8}.
For each $\fg$ and for each nilpotent orbit $O_e$, we give its Bala-Carter label, weighted Dynkin diagram, complex dimension $\dim_\bC O$, the discrete part $A(O)$ of the centraliser $\Flavour(O)$ of $e$ in the compact adjoint group $G_{ad}$, and the Lie algebra $\flavour(O)$ of $\Flavour(O)$.
In the tables, $T_n$ stands for $\u(1)^n$. The precise global structure of $\Flavour(O)$ can be found in \cite{LawtherTesterman}.

In the nilpotent orbit literature, $\Flavour(O)$ is usually denoted by $C$, which is the reductive subgroup of  $C(O)$, the centraliser of $e$ in $(G_{ad})_{\bC}$.  We use $F$ (standing for `flavour') to avoid conflict with the notation for a different group, $\Sommers(O)$, in the nomenclature of \cite{Sommers,Achar}.

For each $\fg$, we give two Hasse diagrams: one including all nilpotent orbits, and another one including only special nilpotent orbits. Two orbits in the same special piece are connected by a dotted line. Rigid orbits are framed in a box. Distinguished orbits in $\fg=X_n$ have a Bala-Carter label that starts with $X_n$, so no other marking is given.
The Spaltenstein duality map for exceptional $\fg$ can be determined uniquely from inspection of the Hasse diagrams except in a few cases, in which case we specify what the duals are in the Table captions.
\vfill\eject
\begin{table}[h!]
\def\Ffour#1{#1}
\[
\rowcolors{0}{}{gray!10}
\begin{array}{c|c|c|c|c}
\text{B-C}& \circ\circ\Rightarrow\circ\circ& \dim & A(O) & \flavour(O) \\
\hline \hline
0 & \Ffour{0000} & 0  & 1 & F_4\\
A_1 & \Ffour{1000} & 16  & 1 & C_3\\
\tilde{A}_1 & \Ffour{0001} & 22  & S_2 & A_3\\
A_1+\tilde{A}_1 & \Ffour{0100} & 28  & 1 & 2A_1\\
A_2 & \Ffour{2000} & 30  & S_2 & A_2\\
\tilde{A}_2 & \Ffour{0002} & 30  & 1 & G_2\\
A_2+\tilde{A}_1 & \Ffour{0010} & 34  & 1 & A_1\\
B_2 & \Ffour{2001} & 36  & S_2 & 2A_1\\
\tilde{A}_2+A_1 & \Ffour{0101} & 36  & 1 & A_1\\
C_3(a_1) & \Ffour{1010} & 38  & S_2 & A_1\\
F_4(a_3) & \Ffour{0200} & 40  & S_4 & 1\\
B_3 & \Ffour{2200} & 42  & 1 & A_1\\
C_3 & \Ffour{1012} & 42  & 1 & A_1\\
F_4(a_2) & \Ffour{0202} & 44  & S_2 & 1\\
F_4(a_1) & \Ffour{2202} & 46  & S_2 & 1\\
F_4 & \Ffour{2222} & 48  & 1 & 1\\
\end{array}
\qquad
\vcenter{\hbox{\xymatrix@=8pt@C=-.5em{
&F_4\sp{d}\\
&F_4(a_1)\sp{d}\\
&F_4(a_2)\sp{dr}\sp{dl}\\
B_3\sp{dr} &&C_3\sp{dl}\\
&F_4(a_3)\ns{d}\\
&C_3(a_1)\ns{dr}\ns{dl}\\
B_2\ns{d} & & \relax\rigid{\tilde A_2+A_1}\ns{dll}\sp{dd}\\
\relax\rigid{A_2+\tilde A_1} \sp{d} \\
A_2 \sp{dr} &&\tilde A_2 \sp{dl} \\
& \relax\rigid{A_1+\tilde A_1} \sp{d}\\
& \relax\rigid{\tilde A_1} \ns{d}\\
&  \relax\rigid{A_1} \sp{d}\\
&\relax\rigid{0}
}
\qquad
\xymatrix@=8pt@C=-.5em{
&F_4\sp{d}\\
&F_4(a_1)\sp{d}\\
&F_4(a_2)\sp{dr}\sp{dl}\\
B_3\sp{dr} &&C_3\sp{dl}\\
&F_4(a_3)\sp{dr}\sp{dl}\\
A_2 \sp{dr} &&\tilde A_2 \sp{dl} \\
& A_1+\tilde A_1 \sp{d}\\
& \tilde A_1 \sp{d}\\
&0
}}}
\]
\caption{Nilpotent orbits of $F_4$, and the two $F_4$ Hasse diagrams. Distinguished orbits have the label $F_4$ or $F_4(a_i)$.
The Spaltenstein map for any orbit is clear from the diagram, except for $d_{BV}(A_2)=B_3$, $d_{LS}(A_2)=C_3$. In the table, 'B-C' stands
for the Bala-Carter label.\label{F4}}
\vspace{2in}
{~}
\end{table}

\def\esix#1#2#3#4#5#6{\overset{\displaystyle #2}{\displaystyle #6#5#4#3#1}}
\def\Esix#1{\expandafter\esix#1\relax}
\def\eseven#1#2#3#4#5#6#7{\overset{\displaystyle \phantom{111}#2\phantom{11}}{\displaystyle #7#6#5#4#3#1}}
\def\Eseven#1{\expandafter\eseven#1\relax}
\def\eeight#1#2#3#4#5#6#7#8{\overset{\displaystyle \phantom{1111}#2\phantom{11}}{\displaystyle #8#7#6#5#4#3#1}}
\def\Eeight#1{\expandafter\eeight#1\relax}

\begin{table}
\[
\rowcolors{0}{}{gray!10}
\begin{array}{c|c|c|c|c}
\text{B-C}& h & \dim & A(O) & \flavour(O) \\
\hline \hline
0 & \Esix{000000} & 0  & 1 & E_6\\
A_1 & \Esix{010000} & 22  & 1 & A_5\\
2A_1 & \Esix{100001} & 32  & 1 & B_3+T_1\\
3A_1 & \Esix{000100} & 40  & 1 & A_2+A_1\\
A_2 & \Esix{020000} & 42  & S_2 & 2A_2\\
A_2+A_1 & \Esix{110001} & 46  & 1 & A_2+T_1\\
2A_2 & \Esix{200002} & 48  & 1 & G_2\\
A_2+2A_1 & \Esix{001010} & 50  & 1 & A_1+T_1\\
A_3 & \Esix{120001} & 52  & 1 & B_2+T_1\\
2A_2+A_1 & \Esix{100101} & 54  & 1 & A_1\\
A_3+A_1 & \Esix{011010} & 56  & 1 & A_1+T_1\\
D_4(a_1) & \Esix{000200} & 58  & S_3 & T_2\\
A_4 & \Esix{220002} & 60  & 1 & A_1+T_1\\
D_4 & \Esix{020200} & 60  & 1 & A_2\\
A_4+A_1 & \Esix{111011} & 62  & 1 & T_1\\
A_5 & \Esix{211012} & 64  & 1 & A_1\\
D_5(a_1) & \Esix{121011} & 64  & 1 & T_1\\
E_6(a_3) & \Esix{200202} & 66  & S_2 & 1\\
D_5 & \Esix{220202} & 68  & 1 & T_1\\
E_6(a_1) & \Esix{222022} & 70  & 1 & 1\\
E_6 & \Esix{222222} & 72  & 1 & 1\\
\end{array}
\quad
\vcenter{\hbox{\xymatrix@=8pt@C=-10pt{
&E_6 \sp{d}\\
&E_6(a_1) \sp{d}\\
&D_5 \sp{d}\\
&E_6(a_3) \ns{dl}\sp{dr}\\
A_5 \sp{d} && D_5(a_1) \sp{dd} \sp{dll}\\
A_4+A_1\sp{d}\\
A_4 \sp{dr} &&D_4\sp{dl} \\
&D_4(a_1) \ns{d}\\
&A_3+A_1 \sp{dl} \ns{dr}\\
A_3 \sp{d} &&\relax\rigid{2A_2+A_1} \sp{d}\sp{dll}\\
A_2+2A_1 \sp{dr} && 2A_2 \sp{dl} \\
&A_1+A_1 \sp{d}\\
&A_2 \ns{d}\\
&\relax\rigid{3A_1}\sp{d}\\
&2A_1\sp{d}\\
&\relax\rigid{A_1}\sp{d}\\
&\relax\rigid{0}
}
\quad
\xymatrix@=8pt@C=-10pt{
&E_6 \sp{d}\\
&E_6(a_1) \sp{d}\\
&D_5 \sp{d}\\
&E_6(a_3) \sp{d}\\
& D_5(a_1) \sp{ddr} \sp{dl}\\
A_4+A_1\sp{d}\\
A_4 \sp{dr} &&D_4\sp{dl} \\
&D_4(a_1) \sp{dl}\sp{dr}\\
A_3\sp{d}&& 2A_2 \sp{ddl}\\
A_2+2A_1 \sp{dr}  \\
&A_1+A_1 \sp{d}\\
&A_2 \sp{d}\\
&2A_1\sp{d}\\
&A_1\sp{d}\\
&0
}}}
\]
\caption{Nilpotent orbits of $E_6$ and the $E_6$ Hasse diagrams.\label{E6}}
\end{table}

\begin{table}
\vspace*{-1em}
\[
\hspace*{-2em}
\rowcolors{0}{}{gray!10}
\begin{array}{c|c|c|c|c}
\text{B-C}& h & \dim & A(O) & \flavour(O) \\
\hline\hline 
0 & \Eseven{0000000} & 0  & 1 & E_7\\
A_1 & \Eseven{1000000} & 34  & 1 & D_6\\
2A_1 & \Eseven{0000010} & 52  & 1 & B_4+A_1\\
(3A_1)'' & \Eseven{0000002} & 54  & 1 & F_4\\
(3A_1)' & \Eseven{0010000} & 64  & 1 & C_3+A_1\\
A_2 & \Eseven{2000000} & 66  & S_2 & A_5\\
4A_1 & \Eseven{0100001} & 70  & 1 & C_3\\
A_2+A_1 & \Eseven{1000010} & 76  & S_2 & A_3+T_1\\
A_2+2A_1 & \Eseven{0001000} & 82  & 1 & 3A_1\\
A_3 & \Eseven{2000010} & 84  & 1 & B_3+A_1\\
2A_2 & \Eseven{0000020} & 84  & 1 & G_2+A_1\\
A_2+3A_1 & \Eseven{0200000} & 84  & 1 & G_2\\
(A_3+A_1)'' & \Eseven{2000002} & 86  & 1 & B_3\\
2A_2+A_1 & \Eseven{0010010} & 90  & 1 & 2A_1\\
(A_3+A_1)' & \Eseven{1001000} & 92  & 1 & 3A_1\\
D_4(a_1) & \Eseven{0020000} & 94  & S_3 & 3A_1\\
A_3+2A_1 & \Eseven{1000101} & 94  & 1 & 2A_1\\
D_4 & \Eseven{2020000} & 96  & 1 & C_3\\
D_4(a_1)+A_1 & \Eseven{0110001} & 96  & S_2 & 2A_1\\
A_3+A_2 & \Eseven{0001010} & 98  & S_2 & A_1+T_1\\
A_4 & \Eseven{2000020} & 100  & S_2 & A_2+T_1\\
A_3+A_2+A_1 & \Eseven{0000200} & 100  & 1 & A_1\\
(A_5)'' & \Eseven{2000022} & 102  & 1 & G_2\\
\end{array}
\quad
\rowcolors{0}{}{gray!10}
\raisebox{1.1em}{$\begin{array}{c|c|c|c|c}
\text{B-C}& h & \dim & A(O) & \flavour(O) \\
\hline \hline
D_4+A_1 & \Eseven{2110001} & 102  & 1 & B_2\\
A_4+A_1 & \Eseven{1001010} & 104  & S_2 & T_2\\
D_5(a_1) & \Eseven{2001010} & 106  & S_2 & A_1+T_1\\
A_4+A_2 & \Eseven{0002000} & 106  & 1 & A_1\\
(A_5)' & \Eseven{1001020} & 108  & 1 & 2A_1\\
A_5+A_1 & \Eseven{1001012} & 108  & 1 & A_1\\
D_5(a_1)+A_1 & \Eseven{2000200} & 108  & 1 & A_1\\
D_6(a_2) & \Eseven{0110102} & 110  & 1 & A_1\\
E_6(a_3) & \Eseven{0020020} & 110  & S_2 & A_1\\
D_5 & \Eseven{2020020} & 112  & 1 & 2A_1\\
E_7(a_5) & \Eseven{0002002} & 112  & S_3 & 1\\
A_6 & \Eseven{0002020} & 114  & 1 & A_1\\
D_5+A_1 & \Eseven{2110110} & 114  & 1 & A_1\\
D_6(a_1) & \Eseven{2110102} & 114  & 1 & A_1\\
E_7(a_4) & \Eseven{2002002} & 116  & S_2 & 1\\
D_6 & \Eseven{2110122} & 118  & 1 & A_1\\
E_6(a_1) & \Eseven{2002020} & 118  & S_2 & T_1\\
E_6 & \Eseven{2022020} & 120  & 1 & A_1\\
E_7(a_3) & \Eseven{2002022} & 120  & S_2 & 1\\
E_7(a_2) & \Eseven{2220202} & 122  & 1 & 1\\
E_7(a_1) & \Eseven{2220222} & 124  & 1 & 1\\
E_7 & \Eseven{2222222} & 126  & 1 & 1\\
\end{array}$}
\]
\caption{Nilpotent orbits of $E_7$.  \label{tableE7}}
\end{table}

\begin{table}
\vspace*{-3em}
\[
\scalebox{.9}{\xymatrix@=8pt{
&E_7\sp{d}\\
&E_7(a_1)\sp{d}\\
&E_7(a_2)\sp{dr}\sp{dl}\\
E_7(a_3)\ns{d}\sp{drr} && E_6 \sp{d}\\
D_6\sp{dr} && E_6(a_1) \sp{dl}\\
&E_7(a_4) \sp{d}\sp{dr}\sp{dl}\\
A_6\sp{dr} & D_6(a_1) \sp{d}\sp{dr} & D_5+A_1\sp{dl}\sp{d} \\
&E_7(a_5)\ns{d}\sp{dr} & D_5\sp{d}\\
&D_6(a_2)\ns{dl} \sp{d} \sp{dr} & E_6(a_3)\sp{d} \ns{dl}\\
A_5+A_1 \sp{d}\sp{dr} &A_5'\sp{d} & D_5(a_1)+A_1 \sp{dl}\sp{d}\\
A_5''\sp{dd}&A_4+A_2\sp{d} & D_5(a_1)\sp{dl}\ns{d}\\
& A_4+A_1\sp{dl}\sp{d} & D_4+A_1\sp{dd}\sp{dl}\\
A_4\sp{dr}&A_3+A_2+A_1\sp{d} \\
&A_3+A_2 \sp{d} & D_4\sp{dd}\\
&D_4(a_1)+A_1\sp{dr}\ns{d} \\
&A_3+2A_1 \sp{d}\sp{dr} & D_4(a_1)\ns{d}\\
& (A_3+A_1)'' \sp{ddl}\sp{d} & \relax\rigid{(A_3+A_1)'}\ns{d}\sp{dl} \\
&A_3\sp{dd} & \relax\rigid{2A_2+A_1} \sp{dll} \sp{d} \\
2A_2 \sp{dr}&&A_2+3A_1 \sp{dl}\\
&\relax\rigid{A_2+2A_1}\sp{d} \\
&A_2+A_1 \ns{dl}\sp{dr}\\
\relax\rigid{4A_1} \sp{d}\sp{drr} &&A_2 \ns{d}\\
(3A_1)'' \sp{dr} &&\relax\rigid{(3A_1)'} \sp{dl} \\
&\relax\rigid{2A_1} \sp{d}\\
&\relax\rigid{A_1}\sp{d}\\
&0
}}
\qquad
\scalebox{.9}{\xymatrix@=8pt{
&E_7\sp{d}\\
&E_7(a_1)\sp{d}\\
&E_7(a_2)\sp{dr}\sp{dl}\\
E_7(a_3)\sp{dr} && E_6 \sp{dl}\\
& E_6(a_1) \sp{d}\\
&E_7(a_4) \sp{d}\sp{dr}\sp{dl}\\
A_6\sp{dr} & D_6(a_1) \sp{d}\sp{dr} & D_5+A_1\sp{dl}\sp{d} \\
&E_7(a_5)\sp{d}\sp{ddl} & D_5\sp{dl}\\
& E_6(a_3)\sp{d} \\
A_5''\sp{dd}& D_5(a_1)+A_1 \sp{d}\sp{dr}\\
&A_4+A_2\sp{d} & D_5(a_1)\sp{dl}\sp{ddd}\\
& A_4+A_1\sp{dl}\sp{d} \\
A_4\sp{dr}&A_3+A_2+A_1\sp{d} \\
&A_3+A_2 \sp{d} & D_4\sp{ddl}\\
&D_4(a_1)+A_1\sp{d} \sp{dl}\\
(A_3+A_1)''\sp{d}\sp{dr}& D_4(a_1)\sp{d}\sp{dr}\sp{dl}\\
2A_2 \sp{dr}&A_3\sp{d} &A_2+3A_1 \sp{dl}\\
&A_2+2A_1\sp{d} \\
&A_2+A_1 \sp{dr}\sp{dl}\\
(3A_1)'' \sp{dr} &&A_2 \sp{dl}\\
&2A_1 \sp{d}\\
&A_1\sp{d}\\
&0
}}
\]
\caption{The Hasse diagrams for $E_7$. Here, $d(A_2)=E_7(a_3)$.\label{E7}}
\end{table}

\begin{table}
\vspace*{-2em}
\[
\rowcolors{0}{}{gray!10}
\hspace*{-2em}
\begin{array}{c|c|c|c|c}
\text{B-C}& h & \dim & A(O) & \flavour(O) \\
\hline \hline
0 & \Eeight{00000000} & 0  & 1 & E_8\\
A_1 & \Eeight{00000001} & 58  & 1 & E_7\\
2A_1 & \Eeight{10000000} & 92  & 1 & B_6\\
3A_1 & \Eeight{00000010} & 112  & 1 & F_4+A_1\\
A_2 & \Eeight{00000002} & 114  & S_2 & E_6\\
4A_1 & \Eeight{01000000} & 128  & 1 & C_4\\
A_2+A_1 & \Eeight{10000001} & 136  & S_2 & A_5\\
A_2+2A_1 & \Eeight{00000100} & 146  & 1 & B_3+A_1\\
A_3 & \Eeight{10000002} & 148  & 1 & B_5\\
A_2+3A_1 & \Eeight{00100000} & 154  & 1 & G_2+A_1\\
2A_2 & \Eeight{20000000} & 156  & S_2 & 2G_2\\
2A_2+A_1 & \Eeight{10000010} & 162  & 1 & G_2+A_1\\
A_3+A_1 & \Eeight{00000101} & 164  & 1 & B_3+A_1\\
D_4(a_1) & \Eeight{00000020} & 166  & S_3 & D_4\\
D_4 & \Eeight{00000022} & 168  & 1 & F_4\\
2A_2+2A_1 & \Eeight{00001000} & 168  & 1 & B_2\\
A_3+2A_1 & \Eeight{00100001} & 172  & 1 & B_2+A_1\\
D_4(a_1)+A_1 & \Eeight{01000010} & 176  & S_3 & 3A_1\\
A_3+A_2 & \Eeight{10000100} & 178  & S_2 & B_2+T_1\\
A_4 & \Eeight{20000002} & 180  & S_2 & A_4\\
A_3+A_2+A_1 & \Eeight{00010000} & 182  & 1 & 2A_1\\
D_4+A_1 & \Eeight{01000012} & 184  & 1 & C_3\\
D_4(a_1)+A_2 & \Eeight{02000000} & 184  & S_2 & A_2\\
\end{array}
\quad
\rowcolors{0}{}{gray!10}
\begin{array}{c|c|c|c|c}
\text{B-C}& h & \dim & A(O) & \flavour(O) \\
\hline \hline
A_4+A_1 & \Eeight{10000101} & 188  & S_2 & A_2+T_1\\
2A_3 & \Eeight{10001000} & 188  & 1 & B_2\\
D_5(a_1) & \Eeight{10000102} & 190  & S_2 & A_3\\
A_4+2A_1 & \Eeight{00010001} & 192  & S_2 & A_1+T_1\\
A_4+A_2 & \Eeight{00000200} & 194  & 1 & 2A_1\\
A_5 & \Eeight{20000101} & 196  & 1 & G_2+A_1\\
D_5(a_1)+A_1 & \Eeight{00010002} & 196  & 1 & 2A_1\\
A_4+A_2+A_1 & \Eeight{00100100} & 196  & 1 & A_1\\
D_4+A_2 & \Eeight{02000002} & 198  & S_2 & A_2\\
E_6(a_3) & \Eeight{20000020} & 198  & S_2 & G_2\\
D_5 & \Eeight{20000022} & 200  & 1 & B_3\\
A_4+A_3 & \Eeight{00010010} & 200  & 1 & A_1\\
A_5+A_1 & \Eeight{10010001} & 202  & 1 & 2A_1\\
D_5(a_1)+A_2 & \Eeight{00100101} & 202  & 1 & A_1\\
D_6(a_2) & \Eeight{01100010} & 204  & S_2 & 2A_1\\
E_6(a_3)+A_1 & \Eeight{10001010} & 204  & S_2 & A_1\\
E_7(a_5) & \Eeight{00010100} & 206  & S_3 & A_1\\
D_5+A_1 & \Eeight{10001012} & 208  & 1 & 2A_1\\
E_8(a_7) & \Eeight{00002000} & 208  & S_5 & 1\\
A_6 & \Eeight{20000200} & 210  & 1 & 2A_1\\
D_6(a_1) & \Eeight{01100012} & 210  & S_2 & 2A_1\\
A_6+A_1 & \Eeight{10010100} & 212  & 1 & A_1\\
E_7(a_4) & \Eeight{00010102} & 212  & S_2 & A_1\\
\end{array}
\]
\caption{Nilpotent orbits of $E_8$ (part 1/2). \label{tableE8-1}}
\end{table}
\begin{table}
\vspace*{-2em}
\[
\rowcolors{0}{}{gray!10}
\begin{array}{c|c|c|c|c}
\text{B-C}& h & \dim & A(O) & \flavour(O) \\
\hline \hline
E_6(a_1) & \Eeight{20000202} & 214  & S_2 & A_2\\
D_5+A_2 & \Eeight{00002002} & 214  & S_2 & T_1\\
D_6 & \Eeight{21100012} & 216  & 1 & B_2\\
E_6 & \Eeight{20000222} & 216  & 1 & G_2\\
D_7(a_2) & \Eeight{10010101} & 216  & S_2 & T_1\\
A_7 & \Eeight{10010110} & 218  & 1 & A_1\\
E_6(a_1)+A_1 & \Eeight{10010102} & 218  & S_2 & T_1\\
E_7(a_3) & \Eeight{20010102} & 220  & S_2 & A_1\\
E_8(b_6) & \Eeight{00020002} & 220  & S_3 & 1\\
D_7(a_1) & \Eeight{20002002} & 222  & S_2 & T_1\\
E_6+A_1 & \Eeight{10010122} & 222  & 1 & A_1\\
E_7(a_2) & \Eeight{01101022} & 224  & 1 & A_1\\
E_8(a_6) & \Eeight{00020020} & 224  & S_3 & 1\\
D_7 & \Eeight{21101101} & 226  & 1 & A_1\\
E_8(b_5) & \Eeight{00020022} & 226  & S_3 & 1\\
E_7(a_1) & \Eeight{21101022} & 228  & 1 & A_1\\
E_8(a_5) & \Eeight{20020020} & 228  & S_2 & 1\\
E_8(b_4) & \Eeight{20020022} & 230  & S_2 & 1\\
E_7 & \Eeight{21101222} & 232  & 1 & A_1\\
E_8(a_4) & \Eeight{20020202} & 232  & S_2 & 1\\
E_8(a_3) & \Eeight{20020222} & 234  & S_2 & 1\\
E_8(a_2) & \Eeight{22202022} & 236  & 1 & 1\\
E_8(a_1) & \Eeight{22202222} & 238  & 1 & 1\\
E_8 & \Eeight{22222222} & 240  & 1 & 1\\
\end{array}
\]
\caption{Nilpotent orbits of $E_8$ (part 2/2). \label{tableE8-2}}
\end{table}

\begin{table}
\vspace*{-5em}
\[
\scalebox{.63}{
\xymatrix@=6pt{
&E_8\sp{d}\\
&E_8(a_1)\sp{d}\\
&E_8(a_2)\sp{d}\\
&E_8(a_3)\ns{dl}\sp{dr}\\
E_7\sp{dr} &&E_8(a_4)\sp{dl}\\
& E_8(b_4) \sp{dr}\sp{dl}\\
E_7(a_1)\sp{d} &&E_8(a_5)\sp{dll}\ns{d} \\
E_8(b_5) \sp{drr}\ns{d} && D_7\sp{d} \\
E_7(a_2)\ns{d}\sp{drr} &&E_8(a_6) \sp{d}\\
E_6+A_1\sp{dd}\sp{dr} && D_7(a_1)\sp{dl}\sp{d}\\
&E_8(b_6)\ns{d}\sp{dr}&E_7(a_3)\sp{d}\ns{dr}\\
E_6\sp{dd}&A_7\sp{dr}&E_6(a_1)+A_1\sp{ddll} \sp{d} &D_6\sp{ddl}\\
&&D_7(a_2)\sp{d}\\
E_6(a_1)\sp{d}&&D_5+A_2\sp{dll}\sp{d}\\
E_7(a_4)\sp{d}\sp{drr} &&A_6+A_1\sp{d} \\
D_6(a_1)\sp{dr}\ns{d}&&A_6\sp{dl}\\
D_5+A_1\sp{d}\sp{dr} &E_8(a_7)\ns{d} \\
D_5\sp{dd}&E_7(a_5)\ns{d}\ns{dr} \\
&E_6(a_3)+A_1\sp{dl}\ns{d}\ns{dr}&D_6(a_2)\ns{d}\ns{dl} \\
E_6(a_3) \ns{dr} \sp{dd} &\relax\rigid{A_5+A_1} \sp{d}\ns{dr} & \relax\rigid{D_5(a_1)+A_2} \ns{d} \sp{dr} \\
& A_5 \sp{dd} & \relax\rigid{A_4+A_3} \sp{d} & D_4+A_2 \ar@{-}@/^8pt/[dlll] \sp{dl} \\
D_5(a_1)+A_1 \sp{dd}\sp{dr} && A_4+A_2+A_1\sp{dl} \\
&A_4+A_2\sp{d} \\
D_5(a_1)\ns{dd}\sp{dr} &A_4+2A_1\sp{d}\ns{dr} \\
&A_4+A_1 \sp{dd}\sp{dr} & \relax\rigid{2A_3} \sp{d} \\
D_4+A_1\sp{drr}\sp{dd} && D_4(a_1)+A_2\ns{d} \\
&A_4\sp{dr} &\relax\rigid{A_3+A_2+A_1} \sp{d} \\
D_4\sp{dd} &  & A_3+A_2\sp{d} \\
&& \relax\rigid{D_4(a_1)+A_1} \ns{d} \\
D_4(a_1) \ns{d} && \relax\rigid{A_3+2A_1} \ns{d}\sp{dll} \\
\relax\rigid{A_3+A_1} \sp{dd}\ns{drr} && \relax\rigid{2A_1+2A_1} \sp{d}\\
&&\relax\rigid{2A_2+A_1} \sp{d}\\
A_3\sp{ddr} & & 2A_2 \ns{d} \\
&&\relax\rigid{A_2+3A_1} \sp{dl} \\
&\relax\rigid{A_2+2A_1} \sp{d}\\
&\relax\rigid{A_2+A_1} \sp{dl}\ns{dr} \\
A_2\ns{dr} &&\relax\rigid{4A_1} \sp{dl} \\
&\relax\rigid{3A_1} \sp{d}\\
&\relax\rigid{2A_1}\sp{d}\\
&\relax\rigid{A_1}\sp{d}\\
&\relax\rigid{0}
}}
\qquad
\scalebox{.63}{
\xymatrix@=6pt{
&E_8\sp{d}\\
&E_8(a_1)\sp{d}\\
&E_8(a_2)\sp{d}\\
&E_8(a_3)\sp{d}\\
&E_8(a_4)\sp{d}\\
& E_8(b_4) \sp{dr}\sp{dl}\\
E_7(a_1)\sp{dr} &&E_8(a_5)\sp{dl} \\
&E_8(b_5) \ar@{-}@/_/[dddl]\sp{d}  \\
&E_8(a_6) \sp{d}\\
& D_7(a_1)\sp{d}\sp{dr}\\
E_6\sp{ddd}&E_8(b_6)\ns{d}\sp{d}&E_7(a_3)\sp{dl}\\
&E_6(a_1)+A_1 \sp{dr}\sp{ddl}\\
&&D_7(a_2)\sp{d}\\
E_6(a_1)\sp{d}&&D_5+A_2\sp{dll}\sp{d}\\
E_7(a_4)\sp{d}\sp{drr} &&A_6+A_1\sp{d} \\
D_6(a_1)\sp{dr}\sp{d}&&A_6\sp{dl}\\
D_5\sp{d}&E_8(a_7)\sp{dr}\sp{dl} \\
E_6(a_3)  \sp{d} & &   D_4+A_2\sp{dll} \sp{d} \\
D_5(a_1)+A_1 \sp{d}\sp{drr} && A_4+A_2+A_1\sp{d} \\
D_5(a_1)\sp{ddd}\sp{ddr}&&A_4+A_2\sp{d} \\
 &&A_4+2A_1\sp{dl} \\
&A_4+A_1 \sp{d}\sp{dr} \\
D_4\ar@{-}@/_1pc/[dddr] & D_4(a_1)+A_2\sp{d}  & A_4\sp{dl} \\
&   A_3+A_2\sp{d} \\
 & D_4(a_1)+A_1 \sp{d} \\
& D_4(a_1) \sp{dl}\sp{dr}\\
A_3\sp{dr} & & 2A_2 \sp{dl} \\
&A_2+2A_1 \sp{d}\\
&A_2+A_1 \sp{d} \\
&3A_1 \sp{d}\\
&2A_1\sp{d}\\
&A_1\sp{d}\\
&0
}}
\]
\caption{The Hasse diagrams for  $E_8$. Here, $d(A_3)=E_7(a_1)$, $d(A_4)=E_7(a_3)$.\label{E8}}
\end{table}

\bibliographystyle{ytphys}
\small\baselineskip=.97\baselineskip
\let\bbb\bibitem\def\bibitem{\itemsep1.5pt\bbb}
\bibliography{ref}

\providecommand{\href}[2]{#2}\begingroup\raggedright\begin{thebibliography}{10}

\bibitem{Henningson:2004dh}
M.~Henningson, ``Self-dual strings in six dimensions: Anomalies, the
  {ADE}-classification, and the world-sheet {WZW}-model,''
  \href{http://dx.doi.org/10.1007/s00220-005-1324-7}{{\em Commun. Math. Phys.}
  {\bfseries 257} (2005) 291--302},
\href{http://arxiv.org/abs/hep-th/0405056}{{\ttfamily arXiv:hep-th/0405056}}.
%%CITATION = HEP-TH/0405056;%%.

\bibitem{Witten:1995zh}
E.~Witten, ``Some comments on string dynamics,''
\href{http://arxiv.org/abs/hep-th/9507121}{{\ttfamily arXiv:hep-th/9507121}}.
%%CITATION = HEP-TH/9507121;%%.

\bibitem{Strominger:1995ac}
A.~Strominger, ``Open {$p$}-branes,''
  \href{http://dx.doi.org/10.1016/0370-2693(96)00712-5}{{\em Phys. Lett.}
  {\bfseries B383} (1996) 44--47},
\href{http://arxiv.org/abs/hep-th/9512059}{{\ttfamily arXiv:hep-th/9512059}}.
%%CITATION = HEP-TH/9512059;%%.

\bibitem{Dasgupta:1995zm}
K.~Dasgupta and S.~Mukhi, ``Orbifolds of {M}-theory,''
  \href{http://dx.doi.org/10.1016/0550-3213(96)00070-3}{{\em Nucl. Phys.}
  {\bfseries B465} (1996) 399--412},
\href{http://arxiv.org/abs/hep-th/9512196}{{\ttfamily arXiv:hep-th/9512196}}.
%%CITATION = HEP-TH/9512196;%%.

\bibitem{Witten:1995em}
E.~Witten, ``Five-branes and {M}-theory on an orbifold,''
  \href{http://dx.doi.org/10.1016/0550-3213(96)00032-6}{{\em Nucl. Phys.}
  {\bfseries B463} (1996) 383--397},
\href{http://arxiv.org/abs/hep-th/9512219}{{\ttfamily arXiv:hep-th/9512219}}.
%%CITATION = HEP-TH/9512219;%%.

\bibitem{Vafa:1997mh}
C.~Vafa, ``Geometric origin of {Montonen-Olive} duality,'' {\em Adv. Theor.
  Math. Phys.} {\bfseries 1} (1998) 158--166,
\href{http://arxiv.org/abs/hep-th/9707131}{{\ttfamily arXiv:hep-th/9707131}}.
%%CITATION = HEP-TH/9707131;%%.

\bibitem{Klemm:1996bj}
A.~Klemm, W.~Lerche, P.~Mayr, C.~Vafa, and N.~P. Warner, ``Self-dual strings
  and {$\mathcal{N}=2$} supersymmetric field theory,''
  \href{http://dx.doi.org/10.1016/0550-3213(96)00353-7}{{\em Nucl. Phys.}
  {\bfseries B477} (1996) 746--766},
\href{http://arxiv.org/abs/hep-th/9604034}{{\ttfamily arXiv:hep-th/9604034}}.
%%CITATION = HEP-TH/9604034;%%.

\bibitem{Witten:1997sc}
E.~Witten, ``Solutions of four-dimensional field theories via {M}-theory,''
  \href{http://dx.doi.org/10.1016/S0550-3213(97)00416-1}{{\em Nucl. Phys.}
  {\bfseries B500} (1997) 3--42},
\href{http://arxiv.org/abs/hep-th/9703166}{{\ttfamily arXiv:hep-th/9703166}}.
%%CITATION = HEP-TH/9703166;%%.

\bibitem{Argyres:2007cn}
P.~C. Argyres and N.~Seiberg, ``{S}-duality in {$\mathcal{N}=2$} supersymmetric
  gauge theories,'' \href{http://dx.doi.org/10.1088/1126-6708/2007/12/088}{{\em
  JHEP} {\bfseries 12} (2007) 088},
\href{http://arxiv.org/abs/0711.0054}{{\ttfamily arXiv:0711.0054 [hep-th]}}.
%%CITATION = 0711.0054;%%.

\bibitem{Argyres:2007tq}
P.~C. Argyres and J.~R. Wittig, ``Infinite coupling duals of {$\mathcal{N}=2$}
  gauge theories and new rank 1 superconformal field theories,''
  \href{http://dx.doi.org/10.1088/1126-6708/2008/01/074}{{\em JHEP} {\bfseries
  01} (2008) 074},
\href{http://arxiv.org/abs/0712.2028}{{\ttfamily arXiv:0712.2028 [hep-th]}}.
%%CITATION = 0712.2028;%%.

\bibitem{Gaiotto:2009we}
D.~Gaiotto, ``{$\mathcal{N}=2$} dualities,''
\href{http://arxiv.org/abs/0904.2715}{{\ttfamily arXiv:0904.2715 [hep-th]}}.
%%CITATION = 0904.2715;%%.

\bibitem{Gaiotto:2009gz}
D.~Gaiotto and J.~Maldacena, ``The gravity duals of {$\mathcal{N}=2$}
  superconformal field theories,''
\href{http://arxiv.org/abs/0904.4466}{{\ttfamily arXiv:0904.4466 [hep-th]}}.
%%CITATION = 0904.4466;%%.

\bibitem{Tachikawa:2009rb}
Y.~Tachikawa, ``Six-dimensional {$D_{N}$} theory and four-dimensional {SO-USp}
  quivers,'' \href{http://dx.doi.org/10.1088/1126-6708/2009/07/067}{{\em JHEP}
  {\bfseries 07} (2009) 067},
\href{http://arxiv.org/abs/0905.4074}{{\ttfamily arXiv:0905.4074 [hep-th]}}.
%%CITATION = 0905.4074;%%.

\bibitem{Benini:2009gi}
F.~Benini, S.~Benvenuti, and Y.~Tachikawa, ``Webs of five-branes and
  {$\mathcal{N}{=}2$} superconformal field theories,''
  \href{http://dx.doi.org/10.1088/1126-6708/2009/09/052}{{\em JHEP} {\bfseries
  09} (2009) 052},
\href{http://arxiv.org/abs/0906.0359}{{\ttfamily arXiv:0906.0359 [hep-th]}}.
%%CITATION = 0906.0359;%%.

\bibitem{Nanopoulos:2009xe}
D.~Nanopoulos and D.~Xie, ``{$\mathcal{N}2$} {SU} quiver with {USp} ends or
  {SU} ends with antisymmetric matter,''
  \href{http://dx.doi.org/10.1088/1126-6708/2009/08/108}{{\em JHEP} {\bfseries
  08} (2009) 108},
\href{http://arxiv.org/abs/0907.1651}{{\ttfamily arXiv:0907.1651 [hep-th]}}.
%%CITATION = 0907.1651;%%.

\bibitem{Gaiotto:2009hg}
D.~Gaiotto, G.~W. Moore, and A.~Neitzke, ``Wall-crossing, {H}itchin systems,
  and the {WKB} approximation,''
\href{http://arxiv.org/abs/0907.3987}{{\ttfamily arXiv:0907.3987 [hep-th]}}.
%%CITATION = 0907.3987;%%.

\bibitem{Nanopoulos:2009uw}
D.~Nanopoulos and D.~Xie, ``{H}itchin equation, singularity, and
  {$\mathcal{N}=2$} superconformal field theories,''
  \href{http://dx.doi.org/10.1007/JHEP03(2010)043}{{\em JHEP} {\bfseries 03}
  (2010) 043},
\href{http://arxiv.org/abs/0911.1990}{{\ttfamily arXiv:0911.1990 [hep-th]}}.
%%CITATION = 0911.1990;%%.

\bibitem{Drukker:2010jp}
N.~Drukker, D.~Gaiotto, and J.~Gomis, ``The virtue of defects in {4D} gauge
  theories and {2D CFTs},''
\href{http://arxiv.org/abs/1003.1112}{{\ttfamily arXiv:1003.1112 [hep-th]}}.
%%CITATION = 1003.1112;%%.

\bibitem{Chacaltana:2010ks}
O.~Chacaltana and J.~Distler, ``Tinkertoys for {G}aiotto duality,''
  \href{http://dx.doi.org/10.1007/JHEP11(2010)099}{{\em JHEP} {\bfseries 1011}
  (2010) 099},
\href{http://arxiv.org/abs/1008.5203}{{\ttfamily arXiv:1008.5203 [hep-th]}}.
%%CITATION = ARXIV:1008.5203;%%.

\bibitem{Chacaltana:2011ze}
O.~Chacaltana and J.~Distler, ``Tinkertoys for the {$D_{N}$} series,''
\href{http://arxiv.org/abs/1106.5410}{{\ttfamily arXiv:1106.5410 [hep-th]}}.
%%CITATION = 1106.5410;%%.

\bibitem{Tachikawa:2010vg}
Y.~Tachikawa, ``{$\mathcal{N}=2$} {S}-duality via outer-automorphism twists,''
  \href{http://dx.doi.org/10.1088/1751-8113/44/18/182001}{{\em J. Phys.}
  {\bfseries A44} (2011) 182001},
\href{http://arxiv.org/abs/1009.0339}{{\ttfamily arXiv:1009.0339 [hep-th]}}.
%%CITATION = 1009.0339;%%.

\bibitem{Nishinaka}
T.~Nishinaka, ``{The gravity duals of SO/USp superconformal quivers},''
\href{http://arxiv.org/abs/1202.6613}{{\ttfamily arXiv:1202.6613 [hep-th]}}.
%%CITATION = 1202.6613;%%.

\bibitem{Hanany:2010qu}
A.~Hanany and N.~Mekareeya, ``Tri-vertices and {$SU(2)$}'s,''
  \href{http://dx.doi.org/10.1007/JHEP02(2011)069}{{\em JHEP} {\bfseries 02}
  (2011) 069},
\href{http://arxiv.org/abs/1012.2119}{{\ttfamily arXiv:1012.2119 [hep-th]}}.
%%CITATION = 1012.2119;%%.

\bibitem{Witten:1982fp}
E.~Witten, ``An {$SU(2)$} anomaly,''
\href{http://dx.doi.org/10.1016/0370-2693(82)90728-6}{{\em Phys.Lett.}
  {\bfseries B117} (1982) 324--328}.
%%CITATION = PHLTA,B117,324;%%.

\bibitem{Tachikawa:2011ch}
Y.~Tachikawa, ``On {S}-duality of {5D} super {Yang-Mills} on {$S^1$},''
  \href{http://dx.doi.org/10.1007/JHEP11(2011)123}{{\em JHEP} {\bfseries 11}
  (2011) 123},
\href{http://arxiv.org/abs/1110.0531}{{\ttfamily arXiv:1110.0531 [hep-th]}}.
%%CITATION = 1110.0531;%%.

\bibitem{Gaiotto:2008sa}
D.~Gaiotto and E.~Witten, ``Supersymmetric boundary conditions in
  {$\mathcal{N}=4$} super {Yang-Mills} theory,''
\href{http://arxiv.org/abs/0804.2902}{{\ttfamily arXiv:0804.2902 [hep-th]}}.
%%CITATION = 0804.2902;%%.

\bibitem{Gaiotto:2008ak}
D.~Gaiotto and E.~Witten, ``{S}-duality of boundary conditions in
  {$\mathcal{N}=4$} super {Yang-Mills} theory,''
\href{http://arxiv.org/abs/0807.3720}{{\ttfamily arXiv:0807.3720 [hep-th]}}.
%%CITATION = 0807.3720;%%.

\bibitem{Sommers}
E.~Sommers, ``Lusztig's canonical quotient and generalized duality,''
  \href{http://dx.doi.org/10.1006/jabr.2001.8868}{{\em J. Algebra} {\bfseries
  243} no.~2, (2001) 790--812},
  \href{http://arxiv.org/abs/math.RT/0104162}{{\ttfamily
  arXiv:math.RT/0104162}}.

\bibitem{AcharSommers}
P.~N. Achar and E.~N. Sommers, ``Local systems on nilpotent orbits and weighted
  {D}ynkin diagrams,''
  \href{http://dx.doi.org/10.1090/S1088-4165-02-00174-7}{{\em Represent.
  Theory} {\bfseries 6} (2002) 190--201},
  \href{http://arxiv.org/abs/math/0201248}{{\ttfamily arXiv:math/0201248}}.

\bibitem{Achar}
P.~N. Achar, ``An order-reversing duality map for conjugacy classes in
  {L}usztig's canonical quotient,''
  \href{http://dx.doi.org/10.1007/s00031-003-0422-x}{{\em Transform. Groups}
  {\bfseries 8} no.~2, (2003) 107--145},
  \href{http://arxiv.org/abs/math.RT/0203082}{{\ttfamily
  arXiv:math.RT/0203082}}.

\bibitem{Gukov:2006jk}
S.~Gukov and E.~Witten, ``Gauge theory, ramification, and the geometric
  {L}anglands program,''
\href{http://arxiv.org/abs/hep-th/0612073}{{\ttfamily arXiv:hep-th/0612073}}.
%%CITATION = HEP-TH/0612073;%%.

\bibitem{Gukov:2008sn}
S.~Gukov and E.~Witten, ``Rigid surface operators,''
\href{http://arxiv.org/abs/0804.1561}{{\ttfamily arXiv:0804.1561 [hep-th]}}.
%%CITATION = 0804.1561;%%.

\bibitem{Wyllard:2010rp}
N.~Wyllard, ``{W}-algebras and surface operators in {$\mathcal{N}=2$} gauge
  theories,'' \href{http://dx.doi.org/10.1088/1751-8113/44/15/155401}{{\em J.
  Phys. A} {\bfseries 44} (2011) 155401},
\href{http://arxiv.org/abs/1011.0289}{{\ttfamily arXiv:1011.0289 [hep-th]}}.
%%CITATION = ARXIV:1011.0289;%%.

\bibitem{Wyllard:2010vi}
N.~Wyllard, ``{Instanton Partition Functions in {$\mathcal{N}=2$} {$SU(N)$}
  Gauge Theories with a General Surface Operator, and their W-algebra Duals},''
  \href{http://dx.doi.org/10.1007/JHEP02(2011)114}{{\em JHEP} {\bfseries 1102}
  (2011) 114},
\href{http://arxiv.org/abs/1012.1355}{{\ttfamily arXiv:1012.1355 [hep-th]}}.
%%CITATION = ARXIV:1012.1355;%%.

\bibitem{Tachikawa:2011dz}
Y.~Tachikawa, ``On {W}-algebras and the symmetries of defects of {6D}
  {$\cN=(2,0)$} theory,'' \href{http://dx.doi.org/10.1007/JHEP03(2011)043}{{\em
  JHEP} {\bfseries 03} (2011) 043},
\href{http://arxiv.org/abs/1102.0076}{{\ttfamily arXiv:1102.0076 [hep-th]}}.
%%CITATION = 1102.0076;%%.

\bibitem{Kanno:2011fw}
H.~Kanno and Y.~Tachikawa, ``Instanton counting with a surface operator and the
  chain-saw quiver,'' \href{http://dx.doi.org/10.1007/JHEP06(2011)119}{{\em
  JHEP} {\bfseries 06} (2011) 119},
\href{http://arxiv.org/abs/1105.0357}{{\ttfamily arXiv:1105.0357 [hep-th]}}.
%%CITATION = 1105.0357;%%.

\bibitem{Bais:1987zk}
F.~A. Bais, P.~Bouwknegt, M.~Surridge, and K.~Schoutens, ``Coset construction
  for extended {V}irasoro algebras,''
\href{http://dx.doi.org/10.1016/0550-3213(88)90632-3}{{\em Nucl. Phys.}
  {\bfseries B304} (1988) 371--391}.
%%CITATION = NUPHA,B304,371;%%.

\bibitem{deBoer:1993iz}
J.~de~Boer and T.~Tjin, ``The relation between quantum {W} algebras and {L}ie
  algebras,'' \href{http://dx.doi.org/10.1007/BF02103279}{{\em Commun. Math.
  Phys.} {\bfseries 160} (1994) 317--332},
\href{http://arxiv.org/abs/hep-th/9302006}{{\ttfamily arXiv:hep-th/9302006}}.
%%CITATION = HEP-TH/9302006;%%.

\bibitem{Spaltenstein}
N.~Spaltenstein, {\em Classes unipotentes et sous-groupes de {B}orel}, vol.~946
  of {\em Lecture Notes in Mathematics}.
\newblock Springer, 1982.

\bibitem{Carter}
R.~W. Carter, {\em Finite groups of {L}ie type -- Conjugacy classes and complex
  characters --}.
\newblock Pure and Applied Mathematics. John Wiley \& Sons, 1985.

\bibitem{CollingwoodMcGovern}
D.~H. Collingwood and W.~M. McGovern, {\em Nilpotent orbits in semisimple {L}ie
  algebras}.
\newblock Van Nostrand, 1993.

\bibitem{McGovern}
W.~M. McGovern, ``The adjoint representation and the adjoint action,'' in {\em
  Algebraic quotients. {T}orus actions and cohomology. {T}he adjoint
  representation and the adjoint action}, vol.~131 of {\em Encyclopaedia Math.
  Sci.}, pp.~159--238.
\newblock Springer, 2002.

\bibitem{Benini:2010uu}
F.~Benini, Y.~Tachikawa, and D.~Xie, ``Mirrors of {3D Sicilian} theories,''
  \href{http://dx.doi.org/10.1007/JHEP09(2010)063}{{\em JHEP} {\bfseries 09}
  (2010) 063},
\href{http://arxiv.org/abs/1007.0992}{{\ttfamily arXiv:1007.0992 [hep-th]}}.
%%CITATION = 1007.0992;%%.

\bibitem{BarbaschVogan}
D.~Barbasch and D.~A. Vogan, Jr., ``Unipotent representations of complex
  semisimple groups,'' \href{http://dx.doi.org/10.2307/1971193}{{\em Ann. of
  Math. (2)} {\bfseries 121} no.~1, (1985) 41--110}.

\bibitem{GraafElashvili}
W.~A. de~Graaf and A.~Elashvili, ``Induced nilpotent orbits of the simple {L}ie
  algebras of exceptional type,'' {\em Georgian Math. J.} {\bfseries 16} no.~2,
  (2009) 257--278, \href{http://arxiv.org/abs/0905.2743}{{\ttfamily
  arXiv:0905.2743 [math.RT]}}.

\bibitem{Vogan1}
D.~A. Vogan, Jr., ``The orbit method and primitive ideals for semisimple {L}ie
  algebras,'' in {\em Lie algebras and related topics ({W}indsor, {O}nt.,
  1984)}, vol.~5 of {\em CMS Conf. Proc.}, pp.~281--316.
\newblock Amer. Math. Soc., Providence, RI, 1986.

\bibitem{Vogan2}
D.~A. Vogan, Jr., ``Dixmier algebras, sheets, and representation theory,'' in
  {\em Operator algebras, unitary representations, enveloping algebras, and
  invariant theory ({P}aris, 1989)}, vol.~92 of {\em Progr. Math.},
  pp.~333--395.
\newblock Birkh\"auser Boston, Boston, MA, 1990.

\bibitem{Fu}
B.~Fu, ``Symplectic resolutions for nilpotent orbits,''
  \href{http://dx.doi.org/10.1007/s00222-002-0260-9}{{\em Invent. Math.}
  {\bfseries 151} no.~1, (2003) 167--186},
  \href{http://arxiv.org/abs/math.AG/0205048}{{\ttfamily
  arXiv:math.AG/0205048}}.

\bibitem{Namikawa}
Y.~Namikawa, ``Induced nilpotent orbits and birational geometry,''
  \href{http://dx.doi.org/10.1016/j.aim.2009.05.001}{{\em Adv. Math.}
  {\bfseries 222} no.~2, (2009) 547--564},
  \href{http://arxiv.org/abs/0809.2320}{{\ttfamily arXiv:0809.2320 [math.AG]}}.

\bibitem{covers1}
K.~Liang and L.~Lu, ``Sheets and rigid orbit covers of exceptional {L}ie
  groups,'' \href{http://dx.doi.org/10.1007/BF02883968}{{\em Chinese Sci.
  Bull.} {\bfseries 43} no.~20, (1998) 1702--1707}.

\bibitem{covers2}
K.~Liang, Z.~Hou, and L.~Lu, ``On sheets of orbit covers for classical
  semisimple {L}ie groups,'' \href{http://dx.doi.org/10.1360/02ys9018}{{\em
  Sci. China Ser. A} {\bfseries 45} no.~2, (2002) 155--164}.

\bibitem{Brieskorn}
E.~Brieskorn, ``Singular elements of semi-simple algebraic groups,'' in {\em
  Actes du {Congr\`es} {I}nternational des {Math\'ematiciens} ({N}ice, 1970),
  {T}ome 2}, pp.~279--284.
\newblock Gauthier-Villars, Paris, 1971.

\bibitem{Slodowy}
P.~Slodowy, {\em Simple Singularities and Simple Algebraic Groups}, vol.~815 of
  {\em Lecture Notes in Mathematics}.
\newblock Springer, Berlin, 1980.

\bibitem{Gaiotto:2011xs}
D.~Gaiotto, G.~W. Moore, and Y.~Tachikawa, ``On {6D $\cN=(2,0)$} theory
  compactified on a {R}iemann surface with finite area,''
\href{http://arxiv.org/abs/1110.2657}{{\ttfamily arXiv:1110.2657 [hep-th]}}.
%%CITATION = 1110.2657;%%.

\bibitem{Braverman:2010ef}
A.~Braverman, B.~Feigin, M.~Finkelberg, and L.~Rybnikov, ``A finite analog of
  the {AGT} relation {I}: Finite {$W$}-algebras and quasimaps' spaces,''
  \href{http://dx.doi.org/10.1007/s00220-011-1300-3}{{\em Commun.Math.Phys.}
  {\bfseries 308} (2011) 457--478},
\href{http://arxiv.org/abs/1008.3655}{{\ttfamily arXiv:1008.3655 [math.AG]}}.
%%CITATION = ARXIV:1008.3655;%%.

\bibitem{Benini:2009mz}
F.~Benini, Y.~Tachikawa, and B.~Wecht, ``Sicilian gauge theories and
  {$\mathcal{N}=1$} dualities,''
  \href{http://dx.doi.org/10.1007/JHEP01(2010)088}{{\em JHEP} {\bfseries 01}
  (2010) 088},
\href{http://arxiv.org/abs/0909.1327}{{\ttfamily arXiv:0909.1327 [hep-th]}}.
%%CITATION = 0909.1327;%%.

\bibitem{Alday:2009qq}
L.~F. Alday, F.~Benini, and Y.~Tachikawa, ``{Liouville/Toda} central charges
  from {M5}-branes,''
  \href{http://dx.doi.org/10.1103/PhysRevLett.105.141601}{{\em Phys. Rev.
  Lett.} {\bfseries 105} (2010) 141601},
\href{http://arxiv.org/abs/0909.4776}{{\ttfamily arXiv:0909.4776 [hep-th]}}.
%%CITATION = 0909.4776;%%.

\bibitem{Harvey:1998bx}
J.~A. Harvey, R.~Minasian, and G.~W. Moore, ``Non-{A}belian tensor-multiplet
  anomalies,'' {\em JHEP} {\bfseries 09} (1998) 004,
\href{http://arxiv.org/abs/hep-th/9808060}{{\ttfamily arXiv:hep-th/9808060}}.
%%CITATION = HEP-TH/9808060;%%.

\bibitem{Intriligator:2000eq}
K.~A. Intriligator, ``Anomaly matching and a {Hopf-Wess-Zumino} term in {6D},
  {$\mathcal{N}=(2,0)$} field theories,''
  \href{http://dx.doi.org/10.1016/S0550-3213(00)00148-6}{{\em Nucl. Phys.}
  {\bfseries B581} (2000) 257--273},
\href{http://arxiv.org/abs/hep-th/0001205}{{\ttfamily arXiv:hep-th/0001205}}.
%%CITATION = HEP-TH/0001205;%%.

\bibitem{Yi:2001bz}
P.~Yi, ``Anomaly of (2,0) theories,''
  \href{http://dx.doi.org/10.1103/PhysRevD.64.106006}{{\em Phys. Rev.}
  {\bfseries D64} (2001) 106006},
\href{http://arxiv.org/abs/hep-th/0106165}{{\ttfamily arXiv:hep-th/0106165}}.
%%CITATION = HEP-TH/0106165;%%.

\bibitem{Donagi:1995cf}
R.~Donagi and E.~Witten, ``Supersymmetric {Yang-Mill}s theory and integrable
  systems,'' \href{http://dx.doi.org/10.1016/0550-3213(95)00609-5}{{\em Nucl.
  Phys.} {\bfseries B460} (1996) 299--334},
\href{http://arxiv.org/abs/hep-th/9510101}{{\ttfamily arXiv:hep-th/9510101}}.
%%CITATION = HEP-TH/9510101;%%.

\bibitem{Seiberg:1996nz}
N.~Seiberg and E.~Witten, ``Gauge dynamics and compactification to three
  dimensions,''
\href{http://arxiv.org/abs/hep-th/9607163}{{\ttfamily arXiv:hep-th/9607163}}.
%%CITATION = HEP-TH/9607163;%%.

\bibitem{Shapere:2008zf}
A.~D. Shapere and Y.~Tachikawa, ``Central charges of {$\mathcal{N}=2$}
  superconformal field theories in four dimensions,''
  \href{http://dx.doi.org/10.1088/1126-6708/2008/09/109}{{\em JHEP} {\bfseries
  09} (2008) 109},
\href{http://arxiv.org/abs/0804.1957}{{\ttfamily arXiv:0804.1957 [hep-th]}}.
%%CITATION = 0804.1957;%%.

\bibitem{GuilleminSternberg}
V.~Guillemin and S.~Sternberg, ``The {G}el{'}fand-{C}etlin system and
  quantization of the complex flag manifolds,''
  \href{http://dx.doi.org/10.1016/0022-1236(83)90092-7}{{\em J. Funct. Anal.}
  {\bfseries 52} no.~1, (1983) 106--128}.

\bibitem{MishchenkoFomenko}
A.~S. Mi{\v{s}}{\v{c}}enko and A.~T. Fomenko, ``Integrability of {E}uler's
  equations on semisimple {L}ie algebras,'' {\em Trudy Sem. Vektor. Tenzor.
  Anal.} no.~19, (1979) 3--94.

\bibitem{Bolsinov}
A.~V. Bolsinov, ``Commutative families of functions related to consistent
  {P}oisson brackets,'' \href{http://dx.doi.org/10.1007/BF00047046}{{\em Acta
  Appl. Math.} {\bfseries 24} no.~3, (1991) 253--274}.

\bibitem{CharbonnelMoreau}
J.-Y. Charbonnel and A.~Moreau, ``The index of centralizers of elements of
  reductive {L}ie algebras,'' {\em Doc. Math.} {\bfseries 15} (2010) 387--421,
  \href{http://arxiv.org/abs/1005.0831}{{\ttfamily arXiv:1005.0831 [math.RT]}}.

\bibitem{Gaiotto:2008nz}
D.~Gaiotto, A.~Neitzke, and Y.~Tachikawa, ``Argyres-{S}eiberg duality and the
  {H}iggs branch,'' \href{http://dx.doi.org/10.1007/s00220-009-0938-6}{{\em
  Commun. Math. Phys.} {\bfseries 294} (2010) 389--410},
\href{http://arxiv.org/abs/0810.4541}{{\ttfamily arXiv:0810.4541 [hep-th]}}.
%%CITATION = 0810.4541;%%.

\bibitem{Tachikawa:2011yr}
Y.~Tachikawa and S.~Terashima, ``{Seiberg-Witten} geometries revisited,''
  \href{http://dx.doi.org/10.1007/JHEP09(2011)010}{{\em JHEP} {\bfseries 09}
  (2011) 010},
\href{http://arxiv.org/abs/1108.2315}{{\ttfamily arXiv:1108.2315 [hep-th]}}.
%%CITATION = 1108.2315;%%.

\bibitem{LawtherTesterman}
R.~Lawther and D.~M. Testerman, ``Centres of centralizers of unipotent elements
  in simple algebraic groups,''
  \href{http://dx.doi.org/10.1090/S0065-9266-10-00594-6}{{\em Mem. Amer. Math.
  Soc.} {\bfseries 210} no.~988, (2011) vi+188}.

\end{thebibliography}\endgroup

\end{document}